\documentclass[a4paper,11pt]{article}
\pdfoutput=1
\usepackage{jheppub}
\usepackage{empheq}
\usepackage{booktabs}
\usepackage{cleveref}
\usepackage[T1]{fontenc}
\usepackage[group-digits=false]{siunitx}
\usepackage[subrefformat=parens]{subcaption}

\usepackage{tabularx}
\newcolumntype{C}[1]{>{\centering\arraybackslash}p{#1}}
\usepackage{caption}

\usepackage{xspace}

\newcommand{\Pp}{\mathrm{p}}
\newcommand{\PH}{\mathrm{H}}
\newcommand{\PW}{\mathrm{W}}
\newcommand{\PWp}{\mathrm{W}^+}
\newcommand{\PWm}{\mathrm{W}^-}

\newcommand{\PZ}{\mathrm{Z}}
\newcommand{\PV}{V}
\newcommand{\PB}{B}
\newcommand{\Pj}{\mathrm{j}}

\newcommand{\Plp}{\ell^+}
\newcommand{\Pnulm}{\nu_\ell}
\newcommand{\Pnu}{\nu}
\newcommand{\Pl}{\ell}
\newcommand{\Pu}{\mathrm{u}}
\newcommand{\Pux}{\bar{\mathrm{u}}}
\newcommand{\Pd}{\mathrm{d}}
\newcommand{\Pdx}{\bar{\mathrm{d}}}
\newcommand{\Ps}{\mathrm{s}}
\newcommand{\Pc}{\mathrm{c}}
\newcommand{\Pb}{\mathrm{b}}
\newcommand{\Pt}{\mathrm{t}}

\newcommand{\MH}{M_\PH}
\newcommand{\MVOS}{M_{V}^\text{OS}}
\newcommand{\MWOS}{M_\PW^\text{OS}}
\newcommand{\MW}{M_\PW}
\newcommand{\MZOS}{M_\PZ^\text{OS}}
\newcommand{\MZ}{M_\PZ}
\newcommand{\Mt}{m_\Pt}

\newcommand{\GH}{\Gamma_\PH}
\newcommand{\GVOS}{\Gamma_{V}^\text{OS}}
\newcommand{\GWOS}{\Gamma_\PW^\text{OS}}

\newcommand{\GZOS}{\Gamma_\PZ^\text{OS}}

\newcommand{\Gt}{\Gamma_\Pt}

\newcommand{\GF}{G_\mu}

\newcommand{\gs}{g_\text{s}}
\newcommand{\as}{\alpha_\text{s}}
\newcommand{\rw}{{\mathrm{w}}}
\newcommand{\rr}{{\mathrm{r}}}
\newcommand{\Mv} {M_{V}}
\newcommand{\Gv} {\Gamma_{V}}
\newcommand{\rT} {\mathrm{T}}

\newcommand{\recola}{\textsc{RECOLA}}
\newcommand{\mocanlo}{\textsc{MoCaNLO}}

\def\refeq#1{\mbox{Eq.~(\ref{#1})}}
\def\refeqq#1#2{\mbox{Eq.~(\ref{#1}\textcolor{blue}{#2})}}
\Crefname{equation}{Eq.}{Eqs.}
\def\refeqs#1{\mbox{\Cref{#1}}}
\def\reffi#1{\mbox{Fig.~\ref{#1}}}
\Crefname{figure}{Fig.}{Figs.}
\def\reffis#1{\mbox{\Cref{#1}}}
\def\refta#1{\mbox{Table~\ref{#1}}}
\Crefname{table}{Table}{Tables}

\Crefname{section}{Section}{Sections}
\def\refse#1{\mbox{Section~\ref{#1}}}
\def\refses#1{\mbox{Sections~\ref{#1}}}

\def\citere#1{\mbox{Ref.~\cite{#1}}}
\def\citeres#1{\mbox{Refs.~\cite{#1}}}

\newcommand{\GeV}{\ensuremath{\,\text{GeV}}\xspace}

\title{Double-pole approximation for leading-order semi-leptonic vector-boson scattering at the LHC}

\def\draftdate{\relax}
\def\mda{\relax}
\def\mua{\relax}
\def\mla{\relax}
\def\draft{
\def\thtystars{******************************}
\def\sixtystars{\thtystars\thtystars}
\typeout{}
\typeout{\sixtystars**}
\typeout{* Draft mode!
         For final version remove \protect\draft\space in source file *}
\typeout{\sixtystars**}
\typeout{}
\def\draftdate{\today\\}
\def\mua{\marginpar[\boldmath\hfil$\uparrow$]%
                   {\boldmath$\uparrow$\hfil}\color{black}%
                    \typeout{marginpar: $\uparrow$}\ignorespaces}
\def\mda{\color{red}\marginpar[\boldmath\hfil$\downarrow$]%
                   {\boldmath$\downarrow$\hfil}%
                    \typeout{marginpar: $\downarrow$}\ignorespaces}
\def\mla{\marginpar[\boldmath\hfil$\rightarrow$]%
                   {\boldmath$\leftarrow $\hfil}%
                    \typeout{marginpar: $\leftrightarrow$}\ignorespaces}
\def\Mua{\marginpar[\boldmath\hfil$\Uparrow$]%
                   {\boldmath$\Uparrow$\hfil}\color{black}%
                    \typeout{marginpar: $\uparrow$}\ignorespaces}
\def\Mda{\color{red}\marginpar[\boldmath\hfil$\Downarrow$]%
                   {\boldmath$\Downarrow$\hfil}%
                    \typeout{marginpar: $\downarrow$}\ignorespaces}
\def\Mla{\marginpar[\boldmath\hfil\textcolor{red}{$\Rightarrow$}]%
                   {\boldmath\textcolor{red}{$\Leftarrow $}\hfil}%
                    \typeout{marginpar: $\leftrightarrow$}\ignorespaces}
\overfullrule 5pt
\oddsidemargin 15mm
\marginparwidth 29mm
}

\author[a]{Ansgar Denner,}
\author[a]{Daniele Lombardi,}
\author[a]{Christopher Schwan}

\affiliation[a]{
Institut für Theoretische Physik und Astrophysik,
Universität W\"urzburg,
Emil-Hilb-Weg 22,
97074 W\"urzburg,
Germany}

\emailAdd{ansgar.denner@uni-wuerzburg.de}
\emailAdd{daniele.lombardi@uni-wuerzburg.de}
\emailAdd{christopher.schwan@uni-wuerzburg.de}

\abstract{Measuring vector-boson scattering beyond the fully-leptonic final state is becoming possible at the
  LHC, which demands to have a solid control on the theory predictions
  for all final states of this class of processes.
  In this work we present a full off-shell leading-order calculation for the process
  $\Pp\Pp \to \Plp \Pnulm + 4\Pj$ in two fiducial regions
which are particularly relevant for its experimental  measurement. In addition to
  the fully electroweak order, i.e.\ $\mathcal{O}(\alpha^6)$, we complement our results with $\mathcal{O}(\as\alpha^5)$
  and $\mathcal{O}(\as^2\alpha^4)$ for inclusive predictions. At $\mathcal{O}(\alpha^6)$ we present
  for the first time a systematic treatment of the process in double-pole approximation and we perform a detailed study of its range
  of validity by considering inclusive and differential predictions compared to the full off-shell calculation.}

\begin{document}
\draftdate
\maketitle
\flushbottom

\section{Introduction}
\label{sec:introduction}
The importance of vector-boson scattering (VBS) processes is well established in the particle physics community,
as confirmed by the great attention that they have already received in the past. The interest in VBS is motivated by its high
sensitivity to the electroweak (EW) sector of the standard model (SM), which offers a unique chance to study the EW
symmetry breaking mechanism and the SM as a whole. This will further
corroborate our knowledge of the SM or point to new physics effects.
All that explains why a huge effort has been put into trying to observe VBS by the experimental community at the Large Hadron Collider (LHC), despite its high background and relatively low cross section compared to other LHC reactions.
In view of the upcoming full Run-3 dataset and
of the future high-luminosity (HL) stage of the LHC, a better understanding of VBS processes will be possible. 
Not only will more accurate VBS measurements be provided in the fully leptonic final state, but also
data for all final states will become available.

The ATLAS and CMS collaborations have already performed many studies on VBS processes
in the fully leptonic final states
for all classes of VBS. More precisely, that includes measurements of
same-sign $\PW$~\cite{ATLAS:2014jzl,CMS:2014mra,ATLAS:2016snd,CMS:2017fhs,ATLAS:2019cbr,ATLAS:2023sua},
$\PW\PZ$~\cite{ATLAS:2018mxa,CMS:2019uys,CMS:2020gfh,ATLAS:2024ini},
$\PZ\PZ$~\cite{CMS:2017zmo,ATLAS:2020nlt,CMS:2020fqz,ATLAS:2023dkz},
$\PW\gamma$~\cite{CMS:2020ypo,CMS:2022yrl,ATLAS:2024bho}, and
$\PZ\gamma$~\cite{ATLAS:2019qhm,CMS:2020ioi,CMS:2021gme,ATLAS:2021pdg,ATLAS:2022nru,ATLAS:2023fxh} scattering,
together with an observation 
of $\PW^+\PW^-$~\cite{CMS:2022woe,ATLAS:2024ett} scattering.
Even first measurements of cross sections for the scattering of
polarised same-sign W~bosons
have been performed~\cite{CMS:2020etf}.
More recently, searches for VBS in the semi-leptonic final state have been pursued as well,
to achieve an understanding of the process as comprehensive as possible.
Searches for EW di-boson production with semi-leptonic final states
in association with a high-mass di-jet system have been conducted by
ATLAS~\cite{ATLAS:2019thr} and CMS~\cite{CMS:2019qfk}, while
just a few years ago a first evidence of semi-leptonic VBS processes has been reported
by CMS~\cite{CMS:2021qzz}.

One of the main goals of the rich programme aiming at precise and accurate measurements for VBS is
constraining new physics effects, to which these processes are very sensitive, since they offer
a direct test of the triple and quartic gauge-boson interactions as
well as the vector-boson couplings to the Higgs boson. SM effective field theory (SMEFT)
has nowadays become a standard tool to study effects beyond the SM (BSM) and set limits on anomalous
gauge couplings, which would be captured starting from dimension-6 and dimension-8 operators. Many
studies within and outside the SMEFT framework have been carried out in the last decade in the
context of VBS, for instance
in~\citeres{Delgado:2017cls,Gomez-Ambrosio:2018pnl,Delgado:2019ucx,Araz:2020zyh,Dedes:2020xmo,Ethier:2021ydt,Bellan:2021dcy}.

Along with this tremendous effort from the experimental community to perform such difficult measurements,
a considerable amount of work has been done over the last two decades to improve theory predictions.
This turned out to be a challenging task, owing to the high-multiplicity final state
and the complex resonance structure of the process.
For quite some time next-to-leading-order (NLO) QCD corrections to the EW production (containing the genuine VBS process) have been
computed~\cite{Jager:2006zc,Jager:2006cp,Bozzi:2007ur,Jager:2009xx,Denner:2012dz,Rauch:2016pai}
and matched to parton showers~\cite{Jager:2011ms,Jager:2013mu,Jager:2013iza,Rauch:2016upa,Rauch:2016jxo,Jager:2018cyo,Jager:2024sij}.
On top of that, NLO QCD corrections to the QCD background have been
calculated~\cite{Melia:2010bm,Melia:2011dw,Greiner:2012im,Campanario:2013qba,Campanario:2014ioa,Campanario:2014dpa}
and complemented by parton-shower-matched computations~\cite{Melia:2011tj}. At this order of
accuracy, VBS can be obtained by many different programs like VBFNLO~\cite{Baglio:2014uba},
or proper event generators like
MadGraph5\_aMC@NLO~\cite{Stelzer:1994ta,Alwall:2014hca}, Sherpa~\cite{Sherpa:2019gpd},
and POWHEG-BOX~\cite{Nason:2004rx,Frixione:2007vw,Alioli:2010xd}. It
is worth mentioning that
QCD corrections were originally obtained in the VBS approximation,
where $s$-channel diagrams as
well as the interference of $t$- and $u$-channel diagrams are neglected. Some studies on the quality
 of this approximation have been performed for instance in~\citeres{Denner:2012dz,Ballestrero:2018anz}.
More recently, NLO EW corrections to fully leptonic VBS have been
addressed~\cite{Biedermann:2016yds,Denner:2019tmn,Denner:2020zit,Denner:2022pwc}
without relying on any approximation,
and in~\citere{Chiesa:2019ulk} an event generator for same-sign
$\PW$-boson scattering within the POWHEG framework
has been constructed, which matches NLO EW corrections
to a QED parton shower and interfaces them to a QCD parton
shower. Besides NLO QCD and EW corrections,
all possible NLO contributions of $\mathcal{O}(\as^n\alpha^m)$ with $n+m=7$ have been evaluated
in~\citere{Biedermann:2017bss} for $\PW^+\PW^+$ scattering and in
\citere{Denner:2021hsa} for
VBS into a pair of $\PZ$~bosons.  A more recent line of research focuses on the definition of
polarised cross sections for VBS processes to disentangle the rate of production
of longitudinally polarised vector bosons. Polarised scattering processes are indeed extremely important
and desirable to be measured, owing to their high sensitivity to EW symmetry breaking and
BSM effects. Some results, where a definition of polarised cross section based on the
pole approximation~\cite{Denner:2005fg,Denner:2000bj} was employed, were obtained for VBS at LO
in~\citeres{Ballestrero:2020qgv,Ballestrero:2017bxn,Ballestrero:2019qoy}.

In the context of precise VBS predictions for the LHC, most of the theory efforts have been directed
so far to fully leptonic final states. That is justified both by the simpler structure of the
process as compared to the ones involving at least one hadronically decaying vector boson, and
by their phenomenological relevance in experiments. Now that projections on the HL-LHC operation
expect to collect measurements for semi-leptonic or even fully-hadronic final states (see for
instance~\citeres{Covarelli:2021gyz,BuarqueFranzosi:2021wrv} and references therein),
the corresponding theory calculations will be needed soon.
In~\citere{Ballestrero:2008gf} three out of five LO contributions to $\Pp\Pp \to \Plp \Pnulm + 4\Pj$ have been calculated, i.e.\
$\mathcal{O}(\alpha^6)$, $\mathcal{O}(\as^2\alpha^4)$ and $\mathcal{O}(\as^4\alpha^2)$, for each of which also
top-resonant backgrounds were taken into account.
In \citeres{Jager:2013mu,Jager:2024sij} the $\mathcal{O}(\as \alpha^6)$ corrections to $\Pp \Pp \to \PWp \PWm \Pj\Pj$ and $\Pp \Pp \to \PWp \PZ \Pj\Pj$ were calculated for semi-leptonic final states and matched to QCD parton showers.
In this paper we calculate the $\mathcal{O}(\alpha^6)$ for
$\Pp\Pp \to \Plp \Pnulm + 4\Pj$ for two different fiducial regions, which are inspired by
recent ATLAS and CMS studies~\cite{ATLAS:2018tav,CMS:2021qzz}.
Furthermore, we provide results for the integrated cross sections at $\mathcal{O}(\as^2\alpha^4)$ and $\mathcal{O}(\as\alpha^5)$ in the same fiducial regions.
The main goal of this manuscript, however, is to provide a double-pole
approximation (DPA) for general processes and to assess the quality of such
an approximation for semi-leptonic VBS.

This paper is organised as follows: A detailed description of the
full off-shell calculation is carried out in~\refse{sec:description-of-the-calculation}.
In \refse{sec:pole-approximation} the main result of this manuscript is presented: a 
systematic prescription to compute predictions in a DPA for arbitrary
processes involving two or three resonant vector bosons is outlined in
full generality.  The quality of this approximation
is discussed in detail in~\refse{sec:numerical-results} for the
case of semi-leptonic VBS. Full off-shell results for the
fiducial cross section for the three LO contributions are presented
in~\refse{sec:fiducial-cross-sections}. Moreover, pole-approximated results are
discussed and compared to the full off-shell calculation at the inclusive level for the purely
EW LO, i.e.\ $\mathcal{O}(\alpha^6)$.  A differential analysis of the accuracy of the DPA is then
performed in~\refse{sec:differential-distributions} by considering the impact of the approximation
on VBS-relevant observables.

\section{Description of the calculation}
\label{sec:description-of-the-calculation}

In this article we investigate the process
\begin{equation}
\Pp\Pp \to \Plp \Pnulm + 4\Pj
\label{eq:process}
\end{equation}
at the LHC.
Using $g$ and $\gs$ to denote the EW and strong coupling constants, respectively,
the amplitude for the process in~\refeq{eq:process} receives contributions at
$\mathcal{O}(g^6)$, $\mathcal{O}(\gs g^5)$, $\mathcal{O}(\gs^2 g^4)$, $\mathcal{O}(\gs^3 g^3)$, and $\mathcal{O}(\gs^4 g^2)$. At the squared-amplitude level also five
different leading-order (LO) contributions are present, namely $\mathcal{O}(\alpha^6)$,
$\mathcal{O}(\as\alpha^5)$, $\mathcal{O}(\as^2\alpha^4)$, $\mathcal{O}(\as^3\alpha^3)$, and $\mathcal{O}(\as^4\alpha^2)$,
once all interference terms are properly accounted for.
Note that all orders except the highest and lowest in $\alpha$ include interferences,
\begin{equation}
  \label{eq:amplitudes}
\begin{split}
\mathcal{O}(\alpha^6)      &= \mathcal{O} (g^6) \cdot \mathcal{O} (g^6) \text{,} \\
\mathcal{O}(\as\alpha^5)   &= \mathcal{O} (\gs g^5) \cdot \mathcal{O} (\gs g^5) + \mathcal{O} (g^6) \cdot \mathcal{O} (\gs^2 g^4) \text{,} \\
\mathcal{O}(\as^2\alpha^4) &= \mathcal{O} (\gs^2 g^4) \cdot
\mathcal{O} (\gs^2 g^4) + \mathcal{O} (\gs g^5)\cdot \mathcal{O} (\gs^3 g^3) + \mathcal{O} (g^6) \cdot \mathcal{O} (\gs^4 g^2) \text{,} \\
\mathcal{O}(\as^3\alpha^3) &= \mathcal{O} (\gs^3 g^3) \cdot
\mathcal{O} (\gs^3 g^3) + \mathcal{O} (\gs^2 g^4) \cdot\mathcal{O}(\gs^4 g^2) \text{,} \\
\mathcal{O}(\as^4\alpha^2) &= \mathcal{O} (\gs^4 g^2) \cdot \mathcal{O} (\gs^4 g^2) \text{,}
\end{split}
\end{equation}
where each order is shown as a sum of squared orders (first term) and possible interferences (second/third term).
We note that orders with an odd power in $\gs$ can only exist for partonic channels with an external gluon.
That practically means that amplitudes at $\mathcal{O} (\gs g^5)$ can be constructed only when exactly one external photon and one gluon
are present, whereas the $\mathcal{O} (\gs^3 g^3)$ can appear with either one external photon and three external gluons, or
one external photon, one external gluon and one internal gluon.
The $\mathcal{O}(\alpha^6)$, $\mathcal{O}(\as^2\alpha^4)$, and $\mathcal{O}(\as^4\alpha^2)$ were already computed in a previous work~\cite{Ballestrero:2008gf}.

\subsection{Fully EW contribution}
\label{sec:fully-ew-calculation}

In this paper we focus on the fully EW LO contribution
at $\mathcal{O}(\alpha^6)$. Indeed, this order is the only one that genuinely contains the semi-leptonic VBS
signal we are interested in, and
all other contributions should be considered as background.
As a first step, we provide an independent calculation of the $\mathcal{O}(\alpha^6)$ by fully accounting
for off-shell effects. We perform all calculations with the in-house program \mocanlo, a multichannel Monte Carlo integrator that has already proven suitable for the
evaluation of processes with high-multiplicity final states and an
intricate resonance structure, like the one discussed in this article.
The code is interfaced with \recola~\cite{Actis:2012qn,
  Actis:2016mpe}, which we use to evaluate all appearing tree-level SM matrix elements.
The calculation described in this section
serves as a baseline for the core results of this manuscript, which assess the quality of the double-pole approximation for this
process for the first time, and which are described in the following section. 

Our exact calculation for the $\mathcal{O}(\alpha^6)$ includes all resonant and non-resonant contributions, together with
diagrams involving the Higgs boson. For all unstable particles, the complex-mass scheme is used
\cite{Denner:1999gp,Denner:2005fg,Denner:2006ic,Denner:2019vbn},
resulting in complex input values for the EW boson masses and the EW mixing angle $\theta_{\rw}$:
\begin{equation}
  \label{eq:complex-mass-scheme}
\mu_V^2 = \Mv^2-{\rm i}\Gv \Mv\quad (V={\PW,\PZ})\,,\qquad 
\cos^2\theta_{\rw} = \frac{\mu_{\PW}^2}{\mu_{\PZ}^2}\,.
\end{equation}

With four jets in the final state, which are defined at LO only in terms of quarks, 
the reaction in~\refeq{eq:process} can be initiated at $\mathcal{O}(\alpha^6)$ only by a quark-pair-induced or by a $\gamma\gamma$-induced channel.
Both channels have been included in our results. Since we do not treat photons in the final state as jets, no partonic channel involving final-state photons contributes
to the LO definition of our process. Moreover, at $\mathcal{O}(\alpha^6)$ no gluon can appear, neither in the initial nor in the final state.

In line with previous VBS calculations, we neglect quark mixing and use a unit quark-mixing matrix. Owing to the unitarity of the latter,
this approximation only affects $s$-channel diagrams at LO, which are anyway suppressed with respect to other contributions (see for instance \citere{Denner:2020zit}).

We discard all partonic channels involving external $b$ quarks; contributions with a bottom quark in the initial
state are suppressed by their PDFs, while bottom quarks in the final states can induce top-quark resonances, which would overwhelm the genuine VBS
signal. To avoid the contamination of our signal by top-quark background, we simply drop these contributions throughout by assuming a perfect $b$-jet tagging and veto.

\begin{figure}
\centering
\begin{subfigure}{0.33\textwidth}
\centering
\includegraphics{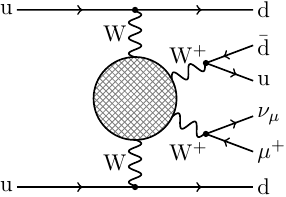}
\caption{VBS $\PWp\PWp$}
\label{fig:born-vbs-wpwp}
\end{subfigure}
\begin{subfigure}{0.33\textwidth}
\centering
\includegraphics{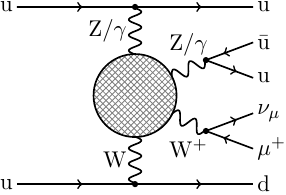}
\caption{VBS $\PWp\PZ$}
\label{fig:born-vbs-wpz}
\end{subfigure}
\begin{subfigure}{0.33\textwidth}
\centering
\includegraphics{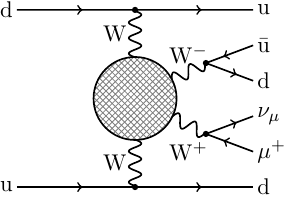}
\caption{VBS $\PWp\PWm$}
\label{fig:born-vbs-wpwm}
\end{subfigure}
\caption{Sample diagrams for the three classes of VBS topologies contributing to the reaction in~\refeq{eq:process}.}
\label{fig:born-vbs-diagrams}
\end{figure}

\begin{figure}
\centering
\begin{subfigure}{0.35\textwidth}
\centering
\includegraphics{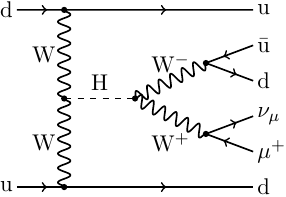}
\caption{VBF with $\PH\rightarrow\PWp\PWm$}
\label{fig:born-higgs-wpwm}
\end{subfigure}
\hspace{20mm}
\begin{subfigure}{0.35\textwidth}
\centering
\includegraphics{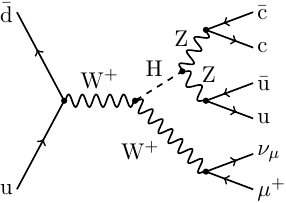}
\caption{Higgs~strahlung with $\PH\rightarrow\PZ\PZ$}
\label{fig:born-higgs-zz}
\end{subfigure}
\caption{Sample diagrams contributing at $\mathcal{O}(\alpha^6)$ mediated by a  Higgs decaying into a pair of gauge bosons.}
\label{fig:born-higgs-diagrams}
\end{figure}

\begin{figure}
\centering
\begin{subfigure}{0.33\textwidth}
\centering
\includegraphics{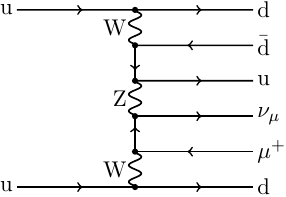}
\caption{non-resonant}
\label{fig:born-non-resonant}
\end{subfigure}
\begin{subfigure}{0.33\textwidth}
\centering
\includegraphics{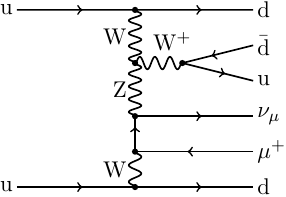}
\caption{singly resonant}
\label{fig:born-single-resonant}
\end{subfigure}
\begin{subfigure}{0.33\textwidth}
\centering
\includegraphics{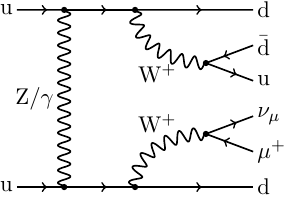}
\caption{doubly resonant}
\label{fig:born-double-resonant}
\end{subfigure}
\caption{Sample diagrams contributing at $\mathcal{O}(\alpha^6)$ for non-VBS topologies involving zero, one, or two resonant gauge bosons.}
\label{fig:born-n-resonant-diagrams}
\end{figure}

\begin{figure}
\centering
\begin{subfigure}{0.33\textwidth}
\centering
\includegraphics{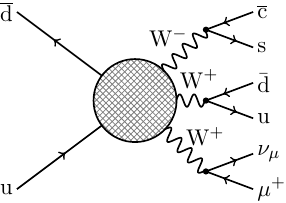}
\caption{$\PWp\PWp\PWm$~production}
\label{fig:born-wpwpwm}
\end{subfigure}
\begin{subfigure}{0.33\textwidth}
\centering
\includegraphics{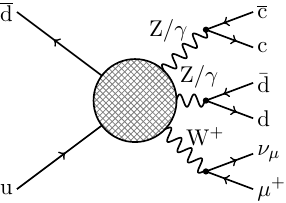}
\caption{$\PWp\PZ\PZ$~production}
\label{fig:born-wpzz}
\end{subfigure}
\begin{subfigure}{0.33\textwidth}
\centering
\includegraphics{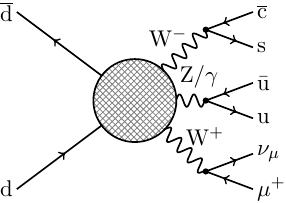}
\caption{$\PWp\PWm\PZ$~production}
\label{fig:born-wpwmz}
\end{subfigure}
\par\bigskip
\begin{subfigure}{0.33\textwidth}
\centering
\includegraphics{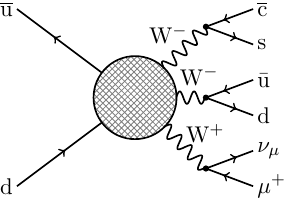}
\caption{$\PWp\PWm\PWm$~production}
\label{fig:born-wpwmwm}
\end{subfigure}
\caption{Sample diagrams of triple gauge-boson production for all combinations allowed by charge conservation,
where the blob stands both for abelian and non-abelian production mechanisms.}
\label{fig:born-vvv-diagrams}
\end{figure}

For semi-leptonic VBS the number of partonic channels and the respective diagram topologies to be considered
is much larger than for fully-leptonic VBS processes, which were computed in Refs.~\cite{Biedermann:2017bss,Denner:2019tmn,Denner:2021csi,Denner:2022pwc}.
The characteristic VBS topology consists of $t$-channel diagrams where two quark lines radiate off two vector bosons with space-like momenta, which
scatter into two vector bosons with time-like momenta, as exemplified
by~\reffis{fig:born-vbs-wpwp,fig:born-vbs-wpz,fig:born-vbs-wpwm}. 
Since in~\refeq{eq:process} only one boson decays leptonically and is forced to be
a $\PWp$ boson, the two time-like vector bosons can be a $\PWp\PWp$ (\reffi{fig:born-vbs-wpwp}), a $\PWp\PZ$ (\reffi{fig:born-vbs-wpz}) and a $\PWp\PWm$ pair (\reffi{fig:born-vbs-wpwm}). In addition to scattering processes through triple or quartic gauge interactions,
a $\PWp\PWm$ and $\PZ\PZ$ pair can also be produced via the decay of
an $s$-channel Higgs boson, which can originate from a vector-boson-fusion (VBF) or Higgs-strahlung mechanism (see~\reffi{fig:born-higgs-wpwm}
and~\reffi{fig:born-higgs-zz}, respectively). 
In the context of a full calculation, on top of the VBS signal all
background topologies of the same order must be accounted
for. Non-VBS-like $t$-channel diagrams are present, and can include
zero (\reffi{fig:born-non-resonant}), one
(\reffi{fig:born-single-resonant}), and two
(\reffi{fig:born-double-resonant}) vector-boson resonances.
Finally, $s$-channel diagrams, where the two incoming quarks are connected by a fermion line, are also allowed for some partonic processes. Within these channels,
topologies involving triple gauge-boson production represent a
relevant part of the irreducible EW background, as illustrated
in~\reffi{fig:born-vvv-diagrams}. 
Owing to charge-conservation
constraints, only four kinds of gauge-boson triplets can be produced, namely $\PWp\PWp\PWm$ (\reffi{fig:born-wpwpwm}), $\PWp\PZ\PZ$ (\reffi{fig:born-wpzz}), $\PWp\PWm\PZ$ (\reffi{fig:born-wpwmz}), and $\PWp\PWm\PWm$ (\reffi{fig:born-wpwmwm}). Their production
can occur via Higgs-strahlung topologies (\reffi{fig:born-higgs-zz}), or by simply connecting three, two, or only one gauge boson to the incoming fermion line.
In the latter case, the reaction proceeds via non-abelian gauge interactions, where an $s$-channel gauge boson decays into
three gauge bosons directly through a quartic vertex, or with a two-step decay chain mediated by triple gauge couplings.

\subsection{Background contributions and their resonance structure}
\label{sec:background}

Furthermore, we evaluated the process in~\refeq{eq:process}
at two more perturbative orders, namely $\mathcal{O}(\as\alpha^5)$ and $\mathcal{O}(\as^2\alpha^4)$.
This allows us to further compare the relative size of the fully LO EW contribution  with some of the
$\as$-enhanced terms. Nevertheless, the new amplitudes entering at these orders
do not contain VBS topologies but rather contribute as a background.

As is clear from~\refeq{eq:amplitudes}, the $\mathcal{O}(\as\alpha^5)$ receives two kinds of contributions. The first one arises from
the product of amplitudes of the same $\mathcal{O}(\gs g^5)$, which requires an incoming photon and an external gluon, either as initial-
or final-state particle. The second contribution to this order is a genuine interference of amplitudes of different coupling orders, i.e.\
$\mathcal{O}(g^6)$ and $\mathcal{O}(\gs^2 g^4)$. In both cases, diagrams with two $s$-channel vector bosons are present and, as for the
fully EW $\mathcal{O}(\alpha^6)$, most of the physics can be captured by a double-pole approximation (DPA), where the momenta of the two
vector-boson resonances are set on~shell (as described in the next section). A pole-approximated computation for this order confirms
such a physical intuition, with the off-shell and on-shell calculations differing by roughly $1\%$ at the integrated level for our definition
of the fiducial region (presented in~\refse{sec:event-selection}). Still, when inspecting results for the individual partonic channels,
differences up to $\sim 50\%$ can be found. Since these large discrepancies affect channels with a tiny relative contribution to the full
cross section, the PA works reasonably well, but one should question
its reliability for the $\mathcal{O}(\as\alpha^5)$ when
considering different selection cuts.
For this reason and owing to its limited
phenomenological relevance, we refrain from showing pole-approximated
results for this contribution and just present numbers
for the integrated cross section in~\refse{sec:fiducial-cross-sections}.

One may wonder whether the DPA can also properly describe the
$\mathcal{O}(\as^2\alpha^4)$ contribution, since doubly-resonant terms
also appear at this order.
However, in this case the DPA leads to differences from the off-shell calculation at the integrated
level of roughly $30$--$40\%$, which can reach more than $70\%$ when the comparison is done for single partonic channels. This outcome suggests
that a large fraction of the background receives a sizeable contribution from singly-resonant topologies.

We first notice that, among the three
different kinds of squared amplitudes entering at $\mathcal{O}(\as^2\alpha^4)$, as shown in~\refeq{eq:amplitudes}, only the first one
can develop double poles. The second term interferes amplitudes belonging to $\mathcal{O}(\gs g^5)$ and $\mathcal{O}(\gs^3 g^3)$.
Since the $\mathcal{O}(\gs g^5)$ can just arise from amplitudes with one external photon and a single external gluon, only amplitudes
of $\mathcal{O}(\gs^3 g^3)$ with one internal and one external gluon contribute, which can have at most one $s$-channel resonance. Similarly,
for the third term the interference of $\mathcal{O}(g^6)$ and $\mathcal{O}(\gs^4 g^2)$ only allows for $\mathcal{O}(\gs^4 g^2)$ amplitudes
with two internal gluons, which can again be at most singly resonant.

\begin{figure}
\centering
\begin{subfigure}{0.33\textwidth}
\centering
\includegraphics{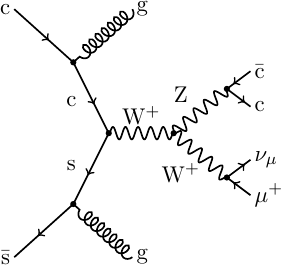}
\label{fig:born-as2a4-dp-1}
\end{subfigure}
\hspace{20mm}
\begin{subfigure}{0.33\textwidth}
\centering
\includegraphics{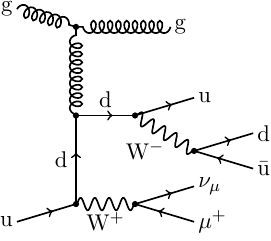}
\label{fig:born-as2a4-dp-2}
\end{subfigure}
\caption{Sample doubly-resonant diagrams contributing at $\mathcal{O}(\as^2\alpha^4)$ for partonic channels involving
  two external gluons (either in the initial or in the final states).}
\label{fig:born-as2a4-dp}
\end{figure}

\begin{figure}
\centering
\begin{subfigure}{0.33\textwidth}
\centering
\includegraphics{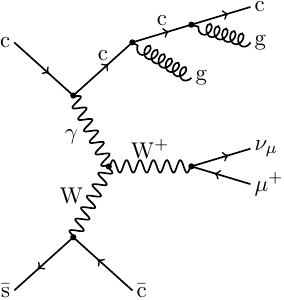}
\label{fig:born-as2a4-sp-1}
\end{subfigure}
\hspace{20mm}
\begin{subfigure}{0.33\textwidth}
\centering
\includegraphics{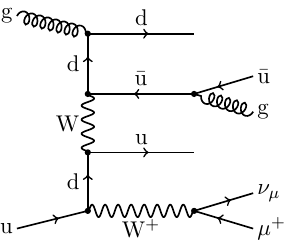}
\label{fig:born-as2a4-sp-2}
\end{subfigure}
\caption{Sample singly-resonant diagrams contributing at $\mathcal{O}(\as^2\alpha^4)$ for partonic channels involving
  two external gluons (either in the initial or in the final states).}
\label{fig:born-as2a4-sp}
\end{figure}

\begin{figure}
\centering
\begin{subfigure}{0.33\textwidth}
\centering
\includegraphics{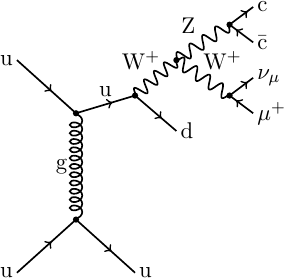}
\label{fig:born-as2a4-q-1}
\end{subfigure}
\hspace{20mm}
\begin{subfigure}{0.33\textwidth}
\centering
\includegraphics{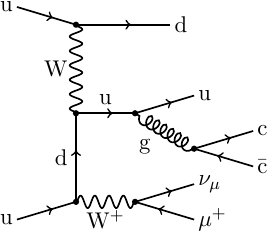}
\label{fig:born-as2a4-q-2}
\end{subfigure}
\caption{Sample doubly- and singly-resonant diagrams contributing at $\mathcal{O}(\as^2\alpha^4)$ for partonic channels involving
  only quarks (and no gluons) as external particles.}
\label{fig:born-as2a4-quarks}
\end{figure}

Doubly-resonant contributions for $\mathcal{O}(\as^2\alpha^4)$ can therefore just be generated when computing the square of amplitudes of the
same order, i.e.\ $\mathcal{O}(\gs^2 g^4)$. The largest contribution 
results from processes with two external gluons, so that the quality of the DPA for this order should be judged on the basis of its ability to
correctly describe these channels. 
By construction, the VBS cuts that we use suppress the QCD
background. As a consequence, many doubly-resonant contributions
of $\mathcal{O}(\gs^2 g^4)$  do not favour two tagging jets needed to
pass the cuts (for examples see \reffi{fig:born-as2a4-dp}). 
On the other hand, there are singly-resonant diagrams (see \reffi{fig:born-as2a4-sp}) involving
vector bosons or photons in the $t$~channel, which are similarly
enhanced as the VBS signal for small momentum transfer, $|t|\ll M^2_{\rm V}$, and thus
contribute a sizeable fraction of the cross section at  $\mathcal{O}(\as^2\alpha^4)$.
For the less important contributions originating from amplitudes with
 no external gluons similar arguments apply (see \reffi{fig:born-as2a4-quarks} for examples
 of doubly- and singly-resonant diagrams for these channels).
Therefore, we do not present pole-approximated results for the
$\mathcal{O}(\as^2\alpha^4)$ contribution but only provide numbers
for the integrated cross section of the off-shell calculation
in~\refse{sec:fiducial-cross-sections}.

\section{Pole approximation}
\label{sec:pole-approximation}

After having discussed in~\refse{sec:description-of-the-calculation} the full off-shell calculation of the process in~\refeq{eq:process},
we now introduce a double-pole approximation~\cite{Denner:2005fg,Denner:2000bj} for the $\mathcal{O}(\alpha^6)$.
At this order, the DPA requires to take into account only those diagrams where at least two specified bosons can become resonant.
Technically, this is achieved by using some features of
\recola~\cite{Actis:2012qn, Actis:2016mpe} to select contributions
involving specific resonances. In this section we
present this approximation in its full generality,
without considering a particular set of selection cuts.
Simplifications of the general case owing to a particular choice of a fiducial phase-space region (defined in \refse{sec:event-selection})
are illustrated in~\refse{sec:pole-approximation-in-specific-setup}.

For convenience and to set up some notation, we recall here some
formulae that are at the core of the pole approximation (PA). We restrict ourselves to the
LO case, which features only
factorisable contributions,%
\footnote{At higher orders, non-factorisable corrections appear \cite{Denner:1997ia,Denner:2000bj,Dittmaier:2014qza,Denner:2019vbn}, with
  a more complicated structure than the one summarized here.} and that is the only one relevant for this work.
Indeed, at LO we can always write the amplitude $\mathcal{M}$ for a process of interest as
\begin{equation}
    \label{eq:amplitude}
      \mathcal{M}=\frac{\mathcal{R}(k^\mathrm{r}_{1},\ldots, k^\mathrm{r}_{n_\mathrm{r}})}{[(k^\mathrm{r}_1)^2-M_1^2]\cdots [(k^\mathrm{r}_{n_\mathrm{r}})^2-M_{n_\mathrm{r}}^2]}+\mathcal{N}(k^\mathrm{r}_{1},\ldots, k^\mathrm{r}_{n_\mathrm{r}})\,,
\end{equation}
where the residue $\mathcal{R}$ summarises all contributions developing resonant parts in the $n_\rr$
resonance momenta $k^\rr_{j_{\mathrm{r}}}$, with $1\le j_{\mathrm{r}}\le
n_\rr$. The propagators of the $n_\rr$ resonances, which have been singled out from
$\mathcal{R}$, are written in terms of the real resonance masses $M_{j_{\mathrm{r}}}$.
The additional contribution $\mathcal{N}$ accounts for all diagrams which are resonant in $n_\rr-1$ or
less resonance momenta.

The PA is based on the pole scheme~\cite{Stuart:1991xk,Aeppli:1993rs,Denner:2019vbn}, which provides a prescription to separate in a gauge-invariant
way the resonant and non-resonant contributions in $\mathcal{M}$. The key idea is to use the gauge invariance
of the  location of the poles of the resonant propagators and of the amplitude residue at the poles, i.e.\ of $\mathcal{R}$
evaluated with on-shell kinematics
$(k^\mathrm{r}_{j_{\mathrm{r}}})^2=M_{j_{\mathrm{r}}}^2$. This allows us to rewrite $\mathcal{M}$ in terms of separately gauge-invariant
terms,
\begin{equation}
  \label{eq:amplitude-pole-scheme}
  \mathcal{M}=\frac{\mathcal{R}(\hat{k}^\mathrm{r}_{1},\ldots, \hat{k}^\mathrm{r}_{n_\mathrm{r}})}{[(k^\mathrm{r}_1)^2-\mu_1^2]\cdots [(k^\mathrm{r}_{n_\mathrm{r}})^2-\mu_{n_\mathrm{r}}^2]}+
  \delta\mathcal{R}(k^\mathrm{r}_{1},\ldots, k^\mathrm{r}_{n_\mathrm{r}})+
  \mathcal{N}(k^\mathrm{r}_{1},\ldots, k^\mathrm{r}_{n_\mathrm{r}})\,,
\end{equation}
where $\mu^2_{j_\mathrm{r}}$, defined
in~\refeq{eq:complex-mass-scheme}, gives the gauge-invariant positions
of the resonant propagator poles. Since only the residue of the
multiple poles
and not $\mathcal{R}$ itself is gauge invariant, the quantity $\mathcal{R}$ has to be computed with on-shell momenta $\hat{k}^\mathrm{r}_{j_\rr}$. Finally, the term $\delta\mathcal{R}$,
which if combined with $\mathcal{N}$ is also gauge invariant, is required so that the full expression of $\mathcal{M}$ is recovered.

Starting from~\refeq{eq:amplitude-pole-scheme}, the pole-approximated
amplitude $\mathcal{M}_{\rm PA}$ is obtained by dropping the
contributions $\delta\mathcal{R}+\mathcal{N}$, which contain less
resonances than the leading term.
They are suppressed with respect
to the leading resonant contributions by a factor $\Gamma_V/\Mv\sim\alpha$ (for $V=\PZ$ or $\PW$) in quantities
that are inclusive in the decay products of all the $n_\rr$ resonances.
The LO expression for $\mathcal{M}_{\rm PA}$ is then obtained as  the first term of~\refeq{eq:amplitude-pole-scheme}:
\begin{equation}
  \label{eq:amplitude-pole-approximation}
  \mathcal{M}^{\rm LO}_{\rm PA}=\frac{\mathcal{R}^{\rm LO}(\hat{k}^\mathrm{r}_{1},\ldots, \hat{k}^\mathrm{r}_{n_\mathrm{r}})}{[(k^\mathrm{r}_1)^2-\mu_1^2]\cdots [(k^\mathrm{r}_{n_\mathrm{r}})^2-\mu_{n_\mathrm{r}}^2]}\,.
\end{equation}
It is worth reminding here that a crucial feature of the PA is that
the off-shell momenta  ${k}^\mathrm{r}_{j_\rr}$ are used for the propagators' denominator of the resonances,
while the on-shell momenta $\hat{k}^\mathrm{r}_{j_\rr}$ for the evaluation of $\mathcal{R}$. Moreover, the off-shell momenta are also used in the definition
of physical observables and at the level of the event selection. 

Starting from~\refeq{eq:amplitude-pole-approximation}, which defines the standard LO PA, the PA for the semi-leptonic VBS considered here
requires some special features, which to our knowledge are needed to be systematically taken into account for the first time:
\begin{enumerate}
\item \label{itm:multiple-pas} \emph{Multiple PAs for a single channel}.
It is possible that a single partonic channel may need multiple PAs to approximate the off-shell matrix element properly.
To illustrate the idea we consider the partonic channel $\Pu \Pdx \to \Plp \Pnulm \Pux \Pd \Pu \Pdx$, which requires at least
the following DPAs:
\begin{enumerate}
\item $\Pu \Pdx \to \PWp(\Plp \Pnulm) \PWm(\Pux \Pd) \Pu \Pdx$ -- ``semi-leptonic opposite-sign WW VBS,''
\item $\Pu \Pdx \to \PWp(\Plp \Pnulm) \PZ(\Pu \Pux)  \Pd \Pdx$ -- ``semi-leptonic WZ VBS,''
\item $\Pu \Pdx \to \PWp(\Plp \Pnulm) \PZ(\Pd \Pdx)  \Pu \Pux$ -- ``semi-leptonic WZ VBS,''
\item $\Pu \Pdx \to \PWp(\Plp \Pnulm) \PWp(\Pu \Pdx) \Pux \Pd$ -- ``semi-leptonic same-sign WW VBS,''
\item $\Pu \Pdx \to \Plp \Pnulm \PZ(\Pux \Pu) \PZ(\Pd \Pdx)$   -- ``fully-hadronic ZZ VBS,''
\item $\Pu \Pdx \to \Plp \Pnulm \PWm(\Pux \Pd) \PWp(\Pu \Pdx)$ -- ``fully-hadronic opposite-sign WW VBS,''
\end{enumerate}
where in our notation particles arising from the decay of an on-shell gauge boson
are enclosed in parentheses, which means the shorthand $\PWp(\Plp \Pnulm)$ denotes the decay $\PWp \to \Plp \Pnulm$
with the $\PWp$ treated on~shell in the PA.
Note that the list of contributions reported above is not complete, and the full set of DPAs that can appear is
discussed in~\refse{sec:all-possible-pas}, where we
present a systematic way of computing a DPA for the process given~\refeq{eq:process}.

\item \emph{``Nested'' PAs}.
We refer to  a PA as ``nested'' if some of the resonant propagators share one or more final-state momenta, meaning they are not independent.
Continuing with the illustrative channel $\Pu \Pdx \to \Plp \Pnulm \Pux \Pd \Pu \Pdx$, we can write down a nested PA with a resonant Higgs and a resonant $\PWm$ boson
as shown in the diagram of~\reffi{fig:born-higgs-zz}.
The Higgs boson is produced in $\Pu \Pdx \to \PH \Pu \Pdx$ (VBF Higgs production and/or associated Higgs production) and subsequently decays as $\PH \to \Plp \Pnulm \PWm$, in which the $\PWm$ boson is produced resonantly and in turn decays hadronically as $\PWm \to \Pux \Pd$.
It is clear that the momenta of the resonant Higgs and the $\PWm$ boson are not independent, which is also reflected in the fact that diagrammatically the Higgs-boson decay $\PH \to \Plp \Pnulm \PWm$ is ``nested'' between its production in $\Pu \Pdx \to \PH \Pu \Pdx$ and the W-boson decay $\PWm \to \Pux \Pd$.

Technically, nested PAs require  to take into account a more complicated diagram selection at the level of the matrix-element generation, since the nested part of the reaction (the
decay of the Higgs boson $\PH \to \Plp \Pnulm \PWm$ in our example) can be attributed both to the ``production'' and the ``decay'' part of the process. This refined diagram selection
has been implemented and made available in \recola~v1.4.4.
Another complication arising for nested PAs  is the on-shell projection, which becomes more involved due to two or more particles, whose momenta are not independent,  required being on~shell.
We describe a general on-shell projection that can be applied with an arbitrary number of resonances, with and without nesting, in~\refse{sec:a-general-on-shell-projection}.

\item \label{itm:overcounting} \emph{Overcounting higher-resonant contributions}.
  In general, it is possible that some pole-approximated matrix elements, where the momenta for all resonances required by the approximation have been set on~shell,
  feature additional propagators that can become resonant. To be more explicit, using again our illustrative channel, each DPA in \cref{itm:multiple-pas} includes a triply-resonant contribution.
  By listing the set of triple-pole approximations (TPA) in the same order of the DPA list presented in \cref{itm:multiple-pas}, we have:
\begin{enumerate}
\item $\Pu \Pdx \to \PWp(\Plp \Pnulm) \PWm(\Pux \Pd) \PWp(\Pu \Pdx)$,
\item $\Pu \Pdx \to \PWp(\Plp \Pnulm) \PZ(\Pu \Pux) \PZ(\Pd \Pdx)$,
\item $\Pu \Pdx \to \PWp(\Plp \Pnulm) \PZ(\Pd \Pdx) \PZ(\Pu \Pux)$,
\item $\Pu \Pdx \to \PWp(\Plp \Pnulm) \PWp(\Pu \Pdx) \PWm(\Pux \Pd)$,
\item $\Pu \Pdx \to \PWp(\Plp \Pnulm) \PZ(\Pux \Pu) \PZ(\Pd \Pdx)$,
\item $\Pu \Pdx \to \PWp(\Plp \Pnulm) \PWm(\Pux \Pd) \PWp(\Pu \Pdx)$.  
\end{enumerate}
It is clear that by simply summing over the set of DPAs in \cref{itm:multiple-pas}, the two distinct TPAs that
appear in the list above are counted three times. 
 \refse{sec:overcounting-of-higher-resonant-contributions} explains a systematic handling of this overcounting issue in more detail.

\item \emph{Singularities}.
  As mentioned at the beginning of~\refse{sec:description-of-the-calculation}, our calculation is performed
  with a prescription in which massive unstable particles have complex masses,
  which regulate the phase-space singularities when resonances become on~shell.
  Within a pole approximation, however, the decay widths entering the definition of the complex masses are usually kept only in resonant propagator denominators,
  while they are set to zero everywhere else.
  This is needed to preserve gauge invariance, since the on-shell projection maps the resonances' invariants to their real masses when replacing
  $\{ k^{\rm r}_j \}_{j=1}^{n_{\rm r}}\rightarrow\{ \hat{k}^{\rm r}_j \}_{j=1}^{n_{\rm r}}$
  in $\mathcal{R}$ in~\refeq{eq:amplitude-pole-approximation}.
  Unfortunately, whenever we have a matrix element in DPA that contains an additional propagator that can become resonant, this procedure reintroduces phase-space
  singularities that were cured by non-zero decay widths.

  This complication, which arises in all partonic channels whose DPAs contain TPAs, requires to use non-zero decay widths in all parts of the pole-approximated
  amplitude in order to be able to numerically evaluate these problematic contributions. Even if not explicitly needed, we decided to adopt this choice
  throughout our calculation, and therefore also for those partonic channels whose DPAs do not develop additional resonant propagators. That means 
  we obtain our results in the complex-mass scheme defined by~\refeq{eq:complex-mass-scheme} also when using the  PA, and not just for the fully off-shell
    computation. Even though this is
  known to formally break gauge invariance, we have verified that the numerical impact of this choice at the level of the fiducial cross section is
  statistically irrelevant and therefore under good numerical control.

\end{enumerate}

\subsection{A systematic double-pole approximation for \texorpdfstring{$\Pp \Pp \to \Plp \Pnulm + 4 \Pj$}{pp -> l+ vl + 4j}}
\label{sec:all-possible-pas}

Our aim here is to write down a systematic DPA of the process defined in~\refeq{eq:process}.
This is a matter of assigning two bosons from the set $\{ \PWp, \PWm, \PZ, \PH \}$ to the given final state by considering all possible $1 \to 2$ and $1 \to 4$ decays.
For $1 \to 4$ decays we may have nested resonances, but we only need to consider decays of heavier bosons into lighter ones, which means the decays $\PH \to \PWp \PWm$, $\PH \to \PZ \PZ$ and $\PZ \to \PWp \PWm$.
All other combinations decay a lighter boson into heavier ones, which prevents the propagators of the inner resonance and of one of its
decay products from being  simultaneously on~shell.
Moreover, decays of the type $1 \to 3$ are not possible at $\mathcal{O} (\alpha^6)$ owing to the absence of gluons and photons in the final state.
In the case of the Higgs boson we also neglect decays of the type $1
\to 2$, which are suppressed by the Yukawa coupling of the Higgs boson to fermions.

Given these organising principles, we divide all DPAs into five classes, depending on whether the two final-state leptons are assigned to
a resonance (\emph{semi-leptonic DPA}) or not  (\emph{fully hadronic
  DPA}), whether only particle pairs are set on~shell (DPA with $1 \to 2$ decays) or
also one group of four final-state particles is assigned to a
resonance (DPA with $1 \to 4$ decay), and finally whether the two resonances are nested or not.
Therefore, we have the ``semi-leptonic DPAs with $1 \to 2$ decays,'' which include all VBS diagrams,
\begin{subequations}
\begin{alignat}{2}
\Pp \Pp &\to \PWp(\Plp \Pnulm) \PWp(\Pj \Pj) \Pj \Pj \text{,}       \label{eq:sl-wpwp} \\
\Pp \Pp &\to \PWp(\Plp \Pnulm) \PZ(\Pj \Pj) \Pj \Pj  \text{,}       \label{eq:sl-wpz} \\
\Pp \Pp &\to \PWp(\Plp \Pnulm) \PWm(\Pj \Pj) \Pj \Pj \text{,}       \label{eq:sl-wpwm} \\
\intertext{the ``semi-leptonic DPAs with $1 \to 4$ decay'', involving Higgs-strahlung configurations or di-boson-like topologies, where one boson further decays into a
  gauge-boson pair,}
\Pp \Pp &\to \PWp(\Plp \Pnulm) \PH( \Pj \Pj \Pj \Pj) \text{,}       & \Pp \Pp &\to \PWp(\Plp \Pnulm) \PZ( \Pj \Pj \Pj \Pj) \text{,} \label{eq:wp-higgs-4} \\
\Pp \Pp &\to \PWp(\Plp \Pnulm) \PWm( \Pj \Pj \Pj \Pj)\text{,}       \label{eq:wp-wm-4} \\
\Pp \Pp &\to \PWp(\Plp \Pnulm \Pj \Pj) \PZ(\Pj \Pj)  \text{,}       \label{eq:dpa-wp-4-z} \\
\Pp \Pp &\to \PWp(\Plp \Pnulm \Pj \Pj) \PWm(\Pj \Pj) \text{,}       \label{eq:dpa-wp-4-wm} \\
\Pp \Pp &\to \PH(\Plp \Pnulm \Pj \Pj) \PWp(\Pj \Pj)  \text{,}       & \Pp \Pp &\to \PZ(\Plp \Pnulm \Pj \Pj) \PWp(\Pj \Pj)  \text{,} \label{eq:dpa-higgs-wp} \\
\Pp \Pp &\to \PH(\Plp \Pnulm \Pj \Pj) \PZ(\Pj \Pj)   \text{,}       & \Pp \Pp &\to \PZ(\Plp \Pnulm \Pj \Pj) \PZ(\Pj \Pj)   \text{,} \label{eq:dpa-higgs-z} \\
\Pp \Pp &\to \PH(\Plp \Pnulm \Pj \Pj) \PWm(\Pj \Pj)  \text{,}       & \Pp \Pp &\to \PZ(\Plp \Pnulm \Pj \Pj) \PWm(\Pj \Pj)  \text{,} \label{eq:dpa-higgs-wm} \\
\intertext{the ``semi-leptonic DPAs with nested $1 \to 4$ decay'', whose on-shell requirements mostly select vector-boson-fusion diagrams,}
\Pp \Pp &\to \PH( \PWp(\Plp \Pnulm) \Pj \Pj) \Pj \Pj \text{,}       & \Pp \Pp &\to \PZ( \PWp(\Plp \Pnulm) \Pj \Pj) \Pj \Pj \text{.} \label{eq:sl-higgs-wp} \\
\Pp \Pp &\to \PH( \Plp \Pnulm \PWm(\Pj \Pj)) \Pj \Pj \text{,}       & \Pp \Pp &\to \PZ( \Plp \Pnulm \PWm(\Pj \Pj)) \Pj \Pj \text{,} \label{eq:sl-higgs-wm} \\
\intertext{the ``fully hadronic DPAs with $1 \to 2$ decays'', comprising diagrams for the production of a gauge-boson pair together with a $\Plp \Pnulm$ pair,}
\Pp \Pp &\to \Plp \Pnulm \PWm(\Pj \Pj) \PWp(\Pj \Pj) \text{,} \quad \label{eq:fh-wpwm} \\
\Pp \Pp &\to \Plp \Pnulm \PWm(\Pj \Pj) \PZ(\Pj \Pj)  \text{,}       \label{eq:fh-wmz} \\
\Pp \Pp &\to \Plp \Pnulm \PWm(\Pj \Pj) \PWm(\Pj \Pj) \text{,}       \label{eq:fh-wmwm} \\
\Pp \Pp &\to \Plp \Pnulm \PZ(\Pj \Pj) \PZ(\Pj \Pj)   \text{,}       \label{eq:fh-zz} \\
\intertext{and finally the ``fully hadronic DPAs with nested $1 \to 4$ decay'', selecting diagrams for the production of a Higgs or $\PZ$ boson accompanied by a $\Plp \Pnulm$ pair,}
\Pp \Pp &\to \Plp \Pnulm \PH( \PWp(\Pj \Pj) \Pj \Pj) \text{,}       & \Pp \Pp &\to \Plp \Pnulm \PZ( \PWp(\Pj \Pj) \Pj \Pj) \text{,} \label{eq:fh-higgs-wp} \\
\Pp \Pp &\to \Plp \Pnulm \PH( \PZ(\Pj \Pj) \Pj \Pj)                                                                        \text{,} \label{eq:fh-higgs-z} \\
\Pp \Pp &\to \Plp \Pnulm \PH( \PWm(\Pj \Pj) \Pj \Pj) \text{,}       & \Pp \Pp &\to \Plp \Pnulm \PZ( \PWm(\Pj \Pj) \Pj \Pj) \text{.} \label{eq:fh-higgs-wm}
\end{alignat}
\label{eq:dpas}
\end{subequations}
Clearly, depending on the charges and the flavours of the particles
present at LO, some classes of DPAs are not allowed for some partonic processes. Moreover,
within a given class, not all combinations are possible. Therefore, our first step is to identify for each partonic channel contributing to the process in~\refeq{eq:process}
how many and which DPAs from the list above are needed to properly approximate the channel itself. Channels where no DPA is possible are simply discarded.

\subsubsection{Overcounting of higher-resonant contributions}
\label{sec:overcounting-of-higher-resonant-contributions}

As outlined at the beginning of this section, by summing up the allowed DPA contributions from~\refeq{eq:dpas} for a given partonic channel
we can overcount triply-resonant contributions contained in DPAs.
The only DPAs never affected by this issue are the ones listed in~\refeqs{eq:wp-wm-4,eq:dpa-wp-4-z,eq:dpa-wp-4-wm}, owing to the relation $M_{\PW} < M_{\PZ}$,
which prevents an on-shell $\PW$~boson from decaying into another on-shell boson.
All other DPAs can potentially contain triply-resonant contributions, whenever the charge and the flavour of the two partons not assigned to an on-shell resonance
allow for that. We can group these TPA contributions into unnested TPAs, which include genuine triple vector-boson production processes,
\begin{subequations}
\begin{align}
\Pp \Pp &\to \PWp(\Plp \Pnulm) \PWp(\Pj \Pj) \PWm(\Pj \Pj)                                                                       \text{,} \label{eq:tpa-wpwpwm} \\
\Pp \Pp &\to \PWp(\Plp \Pnulm) \PWm(\Pj \Pj) \PZ(\Pj \Pj)                                                                        \text{,} \label{eq:tpa-wpwmz} \\
\Pp \Pp &\to \PWp(\Plp \Pnulm) \PWm(\Pj \Pj) \PWm(\Pj \Pj)                                                                       \text{,} \label{eq:tpa-wpwmwm} \\
\Pp \Pp &\to \PWp(\Plp \Pnulm) \PZ(\Pj \Pj) \PZ(\Pj \Pj)                                                                         \text{,} \label{eq:tpa-wpzz} \\
\intertext{and nested TPAs, which mostly comprise Higgs-strahlung
  topologies and  di-boson contributions with a Z~boson further
  decaying into a gauge boson and two fermions,}
\Pp \Pp &\to \PWp(\Plp \Pnulm) \PH( \PWp(\Pj \Pj) \Pj \Pj) \text{,} & \Pp \Pp &\to \PWp(\Plp \Pnulm) \PZ( \PWp(\Pj \Pj) \Pj \Pj) \text{,} \label{eq:tpa-fh-higgs-wm-wp} \\
\Pp \Pp &\to \PWp(\Plp \Pnulm) \PH( \PZ(\Pj \Pj) \Pj \Pj)                                                                        \text{,} \label{eq:tpa-fh-higgs-wm-z} \\
\Pp \Pp &\to \PWp(\Plp \Pnulm) \PH( \PWm(\Pj \Pj) \Pj \Pj) \text{,} & \Pp \Pp &\to \PWp(\Plp \Pnulm) \PZ( \PWm(\Pj \Pj) \Pj \Pj) \text{,} \label{eq:tpa-fh-higgs-wm-wm} \\
\Pp \Pp &\to \PH( \PWp(\Plp \Pnulm) \Pj \Pj) \PWp(\Pj \Pj) \text{,} & \Pp \Pp &\to \PZ( \PWp(\Plp \Pnulm) \Pj \Pj) \PWp(\Pj \Pj) \text{,} \label{eq:tpa-sl-higgs-wp-wp} \\
\Pp \Pp &\to \PH( \PWp(\Plp \Pnulm) \Pj \Pj) \PZ(\Pj \Pj)  \text{,} & \Pp \Pp &\to \PZ( \PWp(\Plp \Pnulm) \Pj \Pj) \PZ(\Pj \Pj)  \text{,} \label{eq:tpa-sl-higgs-wp-z} \\
\Pp \Pp &\to \PH( \PWp(\Plp \Pnulm) \Pj \Pj) \PWm(\Pj \Pj) \text{,} & \Pp \Pp &\to \PZ( \PWp(\Plp \Pnulm) \Pj \Pj) \PWm(\Pj \Pj) \text{,} \label{eq:tpa-sl-higgs-wp-wm} \\
\Pp \Pp &\to \PH( \Plp \Pnulm \PWm(\Pj \Pj)) \PWp(\Pj \Pj) \text{,} & \Pp \Pp &\to \PZ( \Plp \Pnulm \PWm(\Pj \Pj)) \PWp(\Pj \Pj) \text{,} \label{eq:tpa-sl-higgs-wm-wp} \\
\Pp \Pp &\to \PH( \Plp \Pnulm \PWm(\Pj \Pj)) \PZ(\Pj \Pj)  \text{,} & \Pp \Pp &\to \PZ( \Plp \Pnulm \PWm(\Pj \Pj)) \PZ(\Pj \Pj)  \text{,} \label{eq:tpa-sl-higgs-wm-z} \\
\Pp \Pp &\to \PH( \Plp \Pnulm \PWm(\Pj \Pj)) \PWm(\Pj \Pj) \text{,} & \Pp \Pp &\to \PZ( \Plp \Pnulm \PWm(\Pj \Pj)) \PWm(\Pj \Pj) \text{.} \label{eq:tpa-sl-higgs-wm-m}
\end{align}
\label{eq:tpas}
\end{subequations}
It turns out that each of these TPA contributions is contained in three different DPAs (as illustrated in detail in~\refta{tab:overcounting}), which
leads to an overcounting factor of two. Therefore, to avoid this overcounting, for each partonic channel one has to subtract from
the sum of all required DPAs the corresponding TPA twice.
This practical rule holds true also for the cases in \mbox{Eqs.~(\ref{eq:tpa-wpwmwm})} and\mbox{~(\ref{eq:tpa-wpzz})}, where the overcounting is a bit more involved.
Indeed, for these TPAs we can distinguish between two cases, namely where the vector-boson decays are either distinguishable or indistinguishable.

To illustrate how that works, we focus on the TPA in~\refeq{eq:tpa-wpzz}. We consider the partonic channel $\Pp \Pp \to \Plp \Pnulm\Pu \Pux\Pd \Pdx$,
which is part of~\refeq{eq:tpa-wpzz}, i.e.\ $\Pp \Pp \to \PWp(\Plp \Pnulm) \PZ(\Pu \Pux) \PZ(\Pd \Pdx)$, and has distinguishable decays. For this channel
we find the following DPAs
\begin{subequations}
\begin{align}
\Pp \Pp &\to \PWp(\Plp \Pnulm) \PZ(\Pu \Pux)     \Pd \Pdx  \text{,} \\
\Pp \Pp &\to \PWp(\Plp \Pnulm) \PZ(\Pd \Pdx)     \Pu \Pux  \text{,} \\
\Pp \Pp &\to      \Plp \Pnulm  \PZ(\Pd \Pdx) \PZ(\Pu \Pux) \text{,}
\end{align}
\end{subequations}
where each of them contains the TPA of interest once. This can be read off~\refta{tab:overcounting}, where  these DPAs
are referred to as~\refeq{eq:sl-wpz} for the first contribution, again as~\refeq{eq:sl-wpz} for the second contribution and finally as~\refeq{eq:fh-zz} for the third contribution.
In this final-state configuration the overcounting factor of two naturally arises. 
In the case of indistinguishable decays, we can consider for instance the partonic channel $\Pp \Pp \to \Plp \Pnulm \Pu \Pux \Pu \Pux$, where our TPA reads
$\Pp \Pp \to \PWp(\Plp \Pnulm) \PZ(\Pu \Pux) \PZ(\Pu \Pux)$, and where we find the following two DPAs
\begin{subequations}
\begin{align}
\Pp \Pp &\to \PWp(\Plp \Pnulm) \PZ(\Pu \Pux)     \Pu \Pux  \text{,} \label{eq:indistinguishable-wpzz} \\
\Pp \Pp &\to      \Plp \Pnulm  \PZ(\Pu \Pux) \PZ(\Pu \Pux) \text{,} \label{eq:indistinguishable-lpzz}\,
\end{align}
\end{subequations}
which again contain our TPA once each. If we label the indistinguishable particles with indices, then the sum of squared matrix elements
for~\refeq{eq:indistinguishable-wpzz} is identical to
\begin{equation}
\frac{1}{2} \bigl| \mathcal{M} \bigl( \Pp \Pp \to \PWp(\Plp \Pnulm) \PZ(\Pu_1 \Pux_1) \Pu_2 \Pux_2 \bigr) \bigr|^2 + \frac{1}{2} \bigl| \mathcal{M} \bigl( \Pp \Pp \to \PWp(\Plp \Pnulm) \PZ(\Pu_2 \Pux_2) \Pu_1 \Pux_1 \bigr) \bigr|^2 \text{,}
\end{equation}
where the symmetry factors arise from the possibility to attribute to the resonant $\PZ$~boson the two indistinguishable quark pairs $(\Pu_1 \Pux_1)$ and $(\Pu_2 \Pux_2)$.
If we consider the squared matrix element for~\refeq{eq:indistinguishable-lpzz}, we can immediately realise that an identical symmetry factor is needed, due to the
two indistinguishable resonant $\PZ$~bosons. Therefore, since we have three different DPA squared-matrix-element contributions with the same symmetry factors,
 we find again that the corresponding TPA is counted three times.

\begin{table}
  \centering
  \parbox{.45\linewidth}{\centering
\begin{tabular}{@{}l|lll@{}}
\toprule
TPA & DPA1 & DPA2 & DPA3 \\
\midrule
\ref{eq:tpa-wpwpwm} & \ref{eq:sl-wpwp} & \ref{eq:sl-wpwm} & \ref{eq:fh-wpwm} \\
\ref{eq:tpa-wpwmz}  & \ref{eq:sl-wpwm} & \ref{eq:sl-wpz}  & \ref{eq:fh-wmz}  \\
\ref{eq:tpa-wpwmwm} & \ref{eq:sl-wpwm} & \ref{eq:sl-wpwm} & \ref{eq:fh-wmwm} \\
\ref{eq:tpa-wpzz}   & \ref{eq:sl-wpz}  & \ref{eq:sl-wpz}  & \ref{eq:fh-zz}   \\
\bottomrule
\end{tabular}
  \caption*{(a)\label{tab:overcounting_a}}
}
\qquad
\parbox{.45\linewidth}{\centering
\begin{tabular}{@{}l|lll@{}}
\toprule
TPA & DPA1 & DPA2 & DPA3 \\
\midrule
\ref{eq:tpa-fh-higgs-wm-wp} & \ref{eq:sl-wpwp} & \ref{eq:wp-higgs-4} & \ref{eq:fh-higgs-wp} \\
\ref{eq:tpa-fh-higgs-wm-z}  & \ref{eq:sl-wpz}  & \ref{eq:wp-higgs-4} & \ref{eq:fh-higgs-z} \\
\ref{eq:tpa-fh-higgs-wm-wm} & \ref{eq:sl-wpwm} & \ref{eq:wp-higgs-4} & \ref{eq:fh-higgs-wm} \\
\\
\bottomrule
\end{tabular}
\caption*{(b)\label{tab:overcounting_b}}
}
\vspace{0.5cm}

\parbox{.45\linewidth}{\centering
\begin{tabular}{@{}l|llc@{}}
\toprule
TPA & DPA1 & DPA2 & DPA3 \\
\midrule
\ref{eq:tpa-sl-higgs-wp-wp} & \ref{eq:sl-wpwp} & \ref{eq:sl-higgs-wp} & \ref{eq:dpa-higgs-wp} \\
\ref{eq:tpa-sl-higgs-wp-z}  & \ref{eq:sl-wpz}  & \ref{eq:sl-higgs-wp} & \ref{eq:dpa-higgs-z} \\
\ref{eq:tpa-sl-higgs-wp-wm} & \ref{eq:sl-wpwm} & \ref{eq:sl-higgs-wp} & \ref{eq:dpa-higgs-wm} \\
\bottomrule
\end{tabular}
\caption*{(c)\label{tab:overcounting_c}}
}
\qquad
\parbox{.45\linewidth}{\centering
\begin{tabular}{@{}l|llc@{}}
\toprule
TPA & DPA1 & DPA2 & DPA3 \\
\midrule
\ref{eq:tpa-sl-higgs-wm-wp} & \ref{eq:sl-higgs-wm} & \ref{eq:fh-wpwm} & \ref{eq:dpa-higgs-wp} \\
\ref{eq:tpa-sl-higgs-wm-z}  & \ref{eq:sl-higgs-wm} & \ref{eq:fh-wmz}  & \ref{eq:dpa-higgs-z} \\
\ref{eq:tpa-sl-higgs-wm-m}  & \ref{eq:sl-higgs-wm} & \ref{eq:fh-wmwm} & \ref{eq:dpa-higgs-wm} \\
\bottomrule
\end{tabular}
\caption*{(d)\label{tab:overcounting_d}}
}
\caption{Overcounting of TPA contributions in DPAs.
  The left-most column of each table reports the equation number of the TPAs listed in~\refeq{eq:tpas},
  and the following equation numbers in the same line list the DPAs that contain that TPA.
  Every TPA is contained three times in the sum of all DPAs and thus has to be subtracted twice.}
\label{tab:overcounting}
\end{table}

\subsection{A general on-shell projection}
\label{sec:a-general-on-shell-projection}

A PA always requires a mapping of the $n$ off-shell momenta, $\{ k_i \}_{i=1}^n$, which are computed for each phase-space point by the phase-space generator, to on-shell momenta, $\{ \hat{k}_i \}_{i=1}^n$, in which the desired resonances' momenta are on~shell and all other momenta modified as little as possible.
The on-shell momenta are then used for the evaluation of the pole-approximated amplitude
$\mathcal{M}_{\rm PA}$, as made clear in~\refeq{eq:amplitude-pole-approximation}.

Therefore, we need to define a mapping
\begin{equation}
\{ k_i \}_{i=1}^n \mapsto \{ \hat{k}_i \}_{i=1}^n \text{,}
\end{equation}
which fulfils a minimal set of requirements:
\begin{itemize}
\item the external masses are unchanged, meaning $k_i^2 = \hat{k}_i^2 = m^2_i$ for all $1 \le i \le n$;
\item the internal $s$-channel resonances required by the PA are on~shell, meaning that for each of these resonances $j$, $1 \le j \le n_\mathrm{r}$, we need
\begin{equation}
\left( \hat{k}_j^\mathrm{r} \right)^2 = \left( \sum_{i \in R_j} \hat{k}_i \right)^2 = M_j^2 \text{,}
\end{equation}
with $R_j$ the set of indices labelling final-state particles which are decay products of the resonance $j$, which has real mass $M_j$;
\item as many quantities (angles, invariants) as possible, which are not constrained by the equations given above, should be preserved.
\end{itemize}
These conditions do not uniquely define a mapping, and consequently allow for different on-shell projections.
Differences in the results driven by the usage of different projections are expected to be subleading within the accuracy of the PA procedure itself.
By making use of this residual freedom in the definition of the mapping, we construct an on-shell-projection algorithm which is flexible enough to preserve as many invariants as possible, and, on top of that, fully general and applicable  also to the case of nested PAs, introduced in \refse{sec:pole-approximation}.

\paragraph{Terminology}

In the general case the algorithm has to go through $N$ steps, where $N = n_\mathrm{r} + n_s$ is determined by the number of resonances $n_\mathrm{r}$ to be set on~shell and the number of invariants $n_s$ to be preserved.
By default we require $\sqrt{\hat{s}}$ to be preserved by the projection, so that $n_s$ is at least one, and let the user preserve more invariants if needed.
If a single resonance is produced directly from the incoming particles
without any additional particle, e.g.\
Drell--Yan lepton-pair production at LO or Higgs-boson production in
$\mu^+\mu^-$ scattering, the on-shell condition for this resonance and the preservation of the partonic centre-of-mass energy lead to two incompatible equations for the same invariant.
In that case we enforce the on-shell condition for the first resonance and accordingly modify the initial-state momenta $p_1$ and $p_2$ so that $s = (\hat{p}_1 + \hat{p}_2)^2 = M_1^2$; the remaining part of the
algorithm automatically restores momentum conservation, $\hat{p}_1 + \hat{p}_2 = \sum_{i=1}^n \hat{k}_i$.

In the following we make the description of our general on-shell projection more concrete by referring to a specific process, namely  Higgs production in vector-boson fusion,
\begin{equation}
\Pu \Pc \to \PH(\Pl_1 \Pnu_2 \PWm (\Pu_3 \Pdx_4)) \Pu_5 \Pc_6 \text{,}
\label{eq:example}
\end{equation}
in which we added indices to the final-state particles that are referenced in the discussion below.
This example is a case of a DPA ($n_\mathrm{r} = 2$) that is sufficiently general for the description of the concepts of this section.

The algorithm described below makes use of the concept of a \emph{generalised resonance}, which is a set of final-state particles labelled by their indices
\begin{equation}
R = \{i_1, i_2, \ldots, i_m\} \text{,}
\end{equation}
with $m > 1$ so that $R$ contains at least two elements.
A generalised resonance represents either a proper resonance, whose momentum we want to project on~shell, or a set of final-state particles, whose invariant mass computed with off-shell momenta we want to preserve throughout the projection procedure.
The algorithm treats the two cases on the same footing, since the only difference among the two lies in the value to which the invariant is projected, namely a constant real number or a dynamical quantity, respectively.
For the example in~\refeq{eq:example}, we could choose the following generalised resonances,
\begin{subequations}
\begin{align}
R_1 &= \{ 1, 2, 3, 4, 5, 6 \} \text{,} \label{eq:preserve-123456} \\
R_2 &= \{ 1, 2, 3, 4 \} \text{,} \label{eq:project-1234} \\
R_3 &= \{ 3, 4 \} \text{,} \label{eq:preserve-34} \\
R_4 &= \{ 1, 2 \} \text{,} \label{eq:project-12} \\
R_5 &= \{ 5, 6 \} \text{,} \label{eq:preserve-56}
\end{align}
\end{subequations}
which represent the partonic centre of mass, $R_1$, the invariant of the off-shell $\PWm$~boson formed by particles 3 and 4, $R_3$, and the invariant of particles 5 and 6, $R_5$, all of which we want to preserve.
Furthermore, $R_2$ represents the invariant of the Higgs boson and $R_4$ the invariant of the $\PWp$ boson, which we want to project on~shell.

Next, we introduce \emph{spectators}, which are similar to generalised resonances, but contain exactly one index,
\begin{equation}
L_i = \{ i \} \text{,}
\label{eq:spectators}
\end{equation}
and therefore consist of a single-particle state with a given index $i$. As it will become clear below, spectators naturally represent stopping conditions for an algorithm that, starting from  production level, follows the decay chain of all intermediate particles until final states are reached.
In the example of~\refeq{eq:example}, we have
\begin{equation}
L_1 = \{ 1 \} \text{,} \qquad
L_2 = \{ 2 \} \text{,} \qquad
L_3 = \{ 3 \} \text{,} \qquad
L_4 = \{ 4 \} \text{,} \qquad
L_5 = \{ 5 \} \text{,} \qquad
L_6 = \{ 6 \} \text{,}
\end{equation}
namely one spectator for each final-state particle.

Each generalised resonance $R_l$ must decay into either (nested) generalised resonances and/or spectators, and this information is encoded in the list of \emph{decay sets} $D_l$,
\begin{equation}
R_l \to D_l = \{ D^1_l, D^2_l, \ldots, D^d_l \} \text{,}
\label{eq:decay-products}
\end{equation}
whose elements $D^k_l$ in turn can be either  generalised resonances $R$ or spectators $L$.
For every generalised resonance in a decay set $D_l$, there has to be another decay set $D_m$ with $m > l$ so that eventually all spectators are contained in a decay set.
In our example we have the following decay sets:
\begin{subequations}
\begin{align}
R_1 \to D_1 &= \{ R_2, R_5 \} \text{,} \\
R_2 \to D_2 &= \{ R_3, R_4 \} \text{,} \\
R_3 \to D_3 &= \{ L_3, L_4 \} \text{,} \\
R_4 \to D_4 &= \{ L_1, L_2 \} \text{,} \\
R_5 \to D_5 &= \{ L_5, L_6 \} \text{,}
\end{align}
\end{subequations}
where $D_1$ describes the production process of the Higgs boson ($R_2$) and the two-jet system ($R_5$),
whose subsequent decay into the two spectators $L_5$ and $L_6$ is enclosed in $D_5$. Then, $D_2$
denotes the subsequent Higgs decay into a $\PWp$ ($R_3$) and a $\PWm$ ($R_4$) boson, each of which decays into the spectators collected in $D_3$ and $D_4$, respectively.

Finally, each generalised resonance $R_k$ has an associated value $S_k$ that fixes the corresponding invariant after on-shell projection:
\begin{equation}
S_k = \left( \sum_{r \in R_k} \hat{k}_r \right)^2 \text{,}
\label{eq:on-shell-condition}
\end{equation}
and these values are either determined by the PA (and set to the corresponding squared masses of the resonances) or by the phase-space generator, if the corresponding invariant should be preserved.
For spectators, a similar condition holds,
\begin{equation}
m_i^2 = \left( \sum_{r \in L_i} \hat{k}_r \right)^2 = \hat{k}_i^2 \text{,}
\label{eq:on-shell-condition-2}
\end{equation}
where the sum always runs over exactly one element.
Since generalised resonances and spectators are also decay sets, we can rewrite~\refeq{eq:on-shell-condition} and~\refeq{eq:on-shell-condition-2} together in the following equation,
\begin{equation}
S(D) = \left( \sum_{r \in D} \hat{k}_r \right)^2 \text{,}
\end{equation}
where $S(R_k) = S_k$ and $S(L_k) = m_k^2$.
In our example we have
\begin{equation}
S_1 = s_{123456}, \qquad
S_2 = M_{\PH}^2,  \qquad
S_3 = M_{\PW}^2,  \qquad
S_4 = s_{12},     \qquad
S_5 = s_{56},
\end{equation}
which instructs the algorithm to preserve the centre-of-mass energy,
to project the Higgs and the $\PWp$ boson on~shell, and to preserve the $\PWm$-boson invariant and the invariant of particles 5 and 6.

\paragraph{Initialisation}

We now introduce the concept of \emph{resonance information}, which is the data the algorithm needs to generate the on-shell-projected momenta.
More precisely, we define the resonance information as the ordered list
\begin{equation}
\bigl\{ S_l, \,R_l \to D_l \bigr\}_{l=1}^{N}
\label{eq:resonance-information}
\end{equation}
of $N$ generalised resonances, whose corresponding invariants should be set to the value $S_l$, and which decay into the objects represented by the elements of $D_l = \{ D_l^1, D_l^2, \ldots, D_l^{d_l} \}$, where $d_l$ defines the number of decay products for the generalised resonance $R_l$.

The ordering in~\refeq{eq:resonance-information} according to the index $l$ is important, since it corresponds to the way the algorithm proceeds:
if $i, j \in \mathbb{N}$, then either $R_{j + i} \subset R_j$ or $R_{j + i} \cap R_j = \varnothing$.
In other words: generalised resonances that are processed later are either fully nested in previous resonances, or completely disjoint from them (if they are branches from different decay sets).
To summarise the resonance information that we chose for~\refeq{eq:example} in the previous paragraphs, we have:
\begin{equation}
\begin{aligned}
S_1 &= s_{123456} \text{,}\quad & R_1 &= \{ 1, 2, 3, 4, 5, 6 \} &\to\ D_1 &= \{ R_2, R_5 \} \text{,} \\
S_2 &= M_{\PH}^2  \text{,} & R_2 &= \{ 1, 2, 3, 4 \}       &\to\ D_2 &= \{ R_3, R_4 \} \text{,} \\
S_3 &= M_{\PW}^2  \text{,} & R_3 &= \{ 3, 4 \}             &\to\ D_3 &= \{ L_3, L_4 \} \text{,} \\
S_4 &= s_{12}     \text{,} & R_4 &= \{ 1, 2 \}             &\to\ D_4 &= \{ L_1, L_2 \} \text{,} \\
S_5 &= s_{56}     \text{,} & R_5 &= \{ 5, 6 \}             &\to\ D_5 &= \{ L_5, L_6 \} \text{.}
\end{aligned}
\label{eq:example-resonance-information}
\end{equation}
Note that $S_2$, representing the Higgs decay, comes after $S_1$, because the Higgs on-shell momentum is computed at $\ell=1$. On the other hand,  $S_2$ comes before $S_3$ and $S_4$, which describe the decays of the two $\PW$ bosons, whose projected momenta are generated at step $\ell=2$. Notice that $S_5$ could come at any place after $S_1$,  and that $S_2$ and $S_3$ could safely be swapped.
Apart from this freedom, the ordering is fixed.

\paragraph{On-shell projection}

The on-shell-projection algorithm loops over the entries of the resonance information from $l=1$ to $l=N$. At each step $l$ it uses the previously constructed on-shell momentum $\hat{k}_l^\mathrm{r}$ corresponding to the generalised resonance $R_l$ (input) and generates the on-shell-projected momenta corresponding to each element $D_l^k$ of the decay set $D_l$ (output).
The decay set $D_l^k$ is either a spectator, $D_l^k = L_i$, and then the on-shell momentum $\hat{k}_i$ for the final state $i$ has been constructed, or it is another generalised resonance, $D_l^k = R_m$. In the latter case the algorithm constructs the momentum of a nested resonance which eventually decays into spectators in a later step $m > l$.

At each step, first, the off-shell momenta $p_j$ corresponding to the decay set $D_l$ are constructed,
\begin{equation}
p_{j} = \sum_{i \in D_l^j} k_i \text{,}
\end{equation}
together with the off-shell momentum of the generalised resonance:
\begin{equation}
k_l^\mathrm{r} = \sum_{j=1}^{d_l} p_j \text{.}
\end{equation}
Then we boost the momenta $p_j$ into their centre-of-mass frame using the pure boost $\Lambda(k_l^\mathrm{r})$, so that $\tilde{q}_j = \Lambda(k_l^\mathrm{r}) p_j$ and
\begin{equation}
\sum_{j=1}^{d_l} \Lambda(k_l^\mathrm{r}) \, p_j = \sum_{j=1}^{d_l} \tilde{q}_j = \begin{pmatrix} E_l \\ \vec{0} \end{pmatrix} \text{,}
\label{eq:boost}
\end{equation}
where $E_l$ denotes the corresponding centre-of-mass energy.
Because the three-momenta $\{\vec{\tilde{q}}_j\}_{j=1}^{d_l}$ add up to zero in~\refeq{eq:boost}, we are free to rescale them by the same factor $\alpha$,
without destroying spatial-momentum conservation.
This freedom is used to enforce the condition in~\refeq{eq:on-shell-condition} for all generalised resonances and spectators in the list $D_l$, which yields an equation for $\alpha$,
\begin{equation}
\sqrt{S(R_l)} = \sum_{j=1}^{d_l} \sqrt{S(D_l^j) + \alpha^2 \vec{\tilde{q}}_j^{\,2}} \text{,}
\label{eq:on-shell-condition-cms}
\end{equation}
from which we choose the solution closest to $1$. We point out that, in the previous equation, the energies of the on-shell-projected elements of the decay set on the right-hand side have to sum to $\sqrt{S(R_l)}$ and not $E_l$, to consistently match on-shell-projection conditions enforced at previous iterations. In every step $l$ it is possible that~\refeq{eq:on-shell-condition-cms} has no solution, and this is always the case when the on-shell conditions conflict with relations among invariants.
In~\refeq{eq:example}, for example, step $l=2$ imposes the constraint
\begin{equation}
M_{\PH} > M_{\PW} + \sqrt{s_{12}} \text{,}
\end{equation}
but if $\sqrt{s_{12}}$, given by the phase-space generator, is too large, then the on-shell projection is not possible.
In those cases where an on-shell projection can not be constructed, the phase-space point is simply discarded, i.e.\ the
projected matrix element is set to zero.

If instead a value of $\alpha$ is found, we boost the new on-shell momenta $q_j$ back to the laboratory frame. In order to enforce local momentum conservation with respect to the previously computed on-shell momentum $\hat{k}_l^\mathrm{r}$ of $R_l$, this is done using an inverse boost $\Lambda^{-1}(\hat{k}_l^\mathrm{r})$ constructed using $\hat{k}_l^\mathrm{r}$ instead of $k_l^\mathrm{r}$. Therefore, the step $l$ of the algorithm can deliver two kinds of momenta. If $j$ is a spectator, then $D_l^j = L_i = \{ i \}$, with index $i$ the only element of the set, and
\begin{equation}
\hat{k}_i = \Lambda^{-1}(\hat{k}_l^\mathrm{r}) \begin{pmatrix}\sqrt{m_i^2 + \alpha^2 \vec{\tilde{q}}_j^{\,2}} \\ \alpha \vec{\tilde{q}}_j \end{pmatrix}
\label{eq:final-state-momentum-construction}
\end{equation}
is the final on-shell-projected momentum for the particle with index $i$.
Instead, if $j$ denotes a generalised resonance (which must decay further), then $D_l^j = R_i$, with $i$ identifying one of the generalised resonances, and
\begin{equation}
P_i = \Lambda^{-1}(\hat{k}_l^\mathrm{r})\begin{pmatrix}\sqrt{S_i + \alpha^2 \vec{\tilde{q}}_j^{\,2}} \\ \alpha \vec{\tilde{q}}_j \end{pmatrix}
\label{eq:intermediate-momentum-construction}
\end{equation}
is the momentum of a nested resonance, which will be needed in a step $i > l$ of the algorithm.

We note that the construction of the decay momenta,~\refeqs{eq:boost,eq:final-state-momentum-construction,eq:intermediate-momentum-construction}, are a generalisation of some of the ingredients used in the on-shell projection presented in \citere{Denner:2021csi}.
Here~\refeq{eq:on-shell-condition-cms} is the genuinely  new feature allowing us to generate on-shell projections for arbitrary pole approximations.

\paragraph{Preservation of invariants}

As mentioned at the beginning of this section, our algorithm  also offers the freedom
to preserve some invariants to reduce the impact of the on-shell projection in deforming the off-shell
phase-space points. This is done by a proper definition of the resonance information, which must
be specified to initialise the algorithm.
In \refta{tab:preservation-of-invariants} we report the choice that is used to produce the results in \refse{sec:numerical-results}.
We denote with $\PV$ all heavy vector bosons, i.e.\ $\PV \in \{ \PW, \PZ \}$, and with $\PB$  all heavy scalar or vector bosons, i.e.\ $\PB \in \{ \PH, \PW, \PZ \}$. Our choice of preserved invariants
is quite natural when one looks at the different PAs: we pair together leptons or quarks
that do not directly result from an on-shell projected resonance at a given step of the decay chain.
The only case that needs clarification is the DPA with the decay $\PB \to q_1 q_2 q_3 q_4$. In this case
we always pair together quarks belonging to the same generation, preferring combinations where the quarks
$q_1$ and $q_2$ have total electric charge $+1$, $-1$ or $0$ (in this order).
This is motivated by the fact that the contributions involving
resonant W~bosons are usually larger than those involving resonant Z~bosons.
While we could have always fixed two resonances for all DPAs and one
resonance for all TPAs, this was not needed in some cases for the
setups considered in the paper.

\begin{table}
\centering
\small
\setlength{\tabcolsep}{4pt}
\begin{tabular}{@{}llrccc@{}}
\toprule
DPA & equation & \multicolumn{4}{c}{Projected/preserved invariants} \\
\midrule
$\Pp \Pp \to \PWp (\ell^+ \nu_\ell) \PV (q_1 q_2) q_3 q_4$    & \labelcref{eq:sl-wpwp,eq:sl-wpwm,eq:sl-wpz}                                             & $\hat{s}_{\ell^+ \nu_\ell} = M_\PW^2$ & $\hat{s}_{12} = M_\PV^2$ & $s_{34}$ & -- \\
$\Pp \Pp \to \PWp (\ell^+ \nu_\ell) \PB (q_1 q_2 q_3 q_4)$    & \labelcref{eq:wp-higgs-4}--\labelcref{eq:wp-wm-4}                                       & $\hat{s}_{\ell^+ \nu_\ell} = M_\PW^2$ & $\hat{s}_{1234} = M_\PB^2$ & $s_{12}$ & $s_{34}$ \\
$\Pp \Pp \to \PB (\ell^+ \nu_\ell q_1 q_2) \PV (q_3 q_4)$     & \labelcref{eq:dpa-wp-4-z,eq:dpa-wp-4-wm,eq:dpa-higgs-wp,eq:dpa-higgs-z,eq:dpa-higgs-wm} & $\hat{s}_{\ell^+ \nu_\ell 1 2} = M_\PB^2$ & $\hat{s}_{34} = M_\PV^2$ & $s_{\ell^+ \nu_\ell}$ & $s_{12}$ \\
$\Pp \Pp \to \PV ( \PWp (\ell^+ \nu_\ell) q_1 q_2) q_3 q_4$   & \labelcref{eq:sl-higgs-wp}                                                              & $\hat{s}_{\ell^+ \nu_\ell 1 2} = M_\PV^2$ & $\hat{s}_{\ell^+ \nu_\ell} = M_\PW^2$ & $s_{12}$ & $s_{34}$ \\
$\Pp \Pp \to \PV ( \ell^+ \nu_\ell \PWm (q_1 q_2)) q_3 q_4$   & \labelcref{eq:sl-higgs-wm}                                                              & $\hat{s}_{\ell^+ \nu_\ell 1 2} = M_\PV^2$ & $\hat{s}_{1 2} = M_\PW^2$ & $s_{\ell^+\nu_\ell}$ & $s_{34}$ \\
$\Pp \Pp \to \ell^+ \nu_\ell \PV_1 (q_1 q_2) \PV_2 (q_3 q_4)$ & \labelcref{eq:fh-wpwm,eq:fh-wmz,eq:fh-wmwm,eq:fh-zz}                                    & $\hat{s}_{1 2} = M_{\PV_1}^2$ & $\hat{s}_{3 4} = M_{\PV_2}^2$ & $s_{\ell^+\nu_\ell}$ & -- \\
$\Pp \Pp \to \ell^+ \nu_\ell \PB (\PV (q_1 q_2) q_3 q_4)$     & \labelcref{eq:fh-higgs-wp,eq:fh-higgs-z,eq:fh-higgs-wm}                                 & $\hat{s}_{1234} = M_\PB^2$ & $\hat{s}_{1 2} = M_\PV^2$ & $s_{\ell^+ \nu_\ell}$ & $s_{34}$ \\
\midrule
TPA & equation & \multicolumn{4}{c}{Projected/preserved invariants} \\
\midrule
$\Pp \Pp \to \PWp (\ell^+ \nu_\ell) \PV_1 (q_1 q_2) \PV_2 (q_3 q_4)$ & \labelcref{eq:tpa-wpwpwm,eq:tpa-wpwmwm,eq:tpa-wpwmz,eq:tpa-wpzz}             & $\hat{s}_{\ell^+ \nu_\ell} = M_\PW^2$ & $\hat{s}_{1 2} = M_{\PV_1}^2$ & $\hat{s}_{3 4} = M_{\PV_2}^2$ & -- \\
$\Pp \Pp \to \PWp (\ell^+ \nu_\ell) \PB (\PV (q_1 q_2) q_3 q_4)$     & \labelcref{eq:tpa-fh-higgs-wm-wp,eq:tpa-fh-higgs-wm-z,eq:tpa-fh-higgs-wm-wm} & $\hat{s}_{\ell^+ \nu_\ell} = M_\PW^2$ & $\hat{s}_{1234} = M_\PB^2$ & $\hat{s}_{1 2} = M_\PV^2$ & $s_{34}$ \\
$\Pp \Pp \to \PB ( \PWp (\ell^+ \nu_\ell) q_1 q_2) \PV (q_3 q_4)$    & \labelcref{eq:tpa-sl-higgs-wp-wp,eq:tpa-sl-higgs-wp-z,eq:tpa-sl-higgs-wp-wm} & $\hat{s}_{\ell^+ \nu_\ell 1 2} = M_\PB^2$ & $\hat{s}_{\ell^+ \nu_\ell} = M_\PW^2$ & $\hat{s}_{34} = M_\PV^2$ & $s_{12}$ \\
$\Pp \Pp \to \PB ( \ell^+ \nu_\ell \PWm (q_1 q_2)) \PV (q_3 q_4)$    & \labelcref{eq:tpa-sl-higgs-wm-wp,eq:tpa-sl-higgs-wm-z,eq:tpa-sl-higgs-wm-m}  & $\hat{s}_{\ell^+ \nu_\ell 1 2} = M_\PB^2$ & $\hat{s}_{1 2} = M_\PW^2$ & $\hat{s}_{34} = M_\PV^2$ & $s_{\ell^+ \nu_\ell}$ \\
\bottomrule
\end{tabular}
\caption{Choice of projected and preserved invariants for the PAs given in \refse{sec:all-possible-pas} and used to calculate the results in \refse{sec:numerical-results}.
Whenever invariants are given in the form $\hat{s} = M_R^2$, they are
projected to the mass of the resonance $R$, otherwise their off-shell
value is preserved, $\hat{s} = s$, as discussed in
\refse{sec:a-general-on-shell-projection}. The indices of the
invariants $s_{\ldots}$ refer to the corresponding quarks in the final
state as indicated in column 1.}
\label{tab:preservation-of-invariants}
\end{table}

\section{Numerical results}
\label{sec:numerical-results}

To be specific, we consider for the numerical analysis of this section the $\mu^+\Pnu_\mu$
final state, i.e.\ we compute the process
\begin{equation}
\Pp\Pp \to \mu^+ \Pnu_\mu + 4\Pj.
\label{eq:process_mu}
\end{equation}
Since we neglect the charged-lepton mass, our results hold as well for all other lepton flavours,
whenever they are treated as massless (as usually done with the electron).

\subsection{Input parameters}
\label{sec:input-parameters}

Our setup is designed for an LHC run at a centre-of-mass (CM) energy of \SI{13}{\tera\electronvolt}.
We use the NNLO NNPDF3.1luxQED PDF set with $\alpha_\mathrm{s}(\MZ) = 0.118$ \cite{Bertone:2017bme} via LHAPDF~\cite{Buckley:2014ana} in the $N_\text{F}=5$ fixed-flavour scheme for all predictions.
The initial-state collinear splittings are treated by the ${\overline{\rm MS}}$ redefinition of the PDFs.

The central renormalisation and factorisation scales are chosen as the geometric average of the transverse momenta of
the tag jets $\Pj_1$ and $\Pj_2$
\begin{equation}
\mu_\mathrm{R}^\mathrm{central} = \mu_\mathrm{F}^\mathrm{central} = \sqrt{p_\mathrm{T, j_1} \cdot p_\mathrm{T, j_2}} \text{,}
\end{equation}
where the precise definition of tag jets, namely the ones originating from the scattered incoming partons for a typical VBS signal, is given in \refse{sec:event-selection}.
Based on this central scale, we perform a 7-point scale variation of both the renormalisation and factorisation scale, meaning we calculate the observables for the pairs
\begin{equation}
\left( \mu_\mathrm{R}/\mu_\mathrm{R}^\mathrm{central}, \mu_\mathrm{F}/\mu_\mathrm{F}^\mathrm{central} \right) =
\{(0.5,0.5),
(0.5,1),
(1,0.5),
(1,1),
(1,2),
(2,1),
(2,2)\}
\end{equation}
of renormalisation and factorisation scales and use the resulting
envelope to estimate the perturbative (QCD) scale uncertainty. It worth mentioning already at this point that
a 7-point scale variation is strictly performed only for the estimate of the LO QCD uncertainties of the fully off-shell $\mathcal{O}(\as^2\alpha^4)$ and $\mathcal{O}(\as\alpha^5)$.
Indeed, since the $\mathcal{O}(\alpha^6)$ contribution just depends on the factorisation scale, the 7-point scale variation boils down to a
3-point scale variation in this case.

The masses and widths of the massive gauge bosons are taken from the PDG review 2020 \cite{ParticleDataGroup:2020ssz},
\begin{subequations}
\begin{equation}
\begin{aligned}
\MZOS &= \SI{91.1876}{\giga\electronvolt} \text{,} & \qquad
\GZOS &= \SI{2.4952}{\giga\electronvolt}  \text{,} \\
\MWOS &= \SI{80.379}{\giga\electronvolt}  \text{,} &
\GWOS &= \SI{2.085}{\giga\electronvolt}   \text{,}
\end{aligned}
\label{eq:vector-boson-masses}
\end{equation}
and furthermore we set
\begin{equation}
\begin{aligned}
\Mt &= \SI{173}{\giga\electronvolt}     \text{,} & \qquad
\Gt &= \SI{0}{\giga\electronvolt}       \text{,} \\
\MH &= \SI{125}{\giga\electronvolt}     \text{,} &
\GH &= \SI{4.07e-3}{\giga\electronvolt} \text{.}
\end{aligned}
\label{eq:other-masses}
\end{equation}%
\end{subequations}
Without any resonant top quarks in the considered processes, we can set the top-quark width to zero.
The Higgs-boson width is taken from \citere{LHCHiggsCrossSectionWorkingGroup:2013rie}.
From the measured on-shell (OS) values of the masses and widths of the weak vector bosons $V=\PW,\PZ$, we obtain the corresponding pole quantities used in the calculation~\cite{Bardin:1988xt},
\begin{equation}
M_V      = \frac{\MVOS}{\sqrt{1+(\GVOS/\MVOS)^2}} \text{,} \qquad
\Gamma_V = \frac{\GVOS}{\sqrt{1+(\GVOS/\MVOS)^2}} \text{.}
\end{equation}
The pole quantities are then used to initialise complex masses for the unstable particles, as required in the complex-mass scheme~\cite{Denner:1999gp,Denner:2005fg,Denner:2006ic,Denner:2019vbn}.
As a consequence, the EW mixing angle and the related couplings are also complex valued.

We employ the $\GF$ scheme \cite{Denner:2000bj} to define the electromagnetic coupling, which fixes the EW coupling $\alpha$ using the Fermi constant $\GF$ as input parameter via
\begin{equation}
\alpha = \frac{\sqrt{2}}{\pi} \GF \MW^2 \left( 1 - \frac{\MW^2}{\MZ^2} \right)
\qquad \text{with} \qquad
\GF = \SI{1.16638e-5}{\per\giga\electronvolt\squared}
\end{equation}
from the real pole masses.

Our calculation is performed in the 5-flavour scheme, where the bottom quark is treated as massless.
As already motivated in~\refse{sec:description-of-the-calculation}, we exclude all partonic channels
with final-state bottom quarks by assuming a perfect bottom-jet veto.
The contribution of the remaining channels with bottom quarks only in the initial state has been neglected.
Indeed we verified numerically that this contribution is below $10^{-5}$ at $\mathcal{O}(\alpha^6)$ and at the level of $10^{-3}$
(therefore comparable with our  numerical accuracy, see \refta{tab:all-fiducial-cross-sections}) at $\mathcal{O}(\as^2\alpha^4)$
for the setups considered in this paper.

\subsection{Event selection}
\label{sec:event-selection}

The event selection used for this analysis is inspired by the CMS and ATLAS measurements of opposite-sign $\PW$-boson-pair production \cite{ATLAS:2018tav,CMS:2021qzz} and previous works of some of us on VBS \cite{Denner:2022pwc,Denner:2019tmn}.
In our LO definition of the process in~\refeq{eq:process_mu}, where photons are not considered as jets and therefore can not
appear in the final state of any partonic channel, the reconstruction of physical objects significantly simplifies.
QCD partons (quarks, anti-quarks, gluons) are clustered  into jets using the anti-$k_\text{T}$ algorithm \cite{Cacciari:2008gp}.
Only partons with rapidity $|y| < 5$ are considered for recombination, while particles with larger $|y|$ are assumed to be lost in the beam pipe.
The rapidity $y$ and the transverse momentum $p_\mathrm{T}$ of a particle are defined as
\begin{equation}
y = \frac{1}{2} \ln \frac{E + p_z}{E-p_z}, \qquad
p_\mathrm{T} = \sqrt{p_x^2 + p_y^2} \text{,}
\end{equation}
where $E$ is the energy of the particle, $p_z$ the component of its momentum along the beam axis, and $p_x,p_y$ the components perpendicular to the beam axis.

The anti-muon has to fulfil
\begin{equation}
p_{\mathrm{T},\mu} > \SI{30}{\giga\electronvolt} \text{,} \qquad |y_\mu| < 2.4 \text{,}
\label{eq:charged-lepton-cuts}
\end{equation}
while the missing transverse momentum is required to satisfy
\begin{equation}
p_{\mathrm{T}, \text{miss}}  > \SI{30}{\giga\electronvolt}
\label{eq:missing-momentum-cut}
\end{equation}
and is computed as the transverse part of the neutrino momentum at Monte Carlo-truth level, i.e.\ $p_{\mathrm{T},\text{miss}} = p_{\mathrm{T}, \Pnu_\mu}$.
Furthermore the transverse mass of the W boson is bounded by an upper threshold,
\begin{equation}
M_\mathrm{T}^\PW = \sqrt{2 p_{\mathrm{T},\mu} \cdot p_{\mathrm{T}, \Pnu_\mu} (1 - \cos \Delta \phi_{\mu\Pnu_\mu})} < \SI{185}{\giga\electronvolt} \text{.}
\label{eq:transverse-mass-cut}
\end{equation}

The partons are independently clustered with a resolution radius of firstly $R = 0.4$, and secondly of $R = 0.8$.
The resulting objects are called AK4 and AK8 jets, respectively.
AK4 jets must fulfil the conditions
\begin{equation}\label{eq:AK4-jet-cuts}
p_{\mathrm{T}, \Pj_\text{AK4}} > \SI{30}{\giga\electronvolt} \text{,} \qquad
|y_{\Pj_\text{AK4}}|             < 4.7 \text{.}
\end{equation}
In the set of AK8 jets there must be either none (a) or one (b) that fulfils the conditions
\begin{equation} 
  \label{eq:fat-jet-cuts}
p_{\mathrm{T}, \Pj_\text{AK8}} > \SI{200}{\giga\electronvolt} \text{,} \qquad
|y_{\Pj_\text{AK8}}|               < 2.4 \text{,} \qquad
\SI{40}{\giga\electronvolt}      < M_{\Pj_\text{AK8}} < \SI{250}{\giga\electronvolt} \text{,}
\end{equation}
otherwise the event is discarded.
The first case (a) with no AK8 jet and at least four AK4 jets defines the \emph{resolved category},
while the second case (b) with exactly one AK8 jet and at least two AK4 jets defines the \emph{boosted category}.
To remove most of the overlap of AK4 and AK8 jets in the boosted category we demand that
\begin{equation}
\Delta R \left( \Pj_\text{AK4}, \Pj_\text{AK8} \right) > 0.8
\end{equation}
for every AK4 jet.
The distance $\Delta R_{ij}$ for two objects $i$ and $j$ is defined as
\begin{equation}
\Delta R_{ij} = \sqrt{(\Delta \phi_{ij})^2 + (\Delta y_{ij})^2}
\end{equation}
with the azimuthal-angle difference $\Delta \phi_{ij} = \min(|\phi_i - \phi_j|, 2\pi - |\phi_i - \phi_j|)$ and the rapidity difference $\Delta y_{ij} = y_i - y_j$.

In both categories the two AK4 jets with highest invariant mass, called \emph{tag jets}, must obey
\begin{equation}
M_{\Pj_1 \Pj_2} >  \SI{500}{\giga\electronvolt} \text{,} \qquad |\Delta y_{\Pj_1\Pj_2}| > 2.5 \text{.}
\label{eq:vbscuts}
\end{equation}
Furthermore, the tag jet with largest transverse momentum must satisfy
\begin{equation}
  p_{\mathrm{T}, \Pj_1} > \SI{50}{\giga\electronvolt} \text{.}
  \label{eq:leading-tag-jet-pt-cut}
\end{equation}
Finally, we define the hadronically decaying vector boson mass $M_\PV$ as the invariant mass either of the two AK4 jets that are not tag jets and
whose invariant mass is the closest to \SI{85}{\giga\electronvolt} in the resolved category (a) or of the AK8 jet in the boosted category (b).
This invariant mass must fulfil
\begin{equation}
\SI{65}{\giga\electronvolt} < M_\PV < \SI{105}{\giga\electronvolt} \,\, \text{(a)}\quad \text{and} \quad
\SI{70}{\giga\electronvolt} < M_\PV < \SI{115}{\giga\electronvolt} \,\,\text{(b)}
\label{eq:hadronic-on-shell-condition}
\end{equation}
in the resolved and boosted category, respectively.

\subsection{Pole approximation in a VBS-like acceptance region}
\label{sec:pole-approximation-in-specific-setup}

For both setups presented in~\refse{sec:event-selection} the PAs of each partonic channel contributing to our reaction
simplify with respect to the general and systematic case discussed in~\refse{sec:all-possible-pas}.
This is due to the cuts on the invariant mass of the tag jets, $M_{\Pj_1 \Pj_2}$, given in \Cref{eq:vbscuts}, and the invariant mass $M_\PV$ of the pair of jets identified with the hadronic decay of the vector boson, \Cref{eq:hadronic-on-shell-condition}.

By its definition in \Cref{eq:amplitude-pole-approximation}, a PA describes the corresponding fully off-shell matrix element best where its resonances are on~shell.
In this region of phase space, which we denote as $\Omega$, a decent on-shell projection results in on-shell momenta $\hat{k}_i$ that deform the off-shell momenta $k_i$ as little as possible.
Therefore, if the definition of the fiducial phase-space region is such that $\Omega$ is cut out, the application of the pole approximation is not justified.

We use this consistency argument to determine which PAs from the list in \Cref{eq:dpas} to include or not.
This leaves the PAs given in \Cref{eq:sl-wpwp,eq:sl-wpwm,eq:sl-wpz} (that we sometimes denote as \texttt{sl-dpa} category)
and only the Higgs-resonant contribution from \Cref{eq:sl-higgs-wm} (also referred to as \texttt{sl-dpa-h} in the following).
All dropped contributions from \Cref{eq:dpas} fall in three categories:
\begin{enumerate}
\item One resonance $R$ decaying into four quarks: $R \to q_1 q_2 q_3 q_4$.
This concerns the PAs in \Cref{eq:wp-higgs-4,eq:wp-wm-4}, which contain the unnested decays $\PH,\PZ,\PWm \to \Pj\Pj\Pj\Pj$, and in \Cref{eq:fh-higgs-wp,eq:fh-higgs-z,eq:fh-higgs-wm}, which contain the decays $\PH/\PZ \to \PV \Pj \Pj$ with $\PV$ further decaying into two jets, i.e.\ $\PV \to \Pj \Pj$.

For these processes the cut $M_{\Pj_1 \Pj_2} > \SI{500}{\giga\electronvolt}$ on the tag jets leads to the kinematical constraint
\begin{equation}
\sqrt{s_{1234}} > M_{\Pj_1 \Pj_2} > \SI{500}{\giga\electronvolt} \text{,}
\end{equation}
on the off-shell invariant $s_{1234}$ of the four quarks.
Since the on-shell projection sets $\hat{s}_{1234} = M_\PH^2$, $M_\PZ^2,$ or $M_\PW^2$, the phase-space region $\Omega$, in which $s_{1234} \approx \hat{s}_{1234}$, is cut out.
In other words, phase-space points that pass the tag-jet cut are always deformed heavily, in the sense $| (s_{1234} - \hat{s}_{1234}) / \Gamma_B | \gg 1$, which requires these contributions to be dropped.

\item Two resonances $R_1$ and $R_2$, each containing two quarks in its decay products: $R_1 \to q_1 q_2$ (or $R_1 \to \mu^+ \nu_\mu q_1 q_2$) and $R_2 \to q_3 q_4$.
This concerns \Cref{eq:dpa-wp-4-z,eq:dpa-wp-4-wm,eq:dpa-higgs-wp,eq:dpa-higgs-z,eq:dpa-higgs-wm,eq:fh-wpwm,eq:fh-wmz,eq:fh-wmwm,eq:fh-zz}.
For these processes we can distinguish two scenarios:
\begin{enumerate}
\item ``correct'' identification.
This occurs when the invariant $M_\PV$ from \Cref{eq:hadronic-on-shell-condition} corresponds to either the one of the quark pair $q_1 q_2$ or $q_3 q_4$.
Then the invariant $M_{\Pj_1 \Pj_2}$ corresponds to either the one of the quark pair $q_3 q_4$ or $q_1 q_2$, respectively, implying that the region $\Omega$ is cut out, as discussed in the first category.
\item ``incorrect'' identification.
This occurs in the remaining cases, namely when the invariant $M_\PV$ corresponds to either $q_1 q_3$, $q_1 q_4$, $q_2 q_3$, or $q_2 q_4$.
Then the invariant $M_{\Pj_1 \Pj_2}$ corresponds to either the quark pair $q_2 q_4$, $q_2 q_3$, $q_1 q_4$, or $q_1 q_3$, respectively.
Since both cuts act on invariants different from the ones set on~shell, the phase-space region $\Omega$ is not necessarily cut out.
However, by calculating these PAs we find that they are suppressed with respect to the ones in~\Cref{eq:sl-wpwp,eq:sl-wpwm,eq:sl-wpz} by several orders of magnitude, from which we conclude that the contribution from $\Omega$ is negligible.
\end{enumerate}

\item Only two quarks from a resonance associated to a 4-fermion
  decay: $R \to \mu^+\Pnu_\mu q_1 q_2$. 
This concerns \Cref{eq:sl-higgs-wp,eq:sl-higgs-wm}.

In the case of \Cref{eq:sl-higgs-wp} the $\PWp$ boson from the
resonant Higgs decay $\PH \to \PWp \PWm$ is set on~shell, so that the $\PWm$ boson is off~shell.
We can again consider a correct and an incorrect identification.
For the correct identification, the phase-space region $\Omega$ is cut
out by the constraint \refeqq{eq:hadronic-on-shell-condition}{a} on
$M_\PV$, which forces the $\PWm$ boson to be on~shell.
For the incorrect identification, we again find contributions from $\Omega$ suppressed by several orders of magnitude.
When the Higgs boson is replaced by a $\PZ$~boson, i.e.\ $\PZ \to \PWp \PWm$, the suppression is even more enhanced,
because the $\PZ$-boson mass forces the $\PWm$ boson further off~shell.

In the case of \Cref{eq:sl-higgs-wm} the $\PWm$ boson from the Higgs decay $\PH \to \PWp \PWm$ is set on~shell and thus with a correct identification $\Omega$ is not cut out.
Therefore this contribution is included. We note again that the
corresponding case with the decay $\PZ \to \PWp \PWm$ is suppressed
and not included in our results.
\end{enumerate}

Once the contributions listed above are dropped, the overcounting described in \Cref{sec:overcounting-of-higher-resonant-contributions} must be adjusted accordingly.
By looking at \refta{tab:overcounting}, we see that the triply-resonant contributions in \Cref{eq:tpa-wpwpwm,eq:tpa-wpwmwm,eq:tpa-wpwmz,eq:tpa-wpzz}
(sometimes also denoted as \texttt{vvv-tpa} category) are double-counted.
Therefore they have to be subtracted once, instead of twice as in the general case.
For the other cases listed in 
 \refta{tab:overcounting}
there is no overcounting and therefore the TPA contributions in \Cref{eq:tpa-fh-higgs-wm-wp,eq:tpa-fh-higgs-wm-z,eq:tpa-fh-higgs-wm-wm,eq:tpa-sl-higgs-wp-wp,eq:tpa-sl-higgs-wp-z,eq:tpa-sl-higgs-wp-wm,eq:tpa-sl-higgs-wm-wp,eq:tpa-sl-higgs-wm-z,eq:tpa-sl-higgs-wm-m} must not be subtracted.

\subsection{Fiducial cross sections}
\label{sec:fiducial-cross-sections}

In this section we present results at the integrated level for the
reaction in~\refeq{eq:process_mu} for both of the two setups introduced in~\refse{sec:event-selection}. In~\refta{tab:all-fiducial-cross-sections}
we show values for the integrated cross section computed for the three LO
contributions discussed in~\refse{sec:description-of-the-calculation}, namely $\mathcal{O}(\alpha^6)$, $\mathcal{O}(\as\alpha^5)$, and
$\mathcal{O}(\as^2\alpha^4)$.
\begin{table*}
 \centerline{
  \begin{tabular}{|C{2cm}|C{3cm}C{2cm}||C{3cm}C{2cm}|}
    \hline
    &&&&\\[-2.3ex]
           & \multicolumn{2}{c||}{\texttt{resolved-setup}}                       & \multicolumn{2}{c|}{\texttt{boosted-setup}}      \tabularnewline[0.7em]
    Order     & $\sigma_{\textrm{off~shell}}\,[{\rm fb}]$  & $\Delta\,[\%]$  & $\sigma_{\textrm{off~shell}}\,[{\rm fb}]$  & $\Delta\,[\%]$  \tabularnewline[0.7em]
 \hline 
 &&&&\\[-2ex]
 $\mathcal{O}(\alpha^6)$       & 9.042(1)$^{+9.0\%}_{-7.7\%}$ & 22.8  &  2.5070(4)$^{+11.6\%}_{-9.6\%}$ &    21.0              \tabularnewline[0.6em]
 $\mathcal{O}(\as\alpha^5)$    & 0.2952(1)$^{+17.2\%}_{-13.5\%}$ & 0.7  &  0.06920(5)$^{+19.3\%}_{-14.9\%}$ &     0.6         \tabularnewline[0.6em]
 $\mathcal{O}(\as^2\alpha^4)$  & 30.334(5)$^{+36.7\%}_{-24.7\%}$ & 76.5  &  9.338(3)$^{+39.1\%}_{-25.9\%}$ &      78.4        \tabularnewline[0.6em]
 \hline
 &&&&\\[-2ex]
 sum & 39.673(5)$^{+30.2\%}_{-20.8\%}$  & 100.0  & 11.914(3)$^{+33.2\%}_{-22.4\%}$  & 100.0  \tabularnewline[0.6em]
 \hline
  \end{tabular}
 }
 \caption{Fully off-shell LO cross sections (in fb)  for the reaction $\Pp\Pp \to \mu^+ \Pnu_\mu + 4\Pj$, where the three different perturbative
   contributions considered in this paper are shown in separate lines. The results are presented both 
   in the resolved (second and third column) and in the boosted (fourth and fifth column) setup as defined in~\refse{sec:event-selection}.
   Scale uncertainties are shown as percentages, while integration errors are given in parentheses. The third and fifth columns report
   the contribution $\Delta$ in percentage of the specific perturbative order to the full result, given by the sum
 of the three orders and shown in the last line.}\label{tab:all-fiducial-cross-sections}
\end{table*}

Together with numerical uncertainties given in parentheses,
theoretical uncertainties are estimated by a $7$-point scale-variation envelope.
For the $\mathcal{O} (\alpha^6)$ results the cross section has no dependence on the renormalisation scale and thus QCD scale uncertainties
are simply obtained by varying the factorisation scale. This procedure practically results in a $3$-point scale variation. Since the main goal of this manuscript is not to improve on the perturbative QCD accuracy,
we just present scale uncertainties in~\refta{tab:all-fiducial-cross-sections}, but we refrain from showing them elsewhere and especially in
the differential results discussed in the next section.

In the third and fifth columns of~\refta{tab:all-fiducial-cross-sections} we report the percentages of each calculated order to their sum, shown in the last row.
As expected, the $\mathcal{O}(\as^2\alpha^4)$ is
dominating the cross section, with the LO EW contribution containing the VBS signal just amounting to $22.8\%$ and $21.0\%$ in the
resolved and boosted setup, respectively. In both cases, the interference contribution $\mathcal{O}(\as\alpha^5)$ amounts to
less than $1\%$ and is therefore completely negligible from a phenomenological point of view.

In~\refta{tab:fiducial-cross-sections} we report the LO EW results compared to their PA, obtained as described
in~\refses{sec:pole-approximation} and
\ref{sec:pole-approximation-in-specific-setup}. 
\begin{table*}
 \centerline{
  \begin{tabular}{|C{4cm}|C{3cm}C{3cm}C{2.5cm}|}
    \hline
        &&&\\[-2.3ex]
       & $\sigma_{\textrm{off~shell}}\,[{\rm fb}]$   &  $\sigma_{\rm PA}\,[{\rm fb}]$ & $\delta_{\rm PA}\,[\%]$ \tabularnewline[0.7em]
    \hline
     &&&\\[-2ex]
    \texttt{resolved-setup}   &   9.042(1)    &   8.9379(7)     &  $-1.15$              \tabularnewline[0.6em]
     \texttt{boosted-setup}   &   2.5070(4)   &   2.4799(3)     &  $-1.08$              \tabularnewline[0.6em]
 \hline
  \end{tabular}
 }
\caption{
 LO EW cross sections (in fb) in the resolved and boosted setup as defined in~\refse{sec:event-selection} for the reaction $\Pp\Pp \to \mu^+ \Pnu_\mu + 4\Pj$.
 In the first column the fully off-shell result, $\sigma_{\textrm{off~shell}}$,
 is presented, while in the second column the corresponding outcome in PA, $\sigma_{\rm PA}$, is reported.
 Integration errors are shown in parentheses. The last column
 lists the accuracy of the approximation in terms of the relative difference $\delta_{\rm PA}$ (in percentage) between the pole-approximated and off-shell results,
 defined as $\delta_{\rm PA}=(\sigma_{\rm PA}-\sigma_{\textrm{off~shell}})/\sigma_{\textrm{off~shell}}$. 
}\label{tab:fiducial-cross-sections}
\end{table*}
We can immediately see that in both the resolved and the boosted setup the PA describes the fully
off-shell result pretty accurately at the integrated level, with differences within $\sim1\%$, as visible in the third column of~\refta{tab:fiducial-cross-sections}.
The underestimation of the full result is explained
by keeping in mind that the DPA discards diagrams corresponding to single and non-resonant topologies, which at LO give positive contributions at the
squared-amplitude level, if one neglects tiny interference terms. Also the size of the relative difference $\delta_{\rm PA}$
between the full and the approximated cross section is well understood. Indeed, off-shell effects which are not covered by the PA are expected to
be of the order of $\mathcal{O}(\Gamma_V/M_V)\lesssim3\%$ (with
$V=\PZ,\PW$) for the two EW gauge bosons~\cite{Denner:2019vbn}.
Since the fiducial cross sections in the considered setups are largely
inclusive in the decay products of the resonances,
 the quality of the PA meets our expectations for both categories.
Larger differences can appear in particular phase-space regions where non-resonant
effects start to become relevant. These effects are discussed in the next section, when considering differential results.

\begin{table*}
 \centerline{
  \begin{tabular}{|C{2.5cm}|C{2.8cm}C{1.7cm}||C{2.8cm}C{2.5cm}|}
    \hline
    &&&&\\[-2.3ex]
           & \multicolumn{2}{c||}{\texttt{resolved-setup}}                       & \multicolumn{2}{c|}{\texttt{boosted-setup}}      \tabularnewline[0.7em]
    Category     & $\sigma^{(i)}_{\textrm{PA}}\,[{\rm fb}]$  & $\Delta\,[\%]$  & $\sigma^{(i)}_{\textrm{PA}}\,[{\rm fb}]$  & $\Delta\,[\%]$  \tabularnewline[0.7em]
 \hline 
 &&&&\\[-2ex]
 \texttt{sl-dpa}\,$\scriptstyle{(\PW^+\PW^-)}$                 & $3.9697(5)$   & $44.41$   & $1.1225(2)$ & $45.26$     \tabularnewline[0.6em]
 \texttt{sl-dpa}\,$\scriptstyle{(\PW^+\PW^+)}$                 & $3.0497(5)$   & $34.12$   & $0.9243(2)$ & $37.27$    \tabularnewline[0.6em]
 \texttt{sl-dpa}\,$\scriptstyle{(\PW^+\PZ)}$                & $1.6359(2)$   & $18.30$   & $0.38597(6)$  & $15.56$   \tabularnewline[0.6em]
 \hline 
 &&&&\\[-2ex] 
 \texttt{sl-dpa}          &  $8.6552(4)$     & $96.83$   &  $2.4328(3)$                  &  $98.10$      \tabularnewline[0.6em] 
\texttt{sl-dpa-h}        &  $0.28472(3)$    & $3.19$    &  $0.047063(8)$                &  $1.90$      \tabularnewline[0.6em]
\texttt{vvv-tpa}         & $-0.002090(2)$   & $0.02$    &  $-1.30(2)\times10^{-6}$  &  $-0.05\times10^{-4}$      \tabularnewline[0.6em]
 \hline
 &&&&\\[-2ex]
 $\sigma_{\textrm{PA}}=\Sigma_i\sigma^{(i)}_{\textrm{PA}}$            &  $8.9379(7)$     & $100$    &  $2.4799(3)$                   &  $100$  \tabularnewline[0.6em]
 \hline
  \end{tabular}
 }
 \caption{Integrated cross sections $\sigma^{(i)}_{\textrm{PA}}$ (in fb) for the individual PA categories  to the reaction $\Pp\Pp \to \mu^+ \Pnu_\mu + 4\Pj$
   which are considered in our calculation according to the criteria outlined in~\refse{sec:pole-approximation-in-specific-setup}.
   From the third to the fifth line, cross sections for the different VBS processes contributing to \texttt{sl-dpa}
   (giving the sum of the three in the sixth row) are separately shown. 
   The result for the full PA is reported as a reference in the last line and can be recovered by summing the cross sections
   from the sixth to the eighth line.
   Results are presented both 
   in the resolved (second and third column) and in the boosted (fourth and fifth column) setup as defined in~\refse{sec:event-selection}.
   Integration errors are given in parentheses. The third and fifth columns provide
  the percentage contribution $\Delta$ of the specific PA to the full pole-approximated result.}\label{tab:all-PA-cross-sections}
\end{table*}
In~\refta{tab:all-PA-cross-sections} we present results for the integrated cross sections $\sigma^{(i)}_{\textrm{PA}}$ for the individual PA categories that have been included
in our calculation according to the consistency criteria outlined in~\refse{sec:pole-approximation-in-specific-setup}. Their percentage contributions to the
complete pole-approximated prediction $\sigma_{\textrm{PA}}$ are reported in the third and fifth columns for the resolved and boosted setups, respectively. One immediately
notices that the bulk of the PA arises from the \texttt{sl-dpa} category
in~\Cref{eq:sl-wpwp,eq:sl-wpwm,eq:sl-wpz}, which includes the sum of all partonic channels
for which a semi-leptonic DPA with $1 \to 2$ decays is allowed, i.e.\
the genuine VBS contributions. For this category, we also show in the first lines
of the table the
individual contributions of the different VBS processes, namely $\PW^+\PW^-$ of~\Cref{eq:sl-wpwm}, which is
the largest one, $\PW^+\PW^+$ of~\Cref{eq:sl-wpwp}, and $\PW^+\PZ$ of~\Cref{eq:sl-wpz}. The dominance of the
\texttt{sl-dpa} contribution is even more enhanced in the boosted setup, where the selection
cuts further reduce the importance of diagrams with a nested Higgs resonance,
accounted for in the \texttt{sl-dpa-h} category 
of~\Cref{eq:sl-higgs-wm}. This suppression can mostly
be attributed to the transverse momentum cut acting on the fat jet of~\refeq{eq:fat-jet-cuts}: whenever
a Higgs boson is set on~shell, the energy of the two gauge bosons into which it decays is constrained,
and so is the transverse momentum of the possibly reconstructed fat jet arising from one of them.
Nevertheless, the inclusion of the \texttt{sl-dpa-h} category 
is crucial to achieve a complete DPA of the fully off-shell result.
Indeed, if the pole-approximated cross section just included the \texttt{sl-dpa} contribution,
the quality of the PA would degrade from $\delta_{\rm PA}=-1.15\%$ to $\delta_{\rm PA}=-4.28\%$ for
the resolved setup, and from $\delta_{\rm PA}=-1.08\%$ to $\delta_{\rm PA}=-2.96\%$ for the
boosted one.

The Higgs contribution captured by the \texttt{sl-dpa-h} category can be further compared to the \texttt{sl-dpa}\,$\scriptstyle{(\PW^+\PW^-)}$
cross section, which contributes to the same final state. We see that the \texttt{sl-dpa-h} result amounts to only $\delta_{\text{Higgs}}\sim 7\%$ and $\delta_{\text{Higgs}}\sim 4\%$ of the \texttt{sl-dpa}\,$\scriptstyle{(\PW^+\PW^-)}$
one in the resolved and boosted setup, respectively. The first of
these numbers can be compared with the much larger Higgs contribution
of  $\sim 25\%$ that was found in~\citere{Denner:2022pwc} (see Table~$7$) for $\PW^+\PW^-$ scattering with fully leptonic final states.
While this difference can be partially attributed to the different
nature of the process (fully leptonic versus semi-leptonic final
states), it results, in
particular, from the definition of the fiducial regions. While
in~\citere{Denner:2022pwc} no invariant-mass cuts were applied, the invariant-mass cut on the
hadronically decaying vector boson
in~\refeqq{eq:hadronic-on-shell-condition}{a} suppresses
configurations where the Higgs boson and the leptonically
decaying $\PW^+$~boson are simultaneously on~shell, as discussed in~\refse{sec:pole-approximation-in-specific-setup}. Since the \texttt{sl-dpa-h} category
just describes
configurations where the $\PW^-$ and the Higgs boson are on~shell, the
missing contribution with on-shell $\PW^+$ and Higgs bosons roughly
provides a factor of two, raising $\delta_{\text{Higgs}}$ for semi-leptonic VBS from $\sim 7\%$ to $\sim 14\%$, thus rendering this fraction
much closer to the result of~\citere{Denner:2022pwc}.

As a concluding remark, we observe that
the impact of the \texttt{vvv-tpa} of~\Cref{eq:tpa-wpwpwm,eq:tpa-wpwmwm,eq:tpa-wpwmz,eq:tpa-wpzz}
is numerically irrelevant for both setups at the integrated level, when all partonic channels contributing to
$\Pp\Pp \to \mu^+ \Pnu_\mu + 4\Pj$ are summed over.
Nevertheless, their inclusion is indispensable to achieve for all partonic channels a theoretically consistent PA 
which avoids overcounting problems, as described in~\refse{sec:overcounting-of-higher-resonant-contributions}. This is
explicitly illustrated in~\refta{tab:pc-cross-sections}, where exemplary partonic channels are shown, for which the overcounting
of resonant phase-space configurations is numerically sizable. One can see from the fifth column of the table that for these
channels the \texttt{sl-dpa} alone overshoots the off-shell prediction (shown in the second column), and the \texttt{sl-dpa-h}
category (in the sixth column) just provides a small and positive contribution. One definitely needs to subtract the double
counting of triply-resonant contributions to restore the quality of the PA individually for each partonic channel (see values of
$\delta_{\rm PA}$ in fourth column), which otherwise would be degraded to $\delta_{\rm PA}\sim42\%$, $\sim76\%$, and $\sim50\%$,
respectively, for the three channels in the different rows of the table.
\begin{table*}
 \centerline{
  \begin{tabular}{|C{3.0cm}|C{1.8cm}C{1.8cm}C{1.2cm}||C{1.5cm}C{1.5cm}C{1.5cm}|}
    \hline
    &&&&&&\\[-2.3ex]
                      &                                           &                                  &                          &   \texttt{sl-dpa} & \texttt{sl-dpa-h} & \texttt{vvv-tpa} \\
    Partonic channel  & $\sigma_{\textrm{off~shell}}\,[{\rm ab}]$ &$\sigma_{\textrm{PA}}\,[{\rm ab}]$& $\delta_{\rm PA}\,[\%]$  &   $\Delta\,[\%]$   & $\Delta\,[\%]$  & $\Delta\,[\%]$  \tabularnewline[0.7em] 
 \hline 
 &&&&&&\\[-2ex]
 $u\bar{d}\to\mu^+\nu_{\mu}s\bar{s}c\bar{c}$    & $0.1701(5)$   & $0.168(1)$   & $-1.07$   & $143.30$  & $0.23$   & $-43.53$    \tabularnewline[0.6em]
 $d\bar{u}\to\mu^+\nu_{\mu}ss\bar{c}\bar{c}$    & $0.0868(3)$   & $0.0856(9)$   & $-1.45$   & $178.36$  & $0.22$   & $-78.58$    \tabularnewline[0.6em]
 $u\bar{u}\to\mu^+\nu_{\mu}ss\bar{s}\bar{c}$    & $0.0574(1)$   & $0.0560(2)$   & $-2.38$   & $153.86$  & $0.18$   & $-54.05$    \tabularnewline[0.6em]
 \hline
  \end{tabular}
 }
 \caption{Selection of some partonic channels for which the overcounting problem described in~\refse{sec:overcounting-of-higher-resonant-contributions} is numerically sizable.
   The second and the third columns report the integrated cross sections in ab in the resolved setup for the off-shell $\sigma_{\textrm{off~shell}}$ and pole-approximated
   $\sigma_{\textrm{PA}}$ calculation, respectively. Integration
   errors are given in parentheses. The results for each partonic channel are obtained
   by summing over the two channels that are related by interchanging
   $\Pu\leftrightarrow\Pc$ and $\Pd\leftrightarrow\Ps$, and by reweighting them by the appropriate PDF factors.
   In the fourth column the quality of the PA, $\delta_{\rm PA}=(\sigma_{\rm
     PA}-\sigma_{\textrm{off~shell}})/\sigma_{\textrm{off~shell}}$, is shown
   in percentage. The last three columns
   present the percentage contribution $\Delta$ to $\sigma_{\textrm{PA}}$ of the three different PA categories used for the results of this and the following section.
 }\label{tab:pc-cross-sections}
\end{table*}

\subsection{Differential distributions}
\label{sec:differential-distributions}

In this section we study the quality of the PA for the LO EW contribution to the process
in~\refeq{eq:process_mu}
at differential level by examining some distributions which are especially relevant for VBS. All plots
have the following structure. In the main panel the LO EW
cross sections are
presented for the fully off-shell (blue curve) and pole-approximated (red curve) results,
named LO and LO$_{\rm PA}$. Differences
between the two are highlighted in a first ratio panel,
showing the prediction for the PA 
normalised to the LO EW one. In the main panel we also include some of the individual pole-approximated contributions
 used in our calculation, as described in~\refse{sec:pole-approximation-in-specific-setup},
to the full LO$_{\rm PA}$ prediction, which are then presented altogether with the same line colour in a second ratio panel.
Therein the PA is shown again
  with a red curve, used for normalisation. Then we report the sum \texttt{sl-dpa} of the semi-leptonic DPA contributions
  in~\refeqs{eq:sl-wpwp,eq:sl-wpwm,eq:sl-wpz} with a green curve, the Higgs-resonant
  contribution \texttt{sl-dpa-h} in~\refeq{eq:sl-higgs-wm} with an orange curve,
  and the sum \texttt{vvv-tpa} of TPA contributions in~\refeqs{eq:tpa-wpwpwm,eq:tpa-wpwmwm,eq:tpa-wpwmz,eq:tpa-wpzz} with a magenta curve.
  Furthermore, we subdivide the \texttt{sl-dpa} category to show
  the contributions from
  $\PW^+\PW^-$ VBS of~\refeqs{eq:sl-wpwm} in lime,
  $\PW^+\PW^+$ VBS of~\refeqs{eq:sl-wpwp} in cyan, and $\PW^+\PZ$ VBS of~\refeqs{eq:sl-wpz} in violet.
In all pictures of this section, results in the resolved and
boosted categories are shown on the left and on the right side, respectively, for
the same observables (whenever possible) or for observables that are related in the two setups.

We start our discussion by considering a set of distributions involving the two tag jets, namely the two
jets  which are supposed to originate from the scattering of the incoming QCD partons for a typical VBS signature.
More precisely, we let $\Pj_1$ and $\Pj_2$ denote the leading and subleading tag jet, respectively.

\begin{figure}
\centering
\includegraphics[width=0.5\textwidth,page=1]{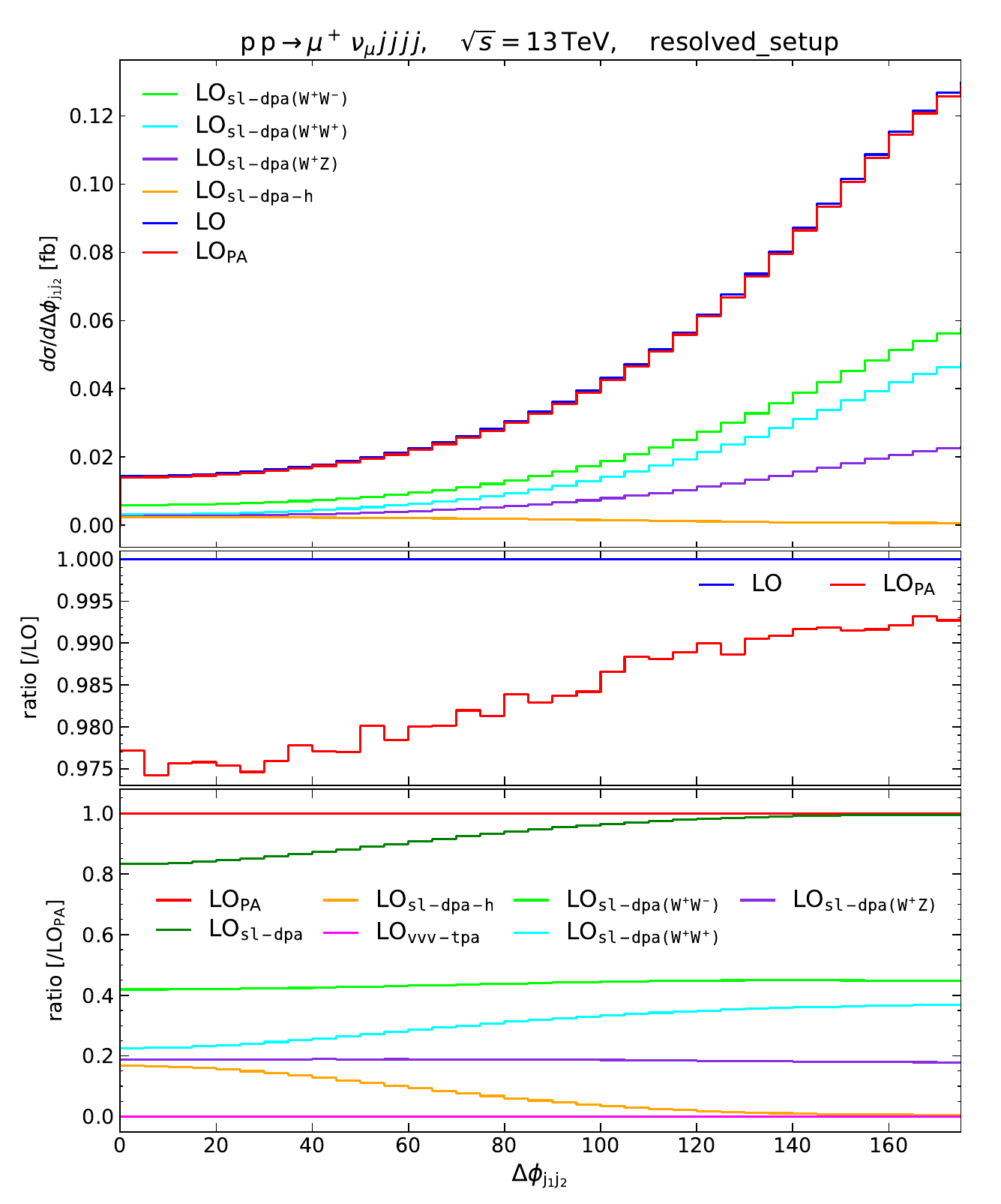}
\includegraphics[width=0.5\textwidth,page=1]{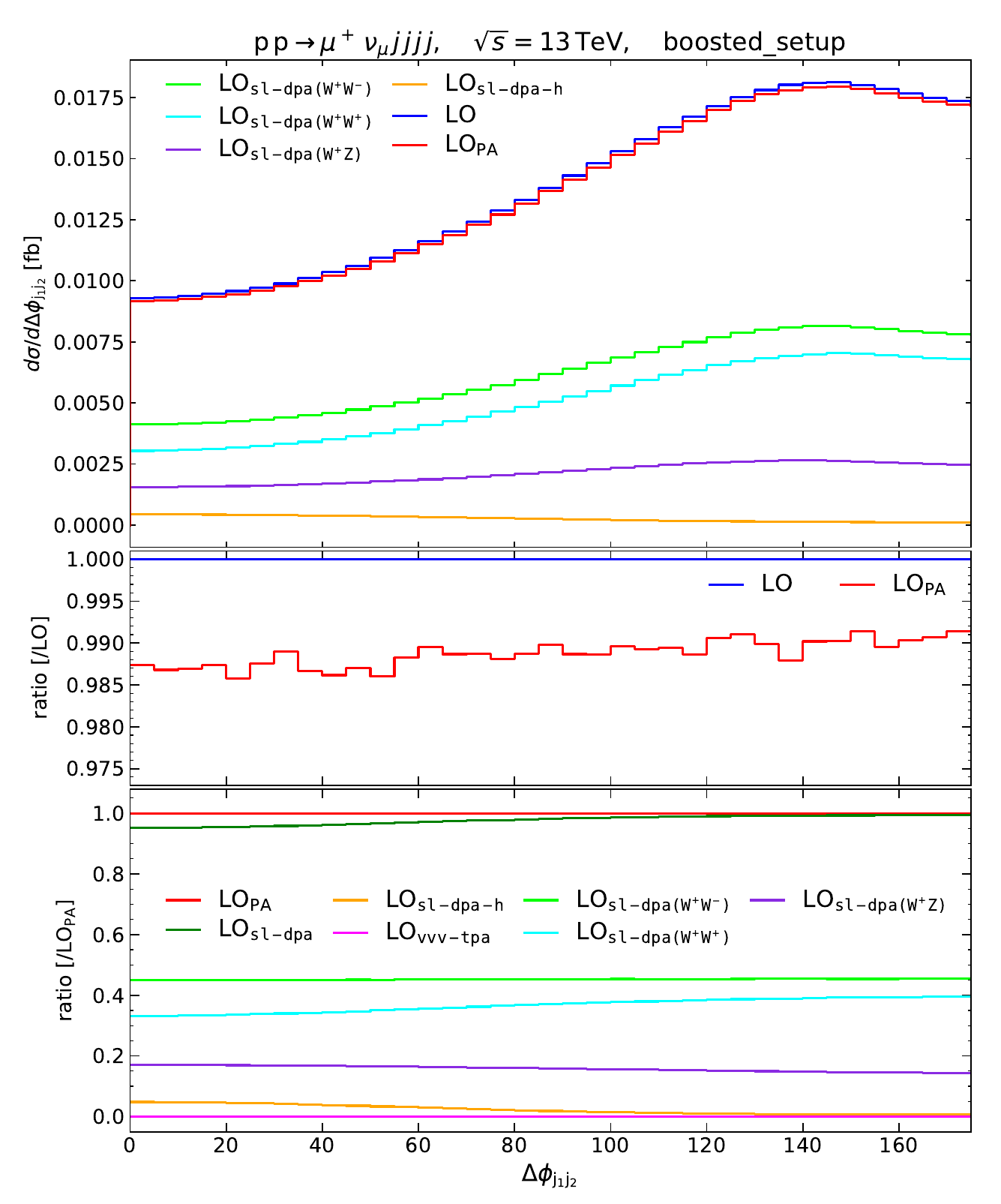}
\caption{Distributions in the azimuthal-angle difference between the two tag jets for the resolved (left) and boosted (right) categories.}
\label{fig:azimuthal-difference-tag-jets}
\end{figure}

Information on the correlation of the two tag jets are typically obtained from the cosine of their
angular distance $\cos{\theta_{\Pj_1\Pj_2}}$ and the azimuthal-angle separation
$\Delta\phi_{\Pj_1\Pj_2}$. In~\reffi{fig:azimuthal-difference-tag-jets} the distribution in the latter
observable is presented. 
For the resolved setup the cross section shows a maximum when the two jets
are maximally azimuthally separated, as already observed in
$\PW\PZ$ scattering in the fully-leptonic final state \cite{Denner:2019tmn}. 
In this region the pole-approximated result differs from the full
calculation by less than $1\%$, with slightly larger differences up to $\sim 2.5\%$ when moving to
smaller angular distances. In the boosted category, the PA
reproduces the off-shell results within roughly $1\%$ over the full spectrum,
but the peak of the distribution is reached
at values smaller than $\Delta\phi_{\Pj_1\Pj_2}=180^\circ$, namely around $\Delta\phi_{\Pj_1\Pj_2}\sim140^{\circ}$. This behaviour can be traced back to the transverse-momentum cut
of~\refeq{eq:fat-jet-cuts} on the reconstructed fat jet. Indeed, in the resolved setup the jet system
arising from the hadronically decaying vector boson is unconstrained, but for an invariant mass cut,
which does not force the jets in any direction. Conversely, in the boosted setup the fat jet
is required to have a relatively high transverse momentum that also affects the directions of the
two tag jets in the overall momentum balance and causes them to prefer kinematic configurations at
$\Delta\phi_{\Pj_1\Pj_2}$ lower than $180^\circ$, as compared to the resolved case.

The different contributions entering the pole-approximated results
show some general features, which are also present
in other observables discussed later in this paper. In line with the results for the integrated
cross sections in~\refta{tab:all-PA-cross-sections}, we observe that
the \texttt{sl-dpa} essentially accounts for the entire ${\rm LO}_{\rm PA}$ result,
especially in the boosted
regime, where the \texttt{sl-dpa-h} category is further suppressed by the transverse-momentum
cut of~\refeq{eq:fat-jet-cuts}, as discussed in the previous section. As far as the individual PAs of the \texttt{sl-dpa}
category are concerned, for most of the observables considered here the respective curves reflect
the hierarchy of the contributions already manifest at the inclusive level (see~\refta{tab:all-PA-cross-sections})
over the full spectrum, with a clear dominance of the $\PW^+\PW^-$ VBS.
Moreover, the \texttt{vvv-tpa}
result, already negligible at the integrated level, amounts to a flat correction with no visible
shape effects in all distributions shown in this manuscript. For the azimuthal-angle separation
$\Delta\phi_{\Pj_1\Pj_2}$ we can see, especially in the resolved regime, that contributions from a
resonant Higgs boson appear as expected for small azimuthal separations of the tag jets, but they completely vanish
for larger separation values.

\begin{figure}
\centering
\includegraphics[width=0.5\textwidth,page=4]{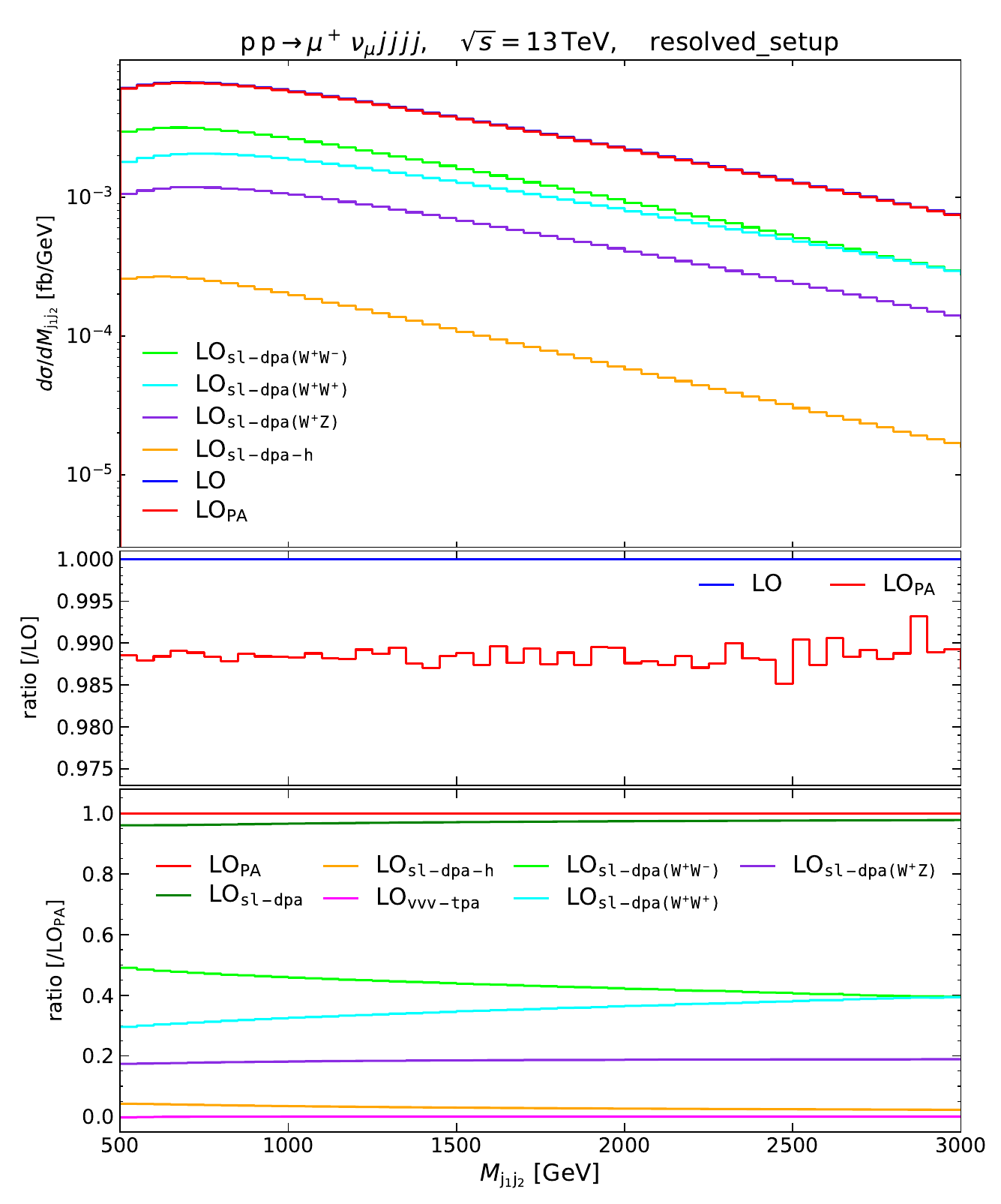}
\includegraphics[width=0.5\textwidth,page=5]{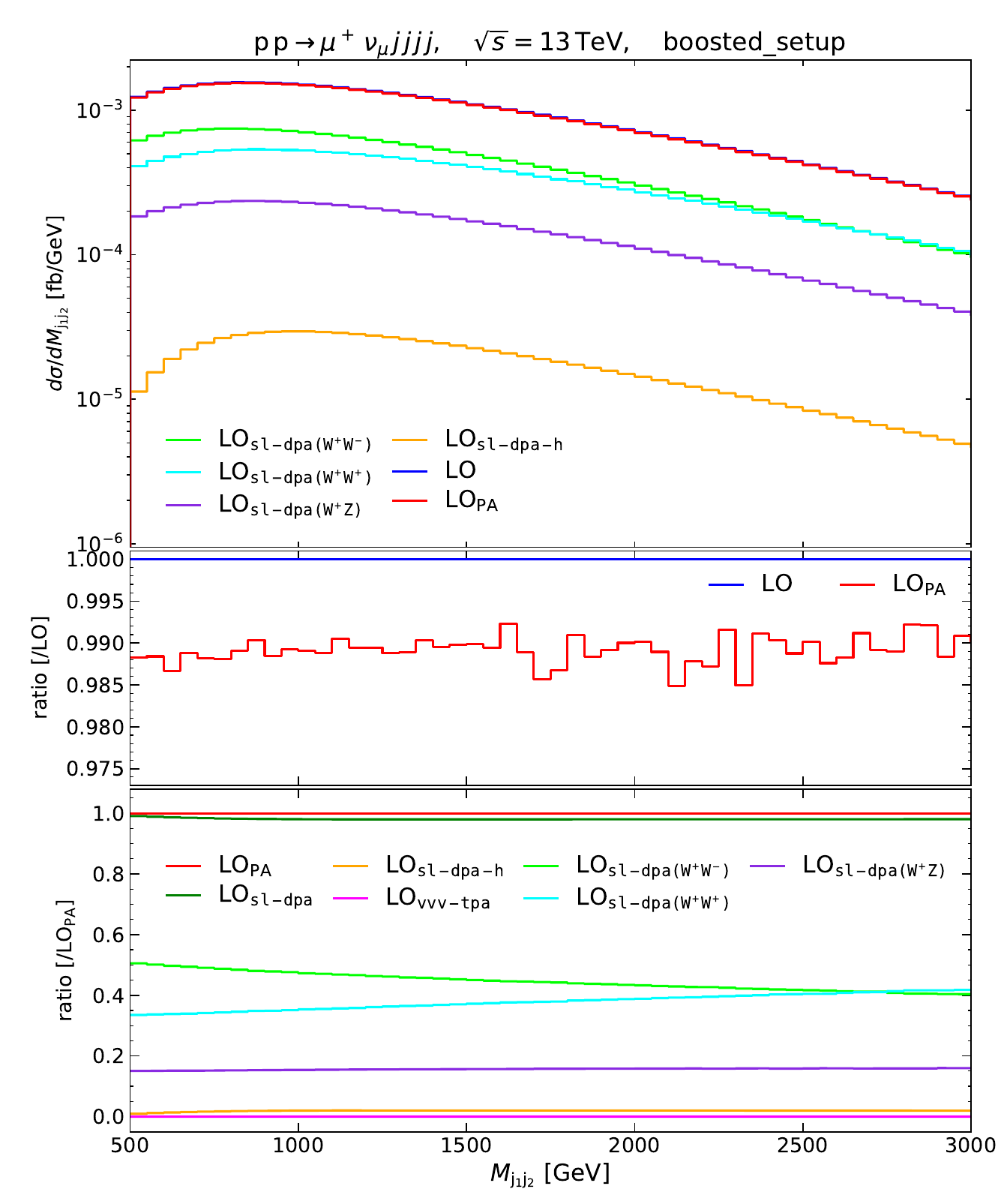}
\caption{Distributions in the invariant mass of the two tag jets for the resolved (left) and boosted (right) categories.}
\label{fig:invariant-mass-tag-jets}
\end{figure}

\begin{figure}
\centering
\includegraphics[width=0.5\textwidth,page=9]{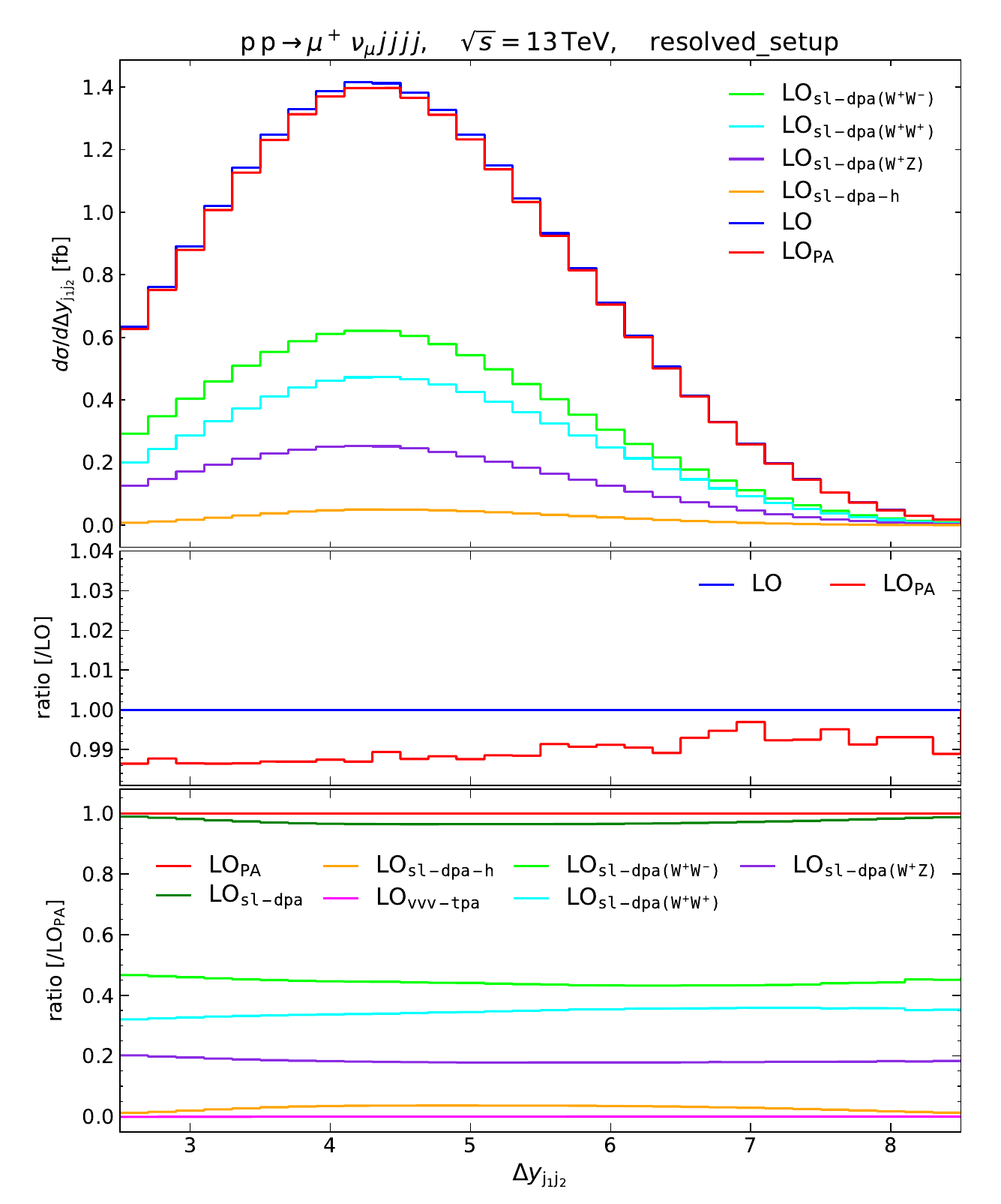}
\includegraphics[width=0.5\textwidth,page=10]{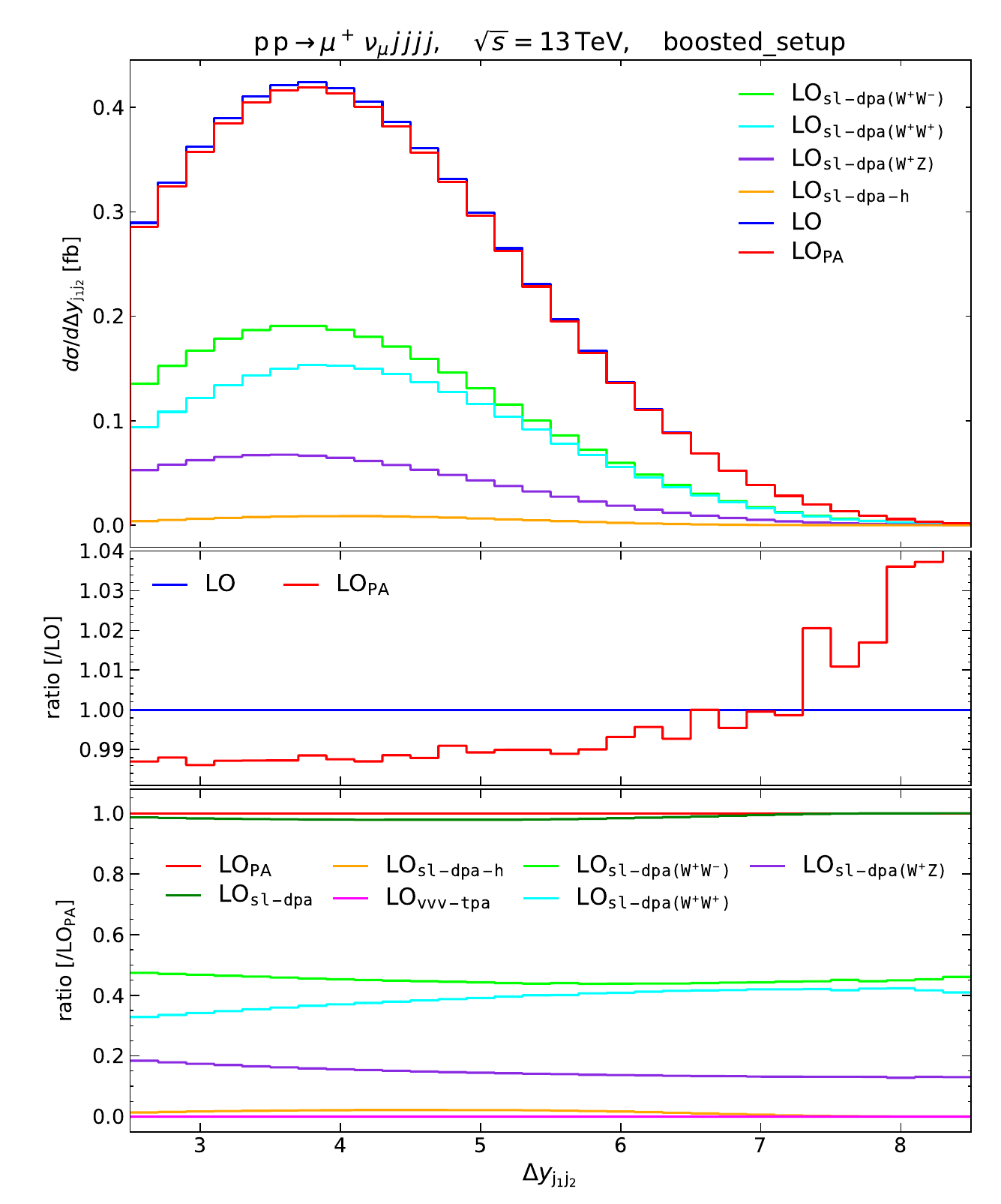}
\caption{Distributions in the rapidity difference between the two tag jets for the resolved (left) and boosted (right) categories.}
\label{fig:rapidity-difference-tag-jets}
\end{figure}

In~\reffis{fig:invariant-mass-tag-jets,fig:rapidity-difference-tag-jets} we consider two
distributions in the invariant mass $M_{\Pj_1\Pj_2}$ and in the rapidity difference
$\Delta y_{\Pj_1\Pj_2}$ of the two tag jets. 
These quantities are strongly correlated and are
typically used to enhance the VBS signature with proper cuts [see~\refeq{eq:vbscuts}].
In~\reffi{fig:invariant-mass-tag-jets} we see that, despite the different overall normalisation
factor, the results for $M_{\Pj_1\Pj_2}$ are fairly similar in the two cut categories both in terms of the shape of the
distributions and in the agreement between the LO and LO$_{\rm PA}$ curves, whose difference is always
well within $1.5\%$. This suggests that for $M_{\Pj_1\Pj_2}$ doubly-resonant topologies dominate
the cross section over the full fiducial phase space. The situation is partly different for
$\Delta y_{\Pj_1\Pj_2}$ in~\reffi{fig:rapidity-difference-tag-jets}. Starting from differences between the
off-shell and pole-approximated results of $\sim 1\%$ in the bulk of the distribution (where
the observable shows a maximum around $\Delta y_{\Pj_1\Pj_2}\sim4$),  the description of the PA
improves even further at larger values. Indeed, for large rapidity separations of the two tag
jets, VBS topologies are known to become more and more relevant, and for those configurations
doubly-resonant diagrams should capture most of the physics. In the boosted setup
this holds true up to values of $\Delta y_{\Pj_1\Pj_2}\sim7.2$, beyond
which the PA starts
overestimating the full calculation; however, this difference is not
statistically significant and within a 
Monte Carlo uncertainty of similar size.

\begin{figure}
\centering
\includegraphics[width=0.5\textwidth,page=7]{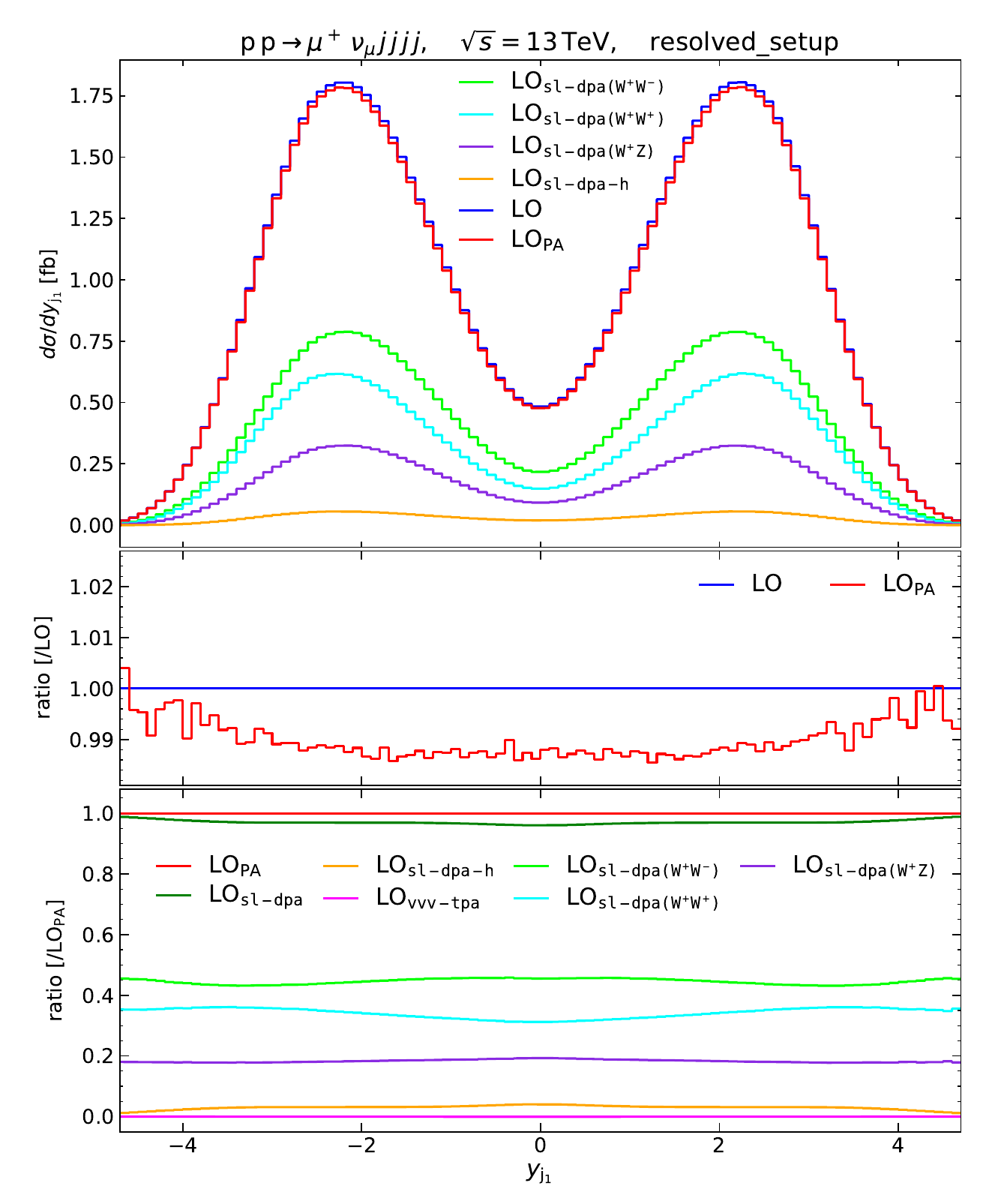}
\includegraphics[width=0.5\textwidth,page=7]{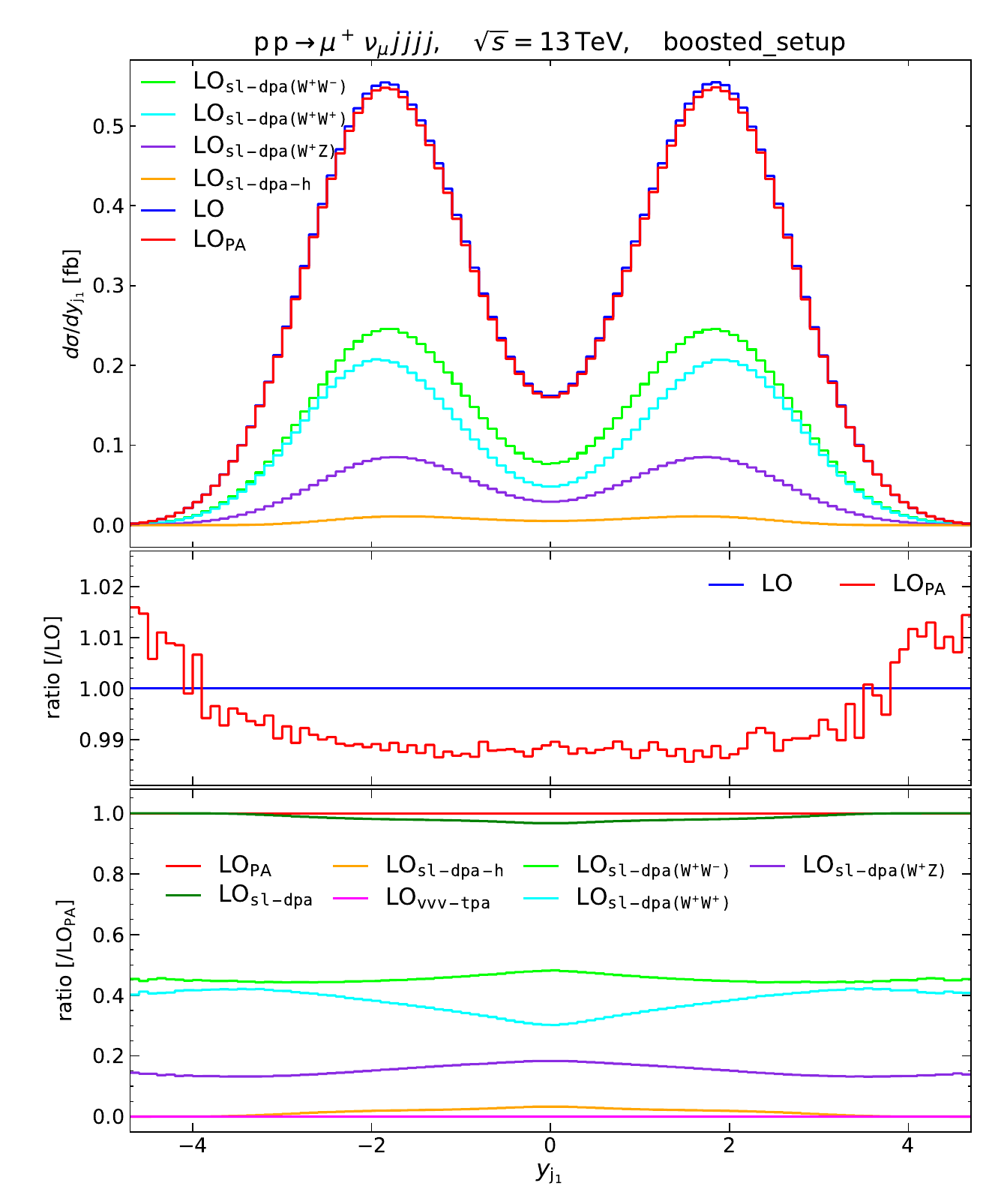}
\caption{Distributions in the rapidity of the leading tag jet for the resolved (left) and boosted (right) categories.}
\label{fig:pseudorapidity-leading-tag-jet}
\end{figure}

We continue our discussion by considering two more distributions for the leading tag jet $\Pj_1$.
In~\reffi{fig:pseudorapidity-leading-tag-jet} the distribution in its rapidity $y_{\Pj_1}$
shows the typical VBS shape, with two symmetric peaks around zero,
where a local minimum is present. 
This behaviour is nicely captured by all of the three PAs contributing to
the \texttt{sl-dpa} category. As far as the complete pole-approximated result is concerned, for the resolved setup we observe
differences in the LO and LO$_{\rm PA}$ curves of $\sim1\%$ in the central region of the distribution,
which slightly get smaller reaching $\sim 0.5\%$ around $|y_{\Pj_1}|\sim4.5$. In the boosted
setup, on the other hand, we still see deviations from this behaviour at large rapidity
values, where the pole-approximated curve overshoots the LO result.
This is, however, not statistically significant,
consistently with what observed for $\Delta y_{\Pj_1\Pj_2}$. Higgs-resonant contributions from the \texttt{sl-dpa-h} category
tend to contribute more in the central rapidity region, especially for the boosted setup, where they
amount to $3$--$5\%$ of the full LO$_{\rm PA}$.

\begin{figure}
\centering
\includegraphics[width=0.5\textwidth,page=13]{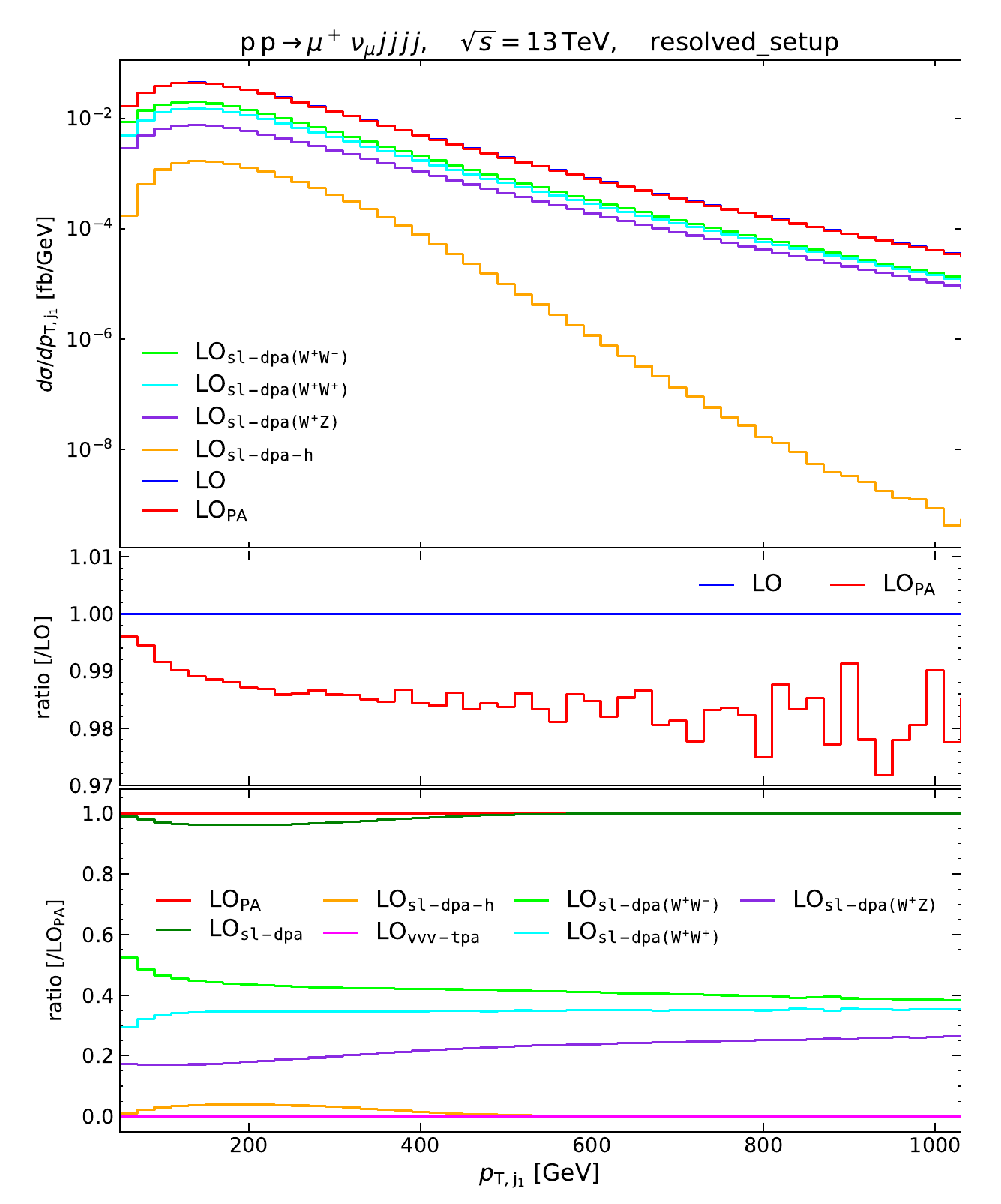}
\includegraphics[width=0.5\textwidth,page=15]{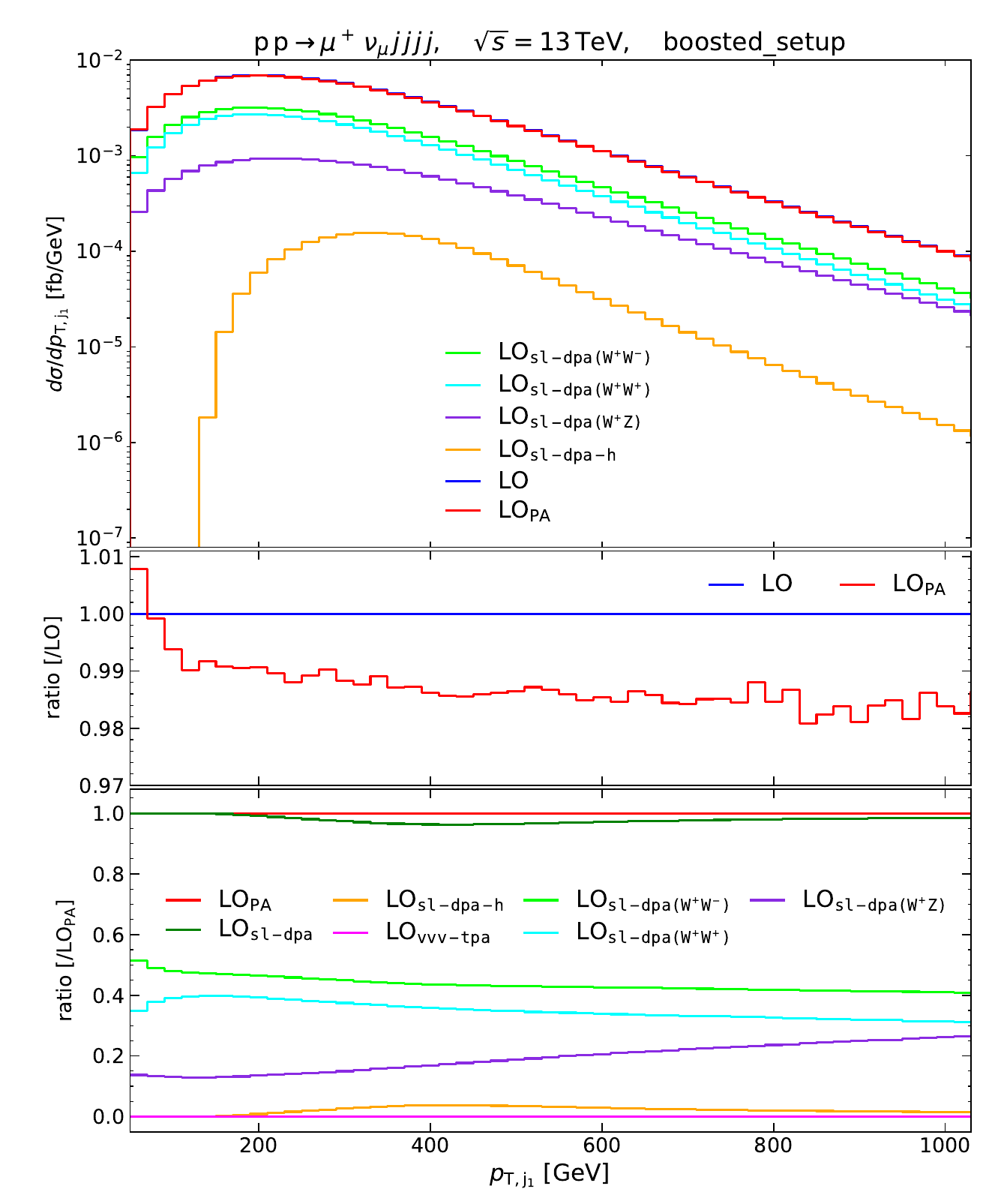}
\caption{Distributions in the transverse momentum of the leading tag jet for the resolved (left) and boosted (right) categories.}
\label{fig:transverse-momentum-leading-tag-jet}
\end{figure}
\begin{figure}
\centering
\includegraphics[width=0.5\textwidth,page=5]{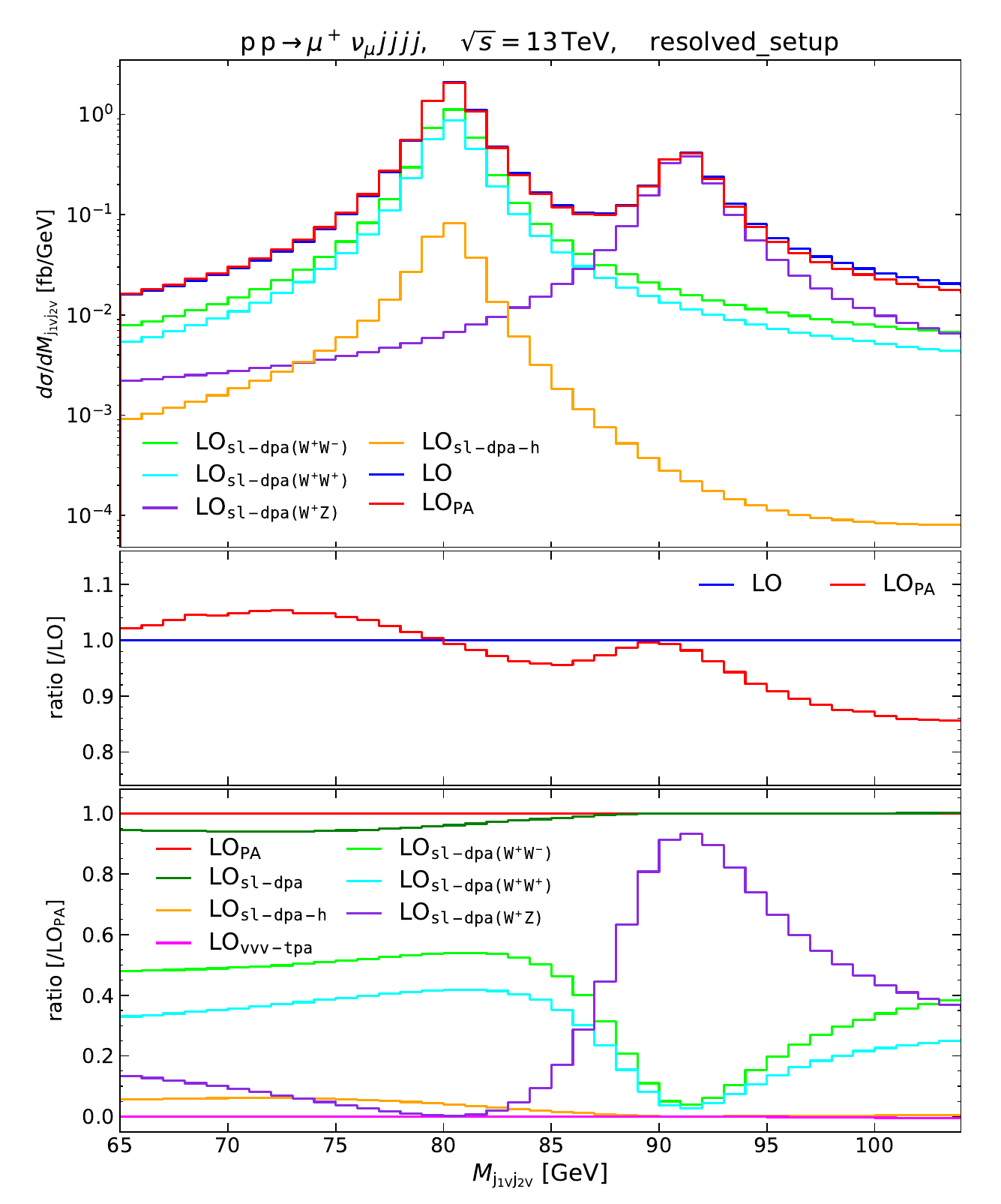}
\includegraphics[width=0.5\textwidth,page=3]{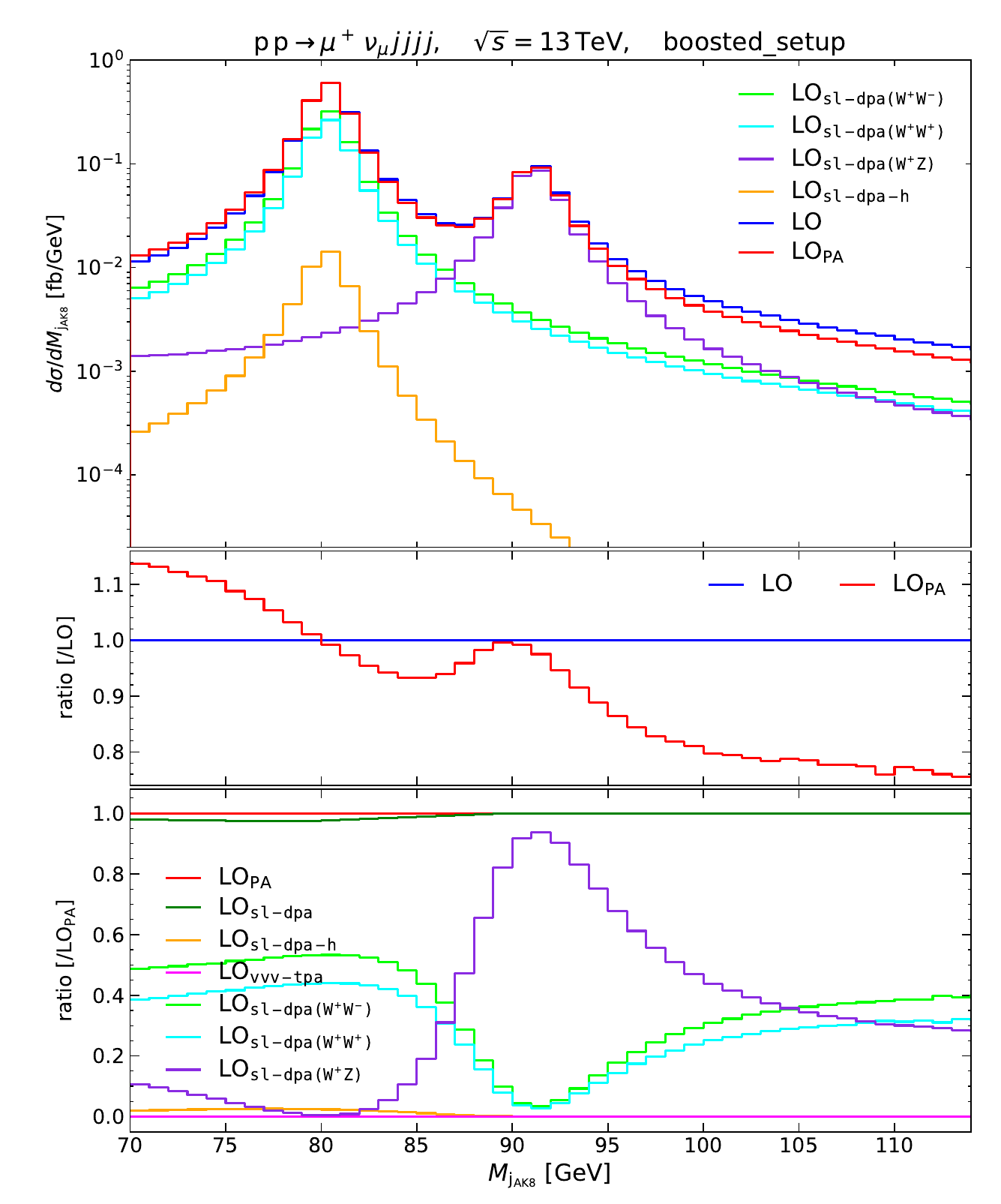}
\caption{Distributions in the invariant mass of the two jets arising from the hadronically decaying vector boson for the resolved category (left)
  and of the fat jet for the boosted category (right).}
\label{fig:invariant-mass-hadronically-decay-boson}
\end{figure}

The distribution in the transverse momentum $p_{{\rm T},\Pj_1}$ of the leading tag jet is
shown in~\reffi{fig:transverse-momentum-leading-tag-jet}. 
In the resolved category, the PA describes
very well the observable in the low-transverse-momentum region close to $p_{{\rm T},\Pj_1}\sim50\,$GeV [see cut in~\refeq{eq:leading-tag-jet-pt-cut}]
within $0.5\%$ and gets progressively
larger at high transverse momentum, even exceeding $2\%$. A similar trend is observed in the
boosted setup, with the exception of the low transverse-momentum region, where the pole-approximated
result exceeds the off-shell calculation by $1\%$.  This phase-space region is
correlated to the one of high tag-jet rapidity, but also of low invariant masses of the hadronically decaying vector boson,
discussed in the following, where a statistically significant overshooting of the PA is also visible in~\reffi{fig:invariant-mass-hadronically-decay-boson}.
For the distribution in $p_{{\rm T},\Pj_1}$, we also see that Higgs-resonant diagrams
play a role at low transverse momentum: the \texttt{sl-dpa-h} result
is peaked around $p_{{\rm T},\Pj_1}\sim180\,$GeV and $p_{{\rm T},\Pj_1}\sim400\,$GeV
in the resolved and boosted setup, respectively, where it contributes up to $5\%$  to the full
pole-approximated prediction. For higher $p_{{\rm T},\Pj_1}$, the \texttt{sl-dpa-h} curve drops much faster
than the other pole-approximated results, and its contribution to the full LO$_{\rm PA}$ quickly becomes negligible.


We now consider two observables which are related to the hadronically decaying vector boson. Its properties can be assessed
via the two non-tag AK4 jets satisfying the invariant-mass condition in~\refeqq{eq:hadronic-on-shell-condition}{a} for the resolved category (that are referred to in the following as $\Pj_{{\rm V},1}$ and $\Pj_{{\rm V},2}$ for the leading and subleading one, respectively),
or via the fat AK8 jet $\Pj_{{\rm AK}8}$, whose invariant mass is also constrained by~\refeqq{eq:hadronic-on-shell-condition}{b}, for the boosted category.
In~\reffi{fig:invariant-mass-hadronically-decay-boson} we present the distribution in the invariant mass
$M_{\Pj_{{\rm V},1}\Pj_{{\rm V},2}}$
of the two jets $\Pj_{{\rm V},1}$ and $\Pj_{{\rm V},2}$ for the resolved category, and
$M_{\Pj_{{\rm AK}8}}$ of the fat AK8 jet $\Pj_{{\rm AK}8}$ for the
boosted one.
As discussed in~\refse{sec:description-of-the-calculation}, for a semi-leptonic VBS-like signature
the pair of the two time-like vector bosons can comprise
a $\PWp\PWp$, a $\PWp\PZ$, and a $\PWp\PWm$ pair, with only one $\PWp$~boson in each pair decaying leptonically.
That explains why the distributions in  $M_{\Pj_{{\rm V},1}\Pj_{{\rm V},2}}$ and $M_{\Pj_{{\rm AK}8}}$
are enhanced both at $\MW$ and at $\MZ$, with a larger cross section for the former value.
This picture is further confirmed by the position of the single peak of the curves showing the separate
pole-approximated contributions to \texttt{sl-dpa}. When looking at the full LO$_{\rm PA}$, we see
that, around the two resonant peaks, the PA clearly provides the best description of the observable,
with deviations from the LO result of roughly $0.5\%$. Away from the two resonant regions, the
quality of the approximation is bound to degrade. Indeed, between the two peaks the PA underestimates
the off-shell result by up to $5\%$, while above the $\MZ$ pole even up to $20$--$25\%$  for far off-shell
values of $M_{\Pj_{{\rm V},1}\Pj_{{\rm V},2}}\sim M_{\Pj_{{\rm AK}8}}\sim105\,$GeV.
Below the $\MW$ pole, the situation is reversed, and the PA overestimates the correct result
by approximately $5\%$ for $M_{\Pj_{{\rm V},1}\Pj_{{\rm V},2}}\sim70\,$GeV in the resolved
setup, and by almost $15\%$ for $M_{\Pj_{{\rm AK}8}}\sim70\,$GeV in
the boosted case. We remind here that small invariant masses $M_{\Pj_{{\rm
      V},1}\Pj_{{\rm V},2}}$ and $M_{\Pj_{{\rm AK}8}}$ appear to be
correlated with small transverse momenta and invariant masses of the
tag jets.
As expected from the
PA requirements, the \texttt{sl-dpa-h} curve is also peaked at the  $\MW$ pole, but with a much smaller
cross section as compared to the \texttt{sl-dpa} case. Its relative contribution slighlty increases when moving down to the
lower off-shell tail, where for the 
resolved setup it reaches a non-negligible $5\%$ on the
pole-approximated prediction.  

\begin{figure}
\centering
\includegraphics[width=0.5\textwidth,page=15]{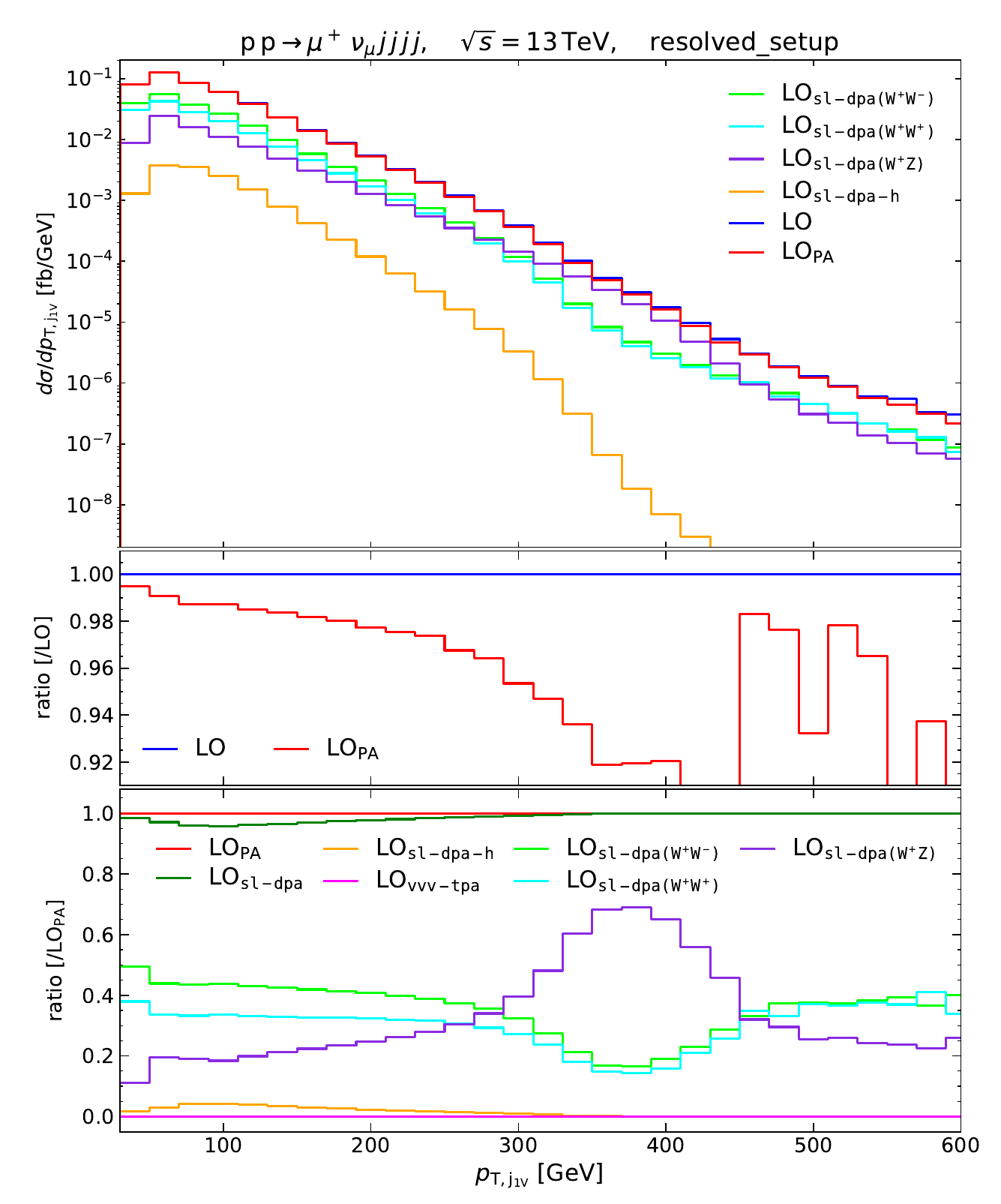}
\includegraphics[width=0.5\textwidth,page=13]{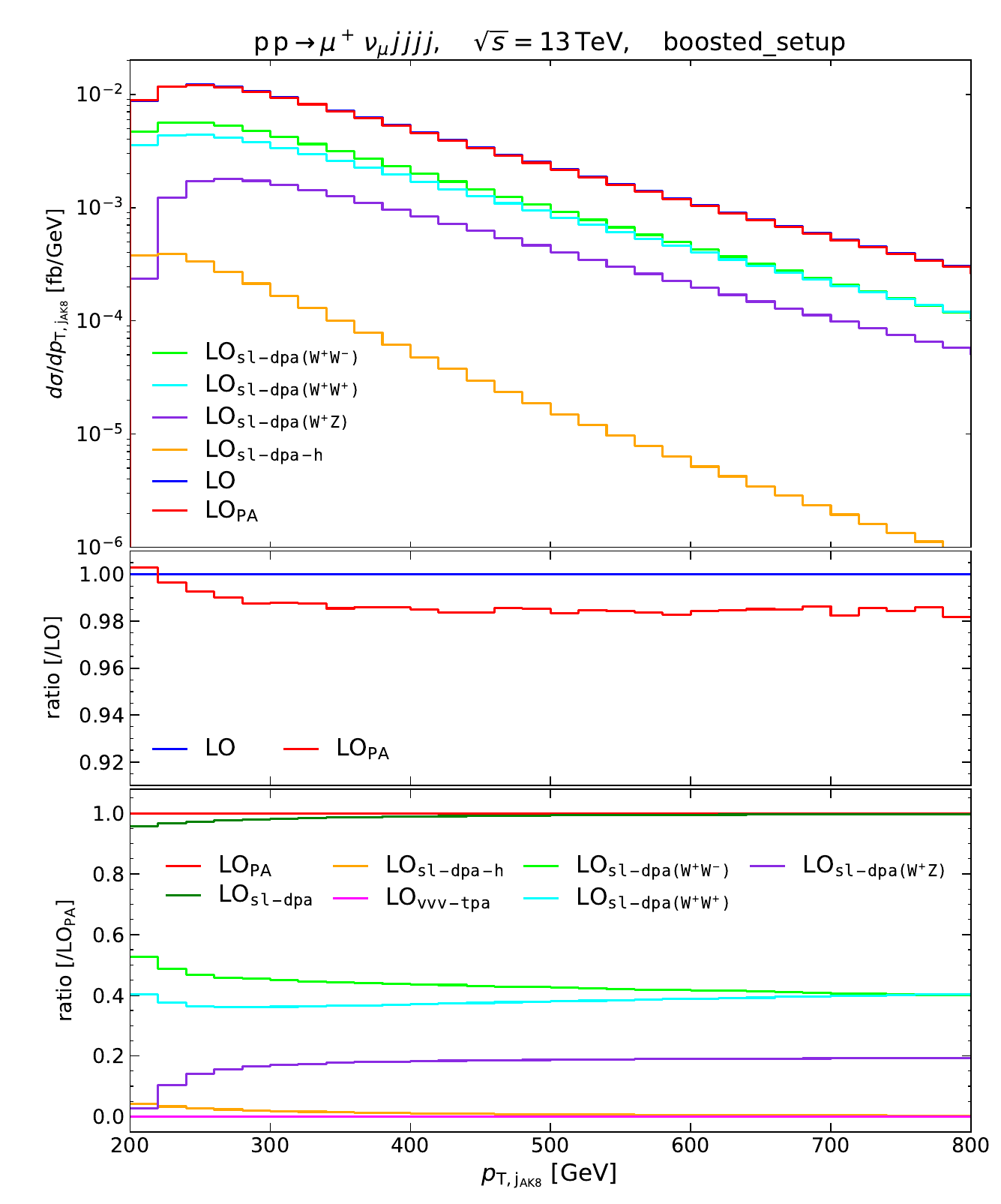}
\caption{Distributions in the transverse momentum of the leading jet arising from the hadronically decaying vector boson for the resolved category (left)
  and of the fat jet for the boosted category (right).}
\label{fig:transverse-momentum-jet-hadronically-decay-boson}
\end{figure}

In~\reffi{fig:transverse-momentum-jet-hadronically-decay-boson} the transverse momentum
$p_{{\rm T},\Pj_{{\rm V},1}}$ of the leading jet $\Pj_{{\rm V},1}$ for the resolved setup and
$p_{{\rm T},\Pj_{{\rm AK}8}}$ of the fat jet $\Pj_{{\rm AK}8}$ for the boosted one are considered.
For the resolved category, the distribution in the transverse momentum
of  $\Pj_{{\rm V},1}$ drops more steeply in the range between $300$
and $450\GeV$. Above $p_{{\rm T},\Pj_{{\rm V},1}}\sim300\GeV$ also the accuracy
of the PA becomes worse, signalling the increasing relevance of
non-resonant contributions. Actually, in this region the cuts
progressively exclude the doubly-resonant contributions, which can be
understood as follows. The invariant mass of the jet pair and their
transverse momenta are related by the equation~\cite{ParticleDataGroup:2020ssz}:
\begin{equation}
  \label{eq:kin}
  M^2_{\Pj_{{\rm V},1}\Pj_{{\rm V},2}}=2p_{{\rm T},\Pj_{{\rm V},1}}\,p_{{\rm T},\Pj_{{\rm V},2}}\,(\cosh\Delta y_{\Pj_{{\rm V},1}\Pj_{{\rm V},2}}-\cos\Delta\phi_{\Pj_{{\rm V},1}\Pj_{{\rm V},2}})\,.
\end{equation}
For small values of $\Delta y_{\Pj_{{\rm V},1}\Pj_{{\rm V},2}}$
and $\Delta \phi_{\Pj_{{\rm V},1}\Pj_{{\rm V},2}}$, we are allowed to perform a Taylor expansion to
obtain the approximate relation: 
\begin{equation}\label{eq:kin-approx}
M^2_{\Pj_{{\rm
      V},1}\Pj_{{\rm V},2}}\sim p_{{\rm T},\Pj_{{\rm V},1}}\,p_{{\rm
    T},\Pj_{{\rm V},2}}\Delta R^2_{\Pj_{{\rm V},1},\Pj_{{\rm V},2}},
\end{equation}
which provides a good approximation even for $\Delta R\sim0.8$.
The momentum of the subleading jet is bounded from below by the first
cut in \refeq{eq:AK4-jet-cuts}. For $p_{\rT,\Pj_{{\rm V},1}}\gtrsim200\GeV$,
the $\Delta R$ distance of the two AK4
jets must be at least equal to the jet-clustering radius $R=0.8$ of
the AK8 jets, since otherwise they would be clustered to an AK8 jet
and the event would not contribute to the resolved category.
Thus we obtain from \refeq{eq:kin-approx} approximative upper bounds
on the transverse momentum of the leading jet resulting from the decay
of a vector boson, specifically  $p_{\rT,\Pj_{{\rm V},1}}\lesssim336\GeV$ and
$p_{\rT,\Pj_{{\rm V},1}}\lesssim433\GeV$ for jets resulting from on-shell
$\PW$ and $\PZ$~bosons by setting $M_{\Pj_{{\rm V},1}\Pj_{{\rm V},2}}=\MW$ and
 $M_{\Pj_{{\rm V},1}\Pj_{{\rm V},2}}=\MZ$, respectively. This explains the steeper
drop of the distributions for the $\PW^+\PW^-$ and $\PW^+\PW^+$ DPAs starting
near $p_{\rT,\Pj_{{\rm V},1}}\sim310\GeV$ and for the $\PW^+\PZ$ DPA
near $p_{\rT,\Pj_{{\rm V},1}}\sim410\GeV$. Owing to the larger mass
of the $\PZ$~boson the fraction of the $\PW^+\PZ$ DPA is enhanced over
those of the $\PW^+\PW^-$ and $\PW^+\PW^+$ DPAs in the interval
$290\GeV<p_{\rT,\Pj_{{\rm V},1}}<450\GeV$. 
As far as the quality of the PA is concerned,
we see for the resolved setup that differences between the LO and LO$_{\rm PA}$ curves start
from $0.5\%$ in the low-energy region and remain below $2\%$ until values of
$p_{{\rm T},\Pj_{{\rm V},1}}\sim200\,$GeV, where then larger deviations are observed. In the boosted
category, the PA describes the $p_{{\rm T},\Pj_{{\rm AK}8}}$ fairly well and discrepancies from
the full result remain below $2\%$ even for relatively high transverse-mometum values of
$p_{{\rm T},\Pj_{{\rm AK}8}}\sim800\,$GeV. As already observed for the distribution in the
transverse momentum of the leading tag jet in the boosted category, for this observable the
PA overestimates the LO result in the first bin of the distribution, even though with deviations
below $0.5\%$.  Higgs-resonant topologies accounted for by the \texttt{sl-dpa-h} category are again relevant
only at low transverse momentum, as already discussed for the other transverse-momentum observables in~\reffi{fig:transverse-momentum-leading-tag-jet}.

\begin{figure}
\centering
\includegraphics[width=0.5\textwidth,page=10]{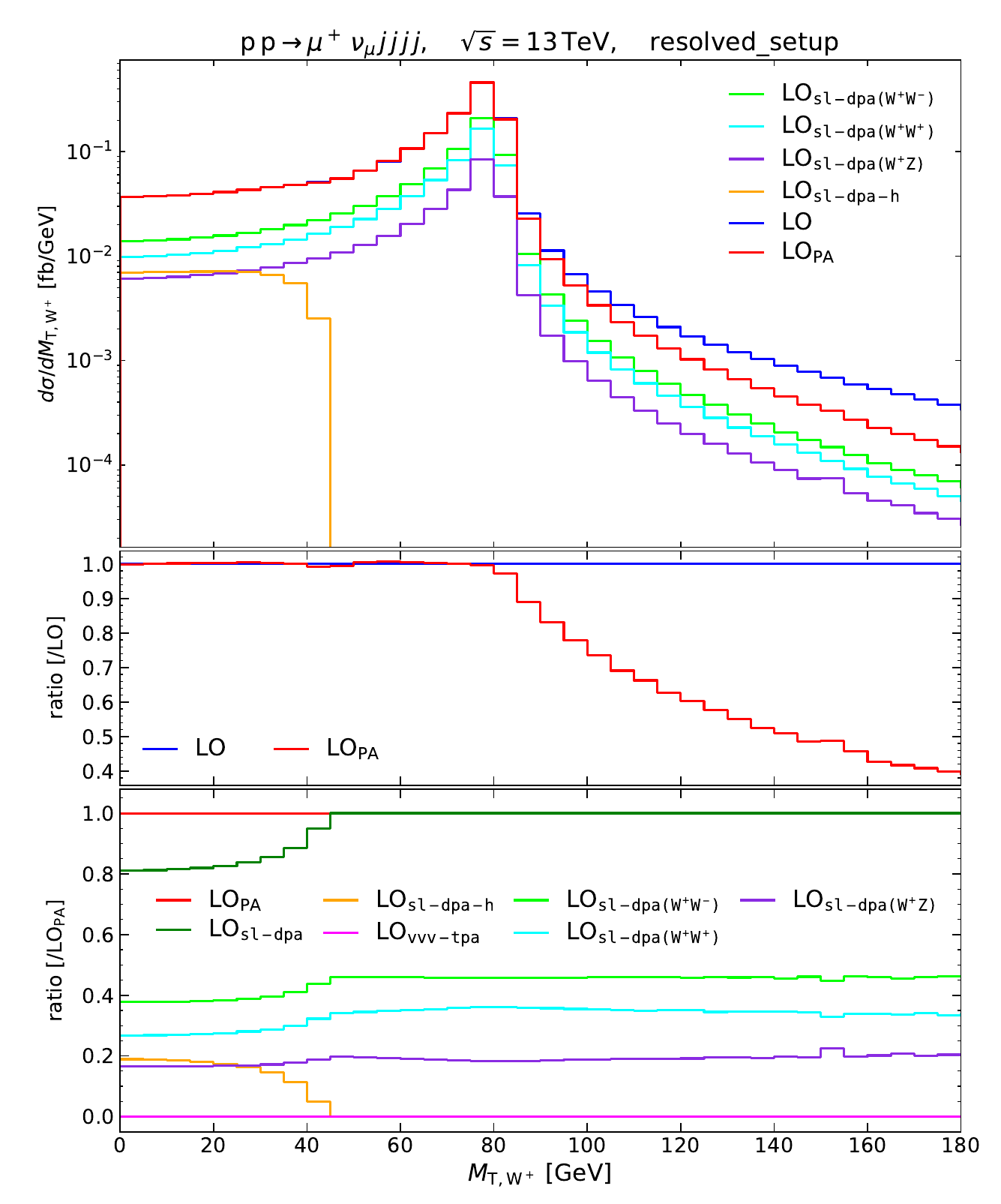}
\includegraphics[width=0.5\textwidth,page=11]{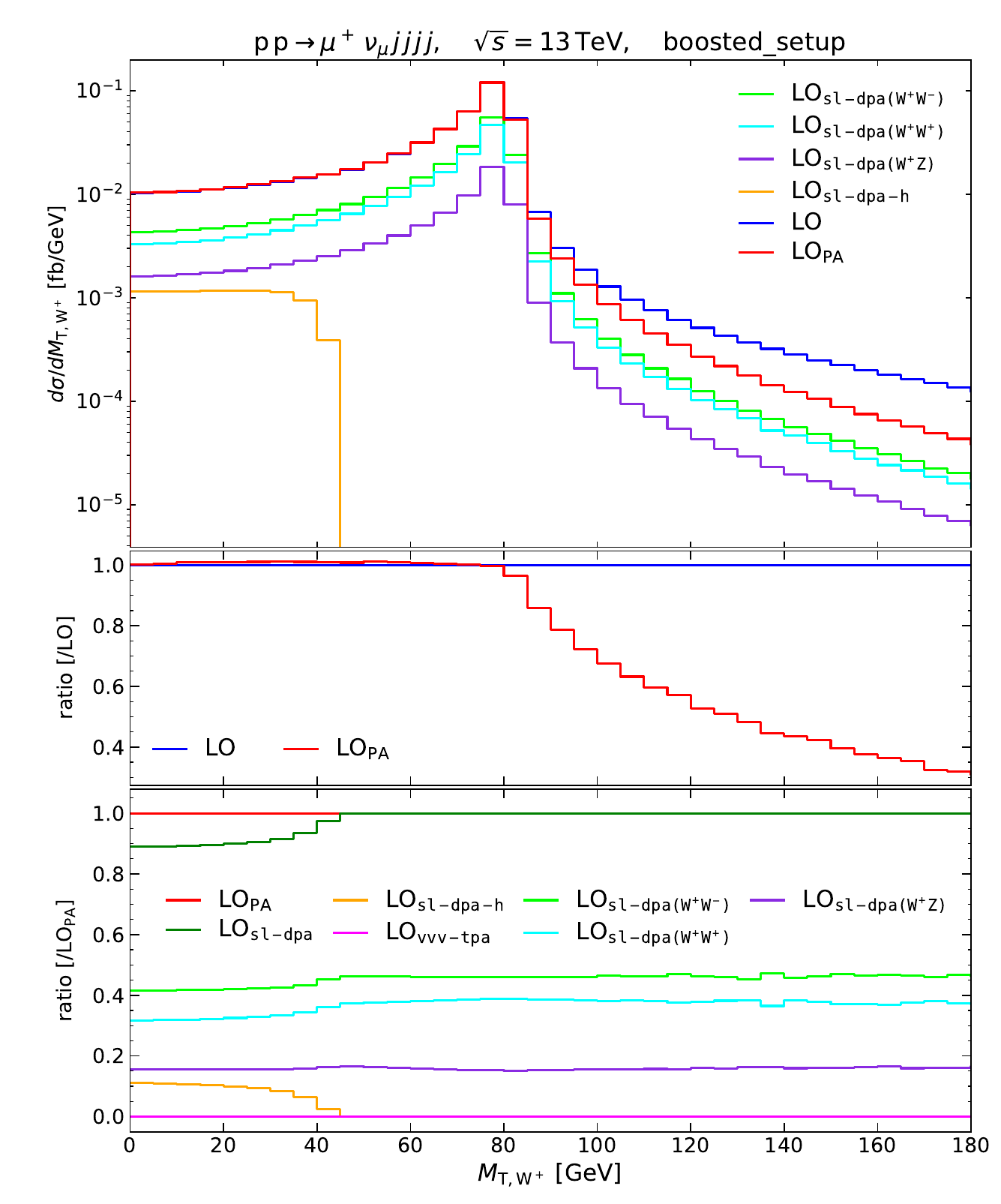}
\caption{Distributions in the transverse mass of the leptonically decaying $\PW^+$ boson for the resolved (left) and boosted (right) categories.}
\label{fig:transverse-mass}
\end{figure}
\begin{figure}
\centering
\includegraphics[width=0.5\textwidth,page=11]{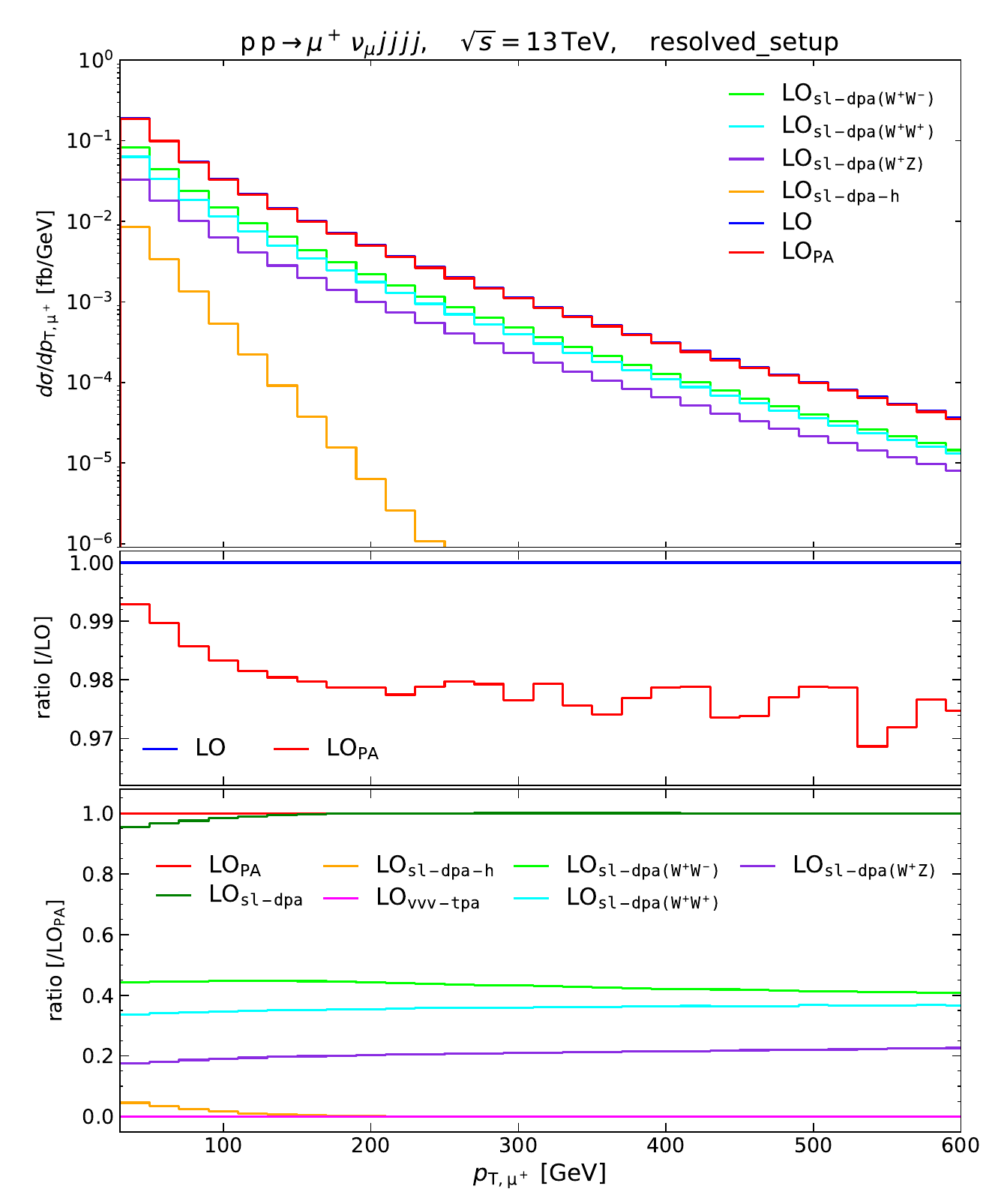}
\includegraphics[width=0.5\textwidth,page=12]{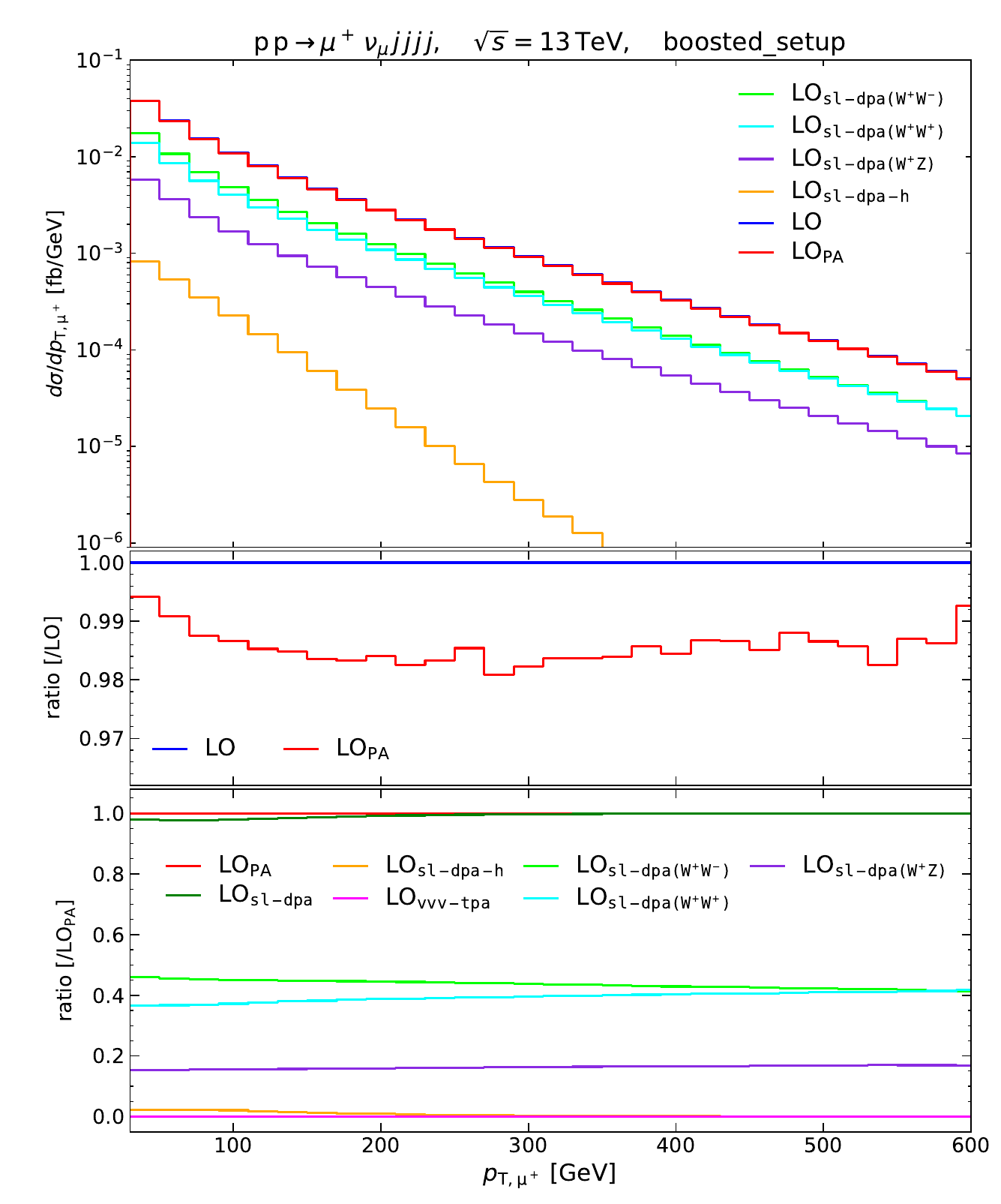}
\caption{Distributions in the transverse momentum of the anti-muon for the resolved (left) and boosted (right) categories.}
\label{fig:transverse-momentum-antimuon}
\end{figure}

We present in~\reffis{fig:transverse-mass,fig:transverse-momentum-antimuon} two
observables related to the leptonically decaying $\PW^+$ boson.
In~\reffi{fig:transverse-mass} the distribution in its transverse mass $M_\mathrm{T}^{\PW^+}$,
defined as in~\refeq{eq:transverse-mass-cut}, is shown. In both the resolved and the boosted categories,
the behaviour of the pole-approximated curves is similar: the PA describes very accurately  the off-shell
result within $1$--$2\%$ up to values of $M_\mathrm{T}^{\PW^+}\sim\MW$, where the well-known Jacobian
peak~\cite{Smith:1983aa},
which plays a central role in the measurement of the $\PW$-boson mass at the hadron collider,
is clearly visible. All the three PAs entering the \texttt{sl-dpa} category correctly reproduce this peak
and, up to a different normalisation,  have the same shape all over the $M_\mathrm{T}^{\PW^+}$ spectrum.
For values above the peak, off-shell contributions become more and more dominant
and we observe an increasing degradation of the PA, with differences reaching $50\%$ already at
$M_\mathrm{T}^{\PW^+}\sim 140\,$GeV in the resolved setup.
The \texttt{sl-dpa-h} contribution amounts to $20\%$ and $10\%$ in the resolved and boosted
setup, respectively, of the full pole-approximated prediction for low transverse-mass values, but
vanishes at $M_\mathrm{T}^{\PW^+}\sim45\,$GeV. This can be easily explained by looking closer at the
on-shell projection in~\refeq{eq:sl-higgs-wm}: since
the Higgs boson and the hadronically decaying $\PW^-$~boson are set on~shell, the invariant mass of
the leptonically decaying $\PW^+$~boson is forced to an off-shell value of roughly
$\sqrt{s_{\textrm{W}^+}}\lesssim45\,$GeV, in agreement with the end point of the distribution for the
\texttt{sl-dpa-h} contribution.

In~\reffi{fig:transverse-momentum-antimuon} we consider the
distribution in the transverse momentum of the anti-muon
$p_{{\rm T},\mu^+}$,
which owes its importance to the fact that the anti-muon is the only final-state particle
that can be measured in a clean way for a semi-leptonic VBS signature. In the very first bin, the PA
and the LO result are very close and just differ by a negative $0.7\%$ and $0.5\%$ in the resolved and boosted categories,
respectively. Indeed, in this low transverse-momentum region, and specifically
for $p_{{\rm T}\mu^+} \sim \MW/2$, a finer binning would allow to observe
 again a Jacobian peak of the distribution, in the vicinity of which
on-shell contributions dominate the cross section.
Then the agreement between the LO and LO$_{\rm PA}$ curves slightly get worse reaching $2\%$ in the
resolved setup and $1.7\%$ in the boosted one starting from roughly
$p_{{\rm T},\mu^+}\sim200\,$GeV. As for other transverse-momentum distributions,
we see also for this observable how the \texttt{sl-dpa-h} contribution impacts the overall LO$_{\rm PA}$
result only for low values, while it drops steeply when moving to higher values. Decays 
into energetic anti-muons are indeed disfavoured by the small off-shell invariant-mass value to which the PA constrains the
$\PW^+$ boson, as discussed above.

%

\begin{figure}
\centering
\includegraphics[width=0.5\textwidth,page=3]{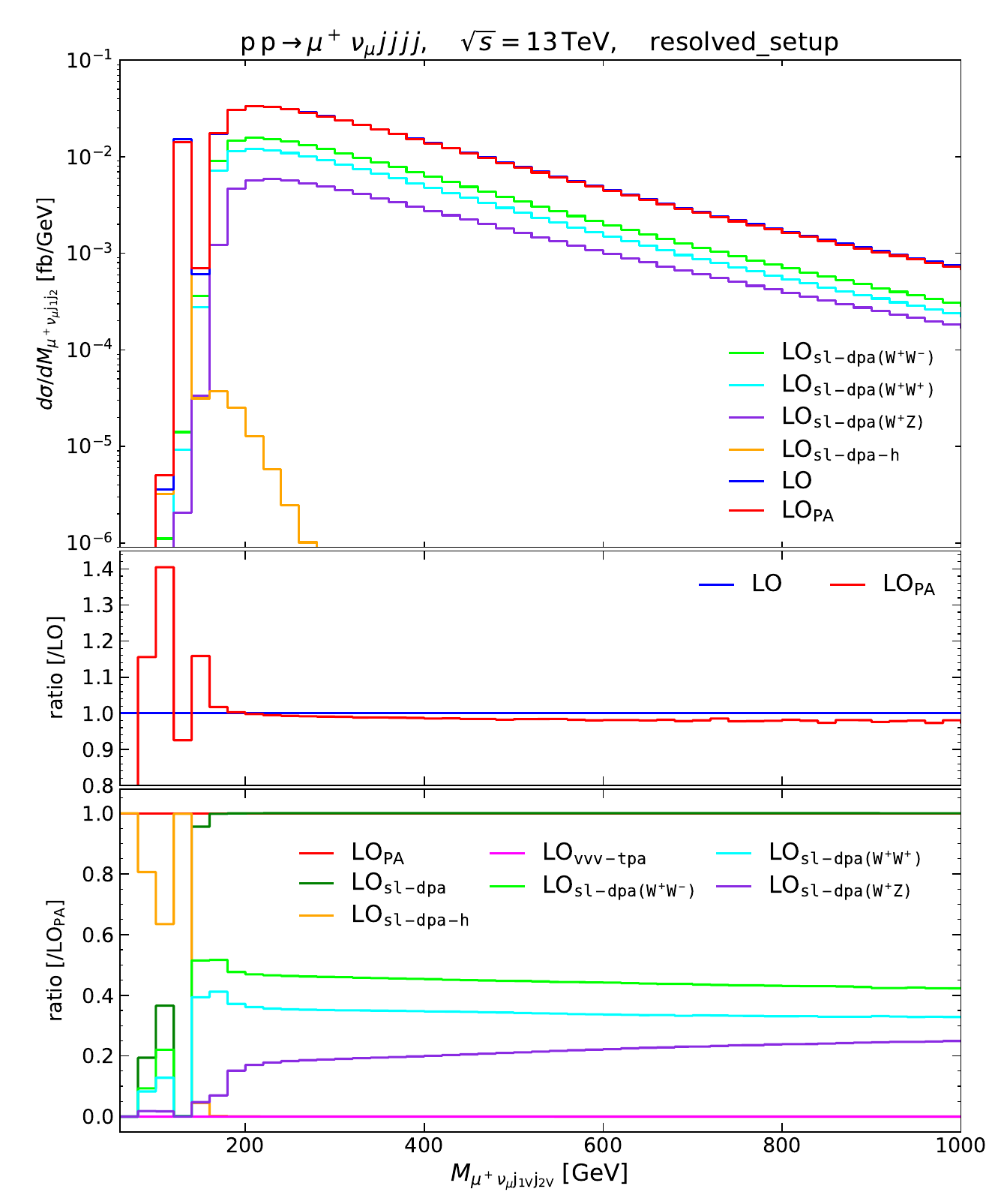}
\includegraphics[width=0.5\textwidth,page=4]{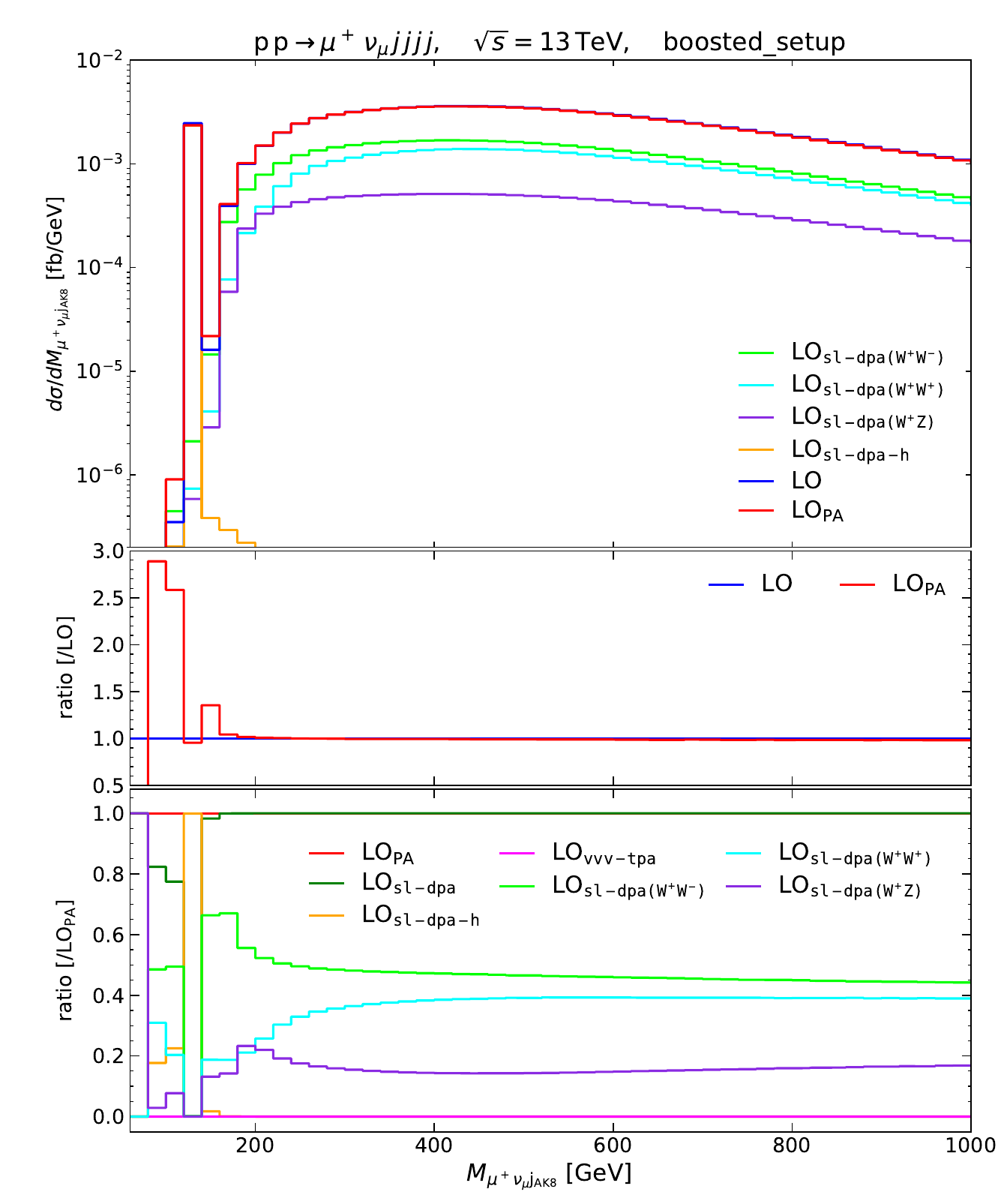}\\
\includegraphics[width=0.5\textwidth,page=18]{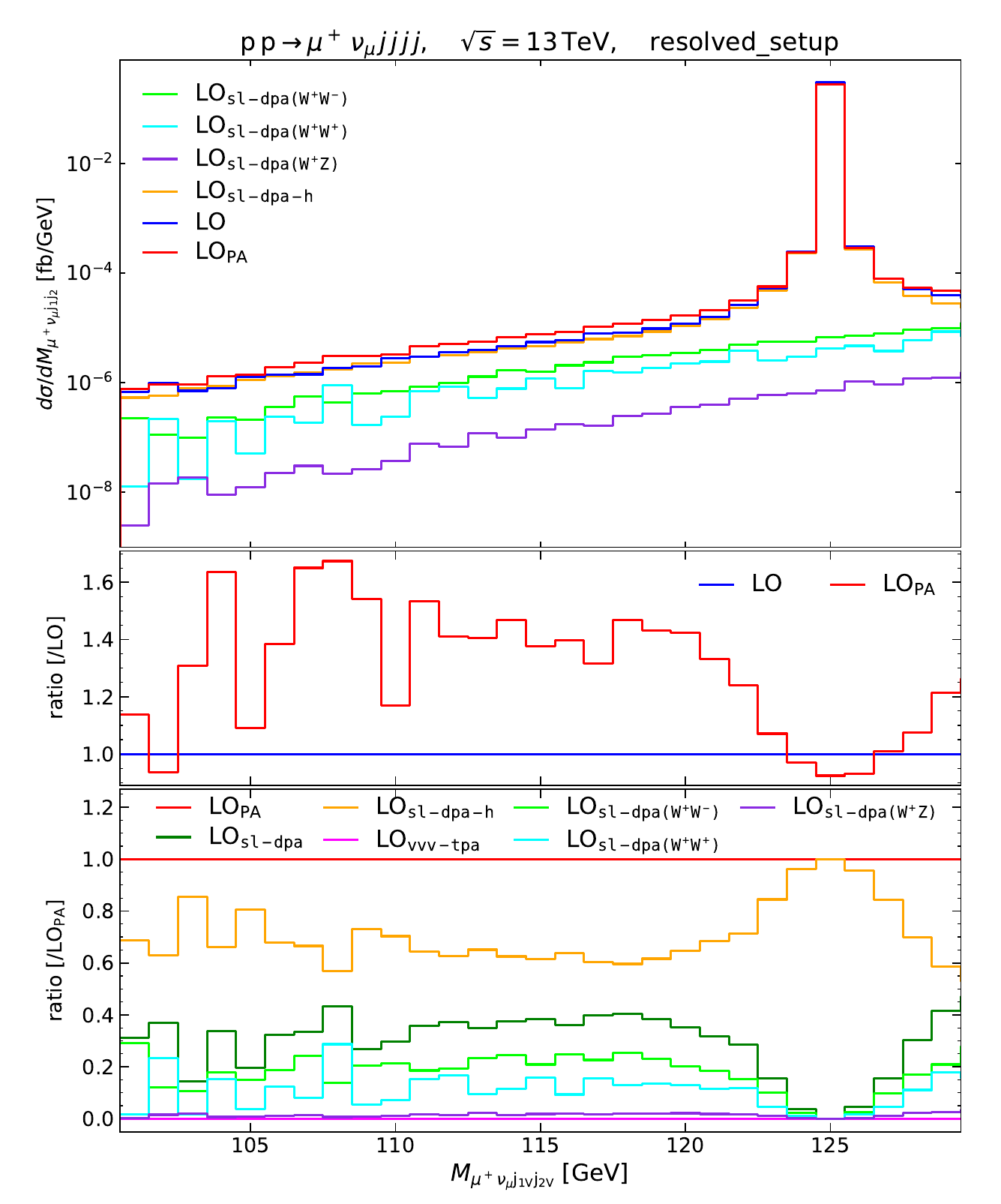}
\includegraphics[width=0.5\textwidth,page=18]{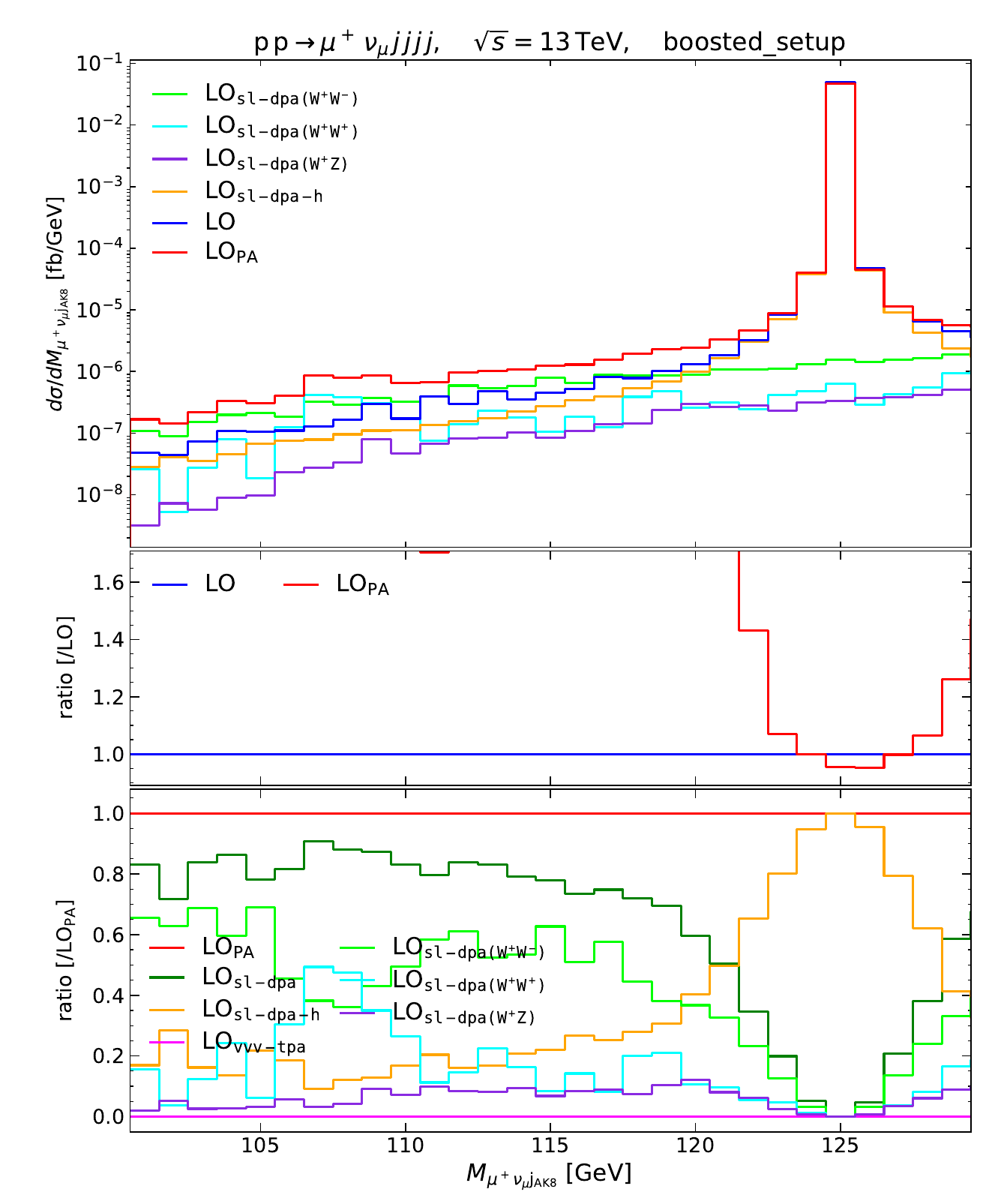}
\caption{Distributions in invariant mass of the two VBS-like vector bosons for the resolved (left) and boosted (right) categories.
  The upper plots show the distributions for values up to $1\,$TeV, while in the lower ones the same observable is presented in
a restricted range of values centered around the Higgs resonance.}
\label{fig:invariant-mass-vv}
\end{figure}

In~\reffi{fig:invariant-mass-vv} we show the distribution in the
invariant mass of the two vector bosons entering a VBS signal, of which one decays leptonically
and the other one into the two non-tag AK4 jets satisfying the invariant-mass condition
in~\refeqq{eq:hadronic-on-shell-condition}{a} for the resolved category
and into the fat AK8 jet $\Pj_{{\rm AK}8}$ for the boosted category.
We refer to the observable as $M_{\mu^+\nu_\mu\Pj_{{\rm V},1}\Pj_{{\rm V},2}}$ and
$M_{\mu^+\nu_\mu\Pj_{{\rm AK}8}}$ for the resolved and boosted category, respectively.
The same observable is presented in~\reffi{fig:invariant-mass-vv} twice: in the upper part the distributions
are shown for values up to  $1\,$TeV, while in the lower part a restricted range of values is selected. 
Above roughly $2\MW$, the off-shell curves are well described by the PA, whose accuracy remains below
 $2\%$ even for very high invariant-mass values. On the other hand, below $2\MW$ both categories
 show how the PA substantially fails in providing a good description of the observables. This can be
 better seen by looking at the lower plots of~\reffi{fig:invariant-mass-vv}. The phase-space regions
 around the Higgs resonance, i.e.\ $M_{\mu^+\nu_\mu\Pj_{{\rm
       V},1}\Pj_{{\rm V},2}}\sim\MH$ and
 $M_{\mu^+\nu_\mu\Pj_{{\rm AK}8}}\sim\MH$, are entirely captured by the \texttt{sl-dpa-h} category,
 which still deviates from the off-shell result by a negative $8\%$ and $5\%$, respectively, in the
 two setups. This difference results from singly-resonant
 contributions, i.e.\ contributions where the Higgs~boson is resonant
 but all other bosons are allowed to be off~shell.%
\footnote{We explicitly verified this statement by comparing the DPA of the dominating
  partonic channel $\Pd\Pu\to
  \PH(\mu^+\nu_\mu\PW^-(\bar\Pc\Ps))\Pu\Pd$ with
  the single-pole approximation for the Higgs boson, i.e.\ $\Pd\Pu\to \PH(\mu^+\nu_\mu\bar\Pc\Ps)\Pu\Pd$. The differences of $-8\%$ and $-5\%$ in the resolved and boosted setup, respectively,
  are larger than $\Gamma_\PW/\MW$ and explained by singly-resonant contributions.
  In the DPA the hadronically decaying $\PW^-$ boson is on shell, and thus the invariant of the $\PW^+$ boson is restricted by $\hat{s}_{\PW^+} \leq (M_\PH - M_\PW)^2$ and far away from $M_\PW^2$.
  In a PA where the $\PW^-$ boson is not set on shell the previous restriction is relaxed, and allows the $\PW^+$-boson propagator to get closer to its pole, without the $\PW^-$~boson being too far off shell.
  }
Then, for yet lower invariant-mass values,
 the pole-approximated predictions become even worse and overestimate the off-shell results by
 several hundred percent in the boosted setup, as expected when probing genuinely non-resonant regions of phase space.

\begin{figure}
\centering
\includegraphics[width=0.5\textwidth,page=17]{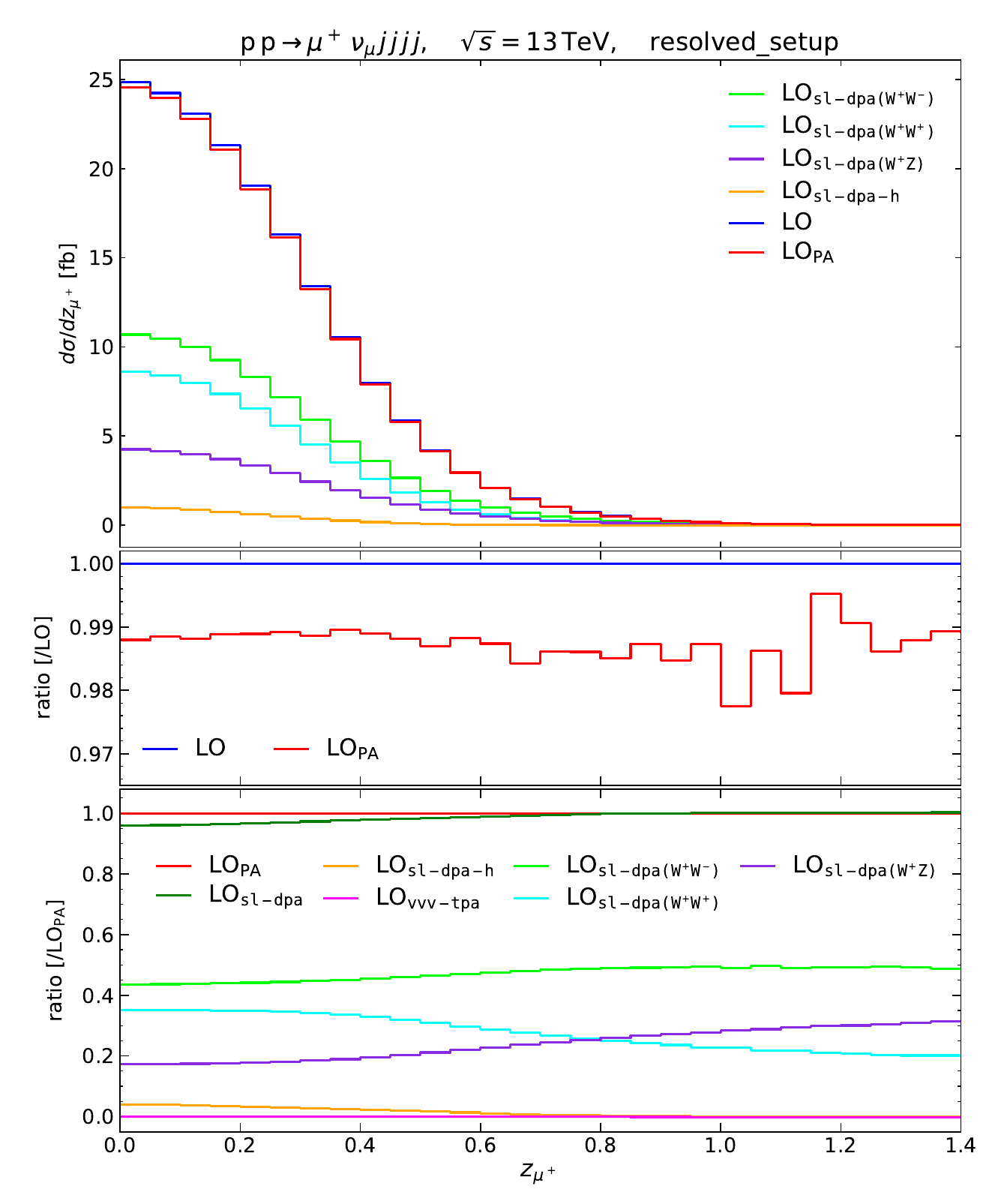}
\includegraphics[width=0.5\textwidth,page=17]{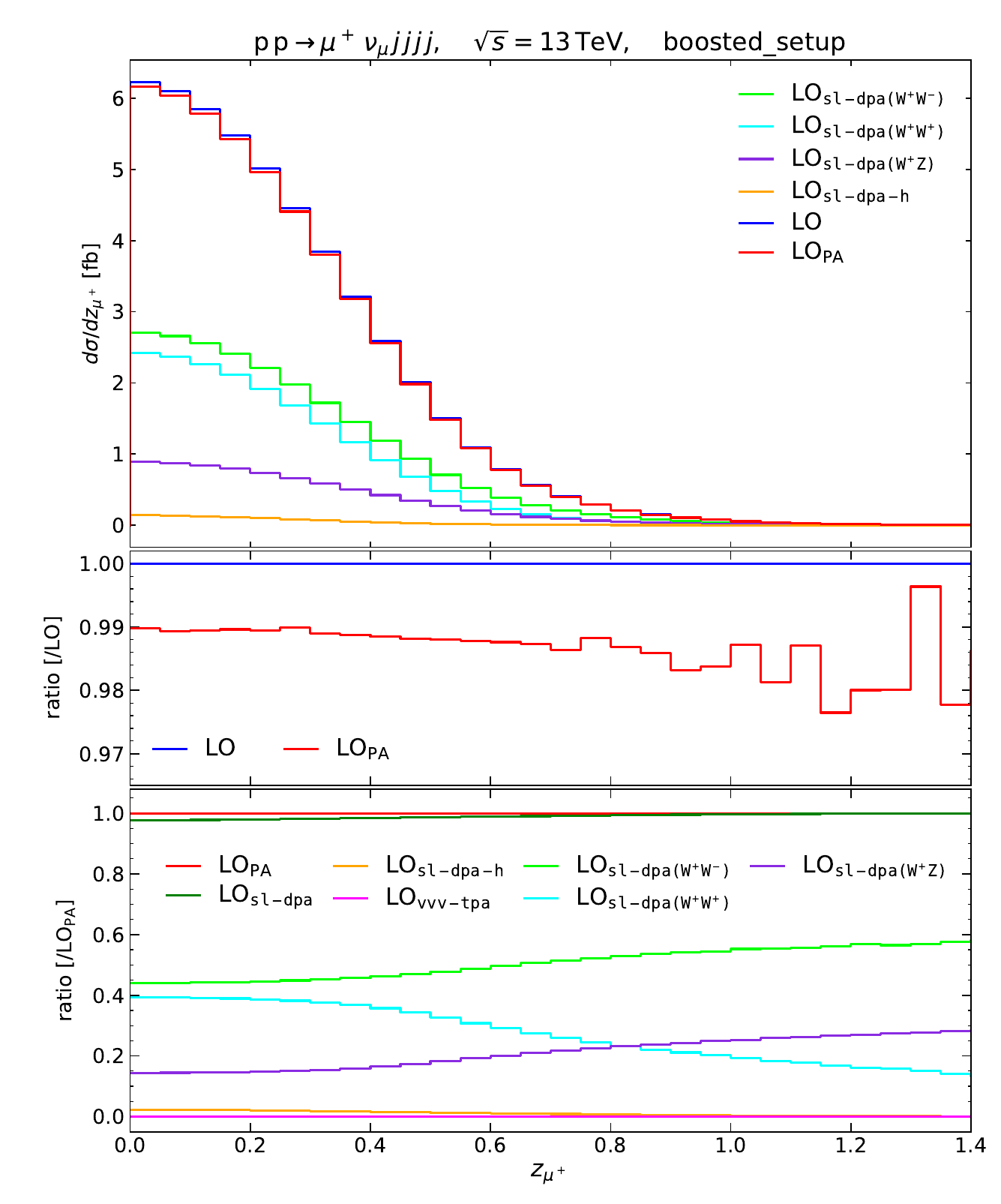}
\caption{Distributions in the Zeppenfeld variable of the charged lepton for the resolved (left) and boosted (right) categories.}
\label{fig:zeppelfeld-variable}
\end{figure}

In~\reffi{fig:zeppelfeld-variable} the distribution in the Zeppenfeld variable $z_{\mu^+}$~\cite{Schissler:2013nga,CMS:2017fhs,CMS:2021qzz} 
for the anti-muon, defined as
\begin{equation}
  z_{\mu^+} = \frac{y_{\mu^+} - \frac{1}{2}\,(y_{\Pj_1}+y_{\Pj_2})}{|y_{\Pj_1}-y_{\Pj_2}|}
  \end{equation}
is shown. 
As is clear from its definition, this observable describes the
relative position in rapidity of the charged lepton with respect to the tag jets. For $z_{\mu^+}=0$, where
the distribution is peaked, the anti-muon rapidity is $y_{\mu^+}=\frac{1}{2}\,(y_{\Pj_1}+y_{\Pj_2})$,
i.e.\ equal to the arithmetic average of the two tag-jet rapidities. This goes along with the picture that, for VBS-like
signatures, the charged lepton is produced between the two tag jets. For values away from zero (in~\reffi{fig:zeppelfeld-variable}
we just show positive values of $z_{\mu^+}$) the distribution steeply decreases, being already strongly suppressed
at values of $z_{\mu^+}\sim0.5$, which correspond to $y_{\mu^+}\sim y_{\Pj_1}$ or $\sim y_{\Pj_2}$
(if $y_{\Pj_1}>y_{\Pj_2}$ or  $y_{\Pj_2}>y_{\Pj_1}$, respectively),
namely to the anti-muon being close to the tag jet with the larger
rapidity in the event. Comparing the PA with the full
calculation, we see that the doubly-resonant contributions can account for most of the physics over the full
range of the observable, with deviations from the off-shell results
constantly around $1\%$, where the distribution is sizeable. If we look at the separate
PAs, we notice that the relative importance of the \texttt{sl-dpa-h}
category is significant only for small values of $z_{\mu^+}$.
Around $z_{\mu^+}\sim 0.85$ we also observe that the $\PW^+\PZ$ contribution to the \texttt{sl-dpa} category exceeds the $\PW^+\PW^+$ one.

\section{Conclusions}
\label{sec:conclusions}

In this work we studied the process $\Pp\Pp \to \Plp \Pnulm + 4\Pj$, which includes semi-leptonic
VBS signatures as well as various sources of irreducible background. We have performed a full off-shell
LO calculation for three perturbative contributions, namely $\mathcal{O}(\alpha^6)$,
$\mathcal{O}(\as\alpha^5)$, and $\mathcal{O}(\as^2\alpha^4)$. All resonant and non-resonant effects,
together with interference terms, have been exactly retained. All partonic channels including the
photon-induced ones have been considered with the exception of the bottom-initiated ones.
These terms are indeed suppressed by the bottom PDFs. We also excluded channels with
bottom quarks in the final state, assuming a perfect $\Pb$-jet tagging and veto, to avoid contaminations
from processes with a resonant top~quark. For all three LO contributions we presented results
for the inclusive cross sections. For the purely electroweak LO, i.e.\ $\mathcal{O}(\alpha^6)$,
which is the only one allowing
for a genuine VBS signal, we carried out a more detailed study by inspecting many differential
distributions. All of the results of this paper have been obtained in two different fiducial
regions, namely the \emph{resolved} and \emph{boosted} setups,
which are inspired by recent ATLAS and CMS studies \cite{ATLAS:2018tav,CMS:2021qzz} and
are tuned to semi-leptonic VBS searches.

The calculation of the LO electroweak contribution has also been performed in the double-pole
approximation. To our knowledge, this is the first time a pole approximation is applied to such
a complicated process, which required to account for many new features and issues. A systematic
treatment of semi-leptonic VBS in the pole approximation has been devised, which is able
to deal with nested decays in a fully general way. For that purpose, a new and flexible on-shell
projection has been developed. To achieve a
proper description of all partonic channels at the inclusive level, another crucial problem
to be considered has been the subtraction of
the overcounting of resonant  regions by overlapping pole approximations.

This powerful machinery has then been applied to semi-leptonic VBS for the two fiducial setups of
interest. We carefully described how our formalism simplifies in this case,
and we outlined some general criteria that should be taken into account when applying
our systematic pole approximation in a fiducial region defined by a specific set of cuts. Our pole-approximated results
describe the inclusive cross section very well, by underestimating the full off-shell calculation
by only roughly $1\%$ both in the resolved and boosted setup. To reach this level of agreement the inclusion of
doubly-resonant contributions involving an on-shell Higgs boson turned out to be essential. At the differential
level our LO double-pole approximation behaves as expected. For observables that are
inclusive in the decay products of the resonant vector bosons,
as the ones related to tag jets, the pole-approximated result differs from the full off-shell one
by $1$--$3\%$ over the entire spectrum. For other distributions, like the one in the invariant mass
of the hadronically decaying vector boson or the transverse mass of the leptonically decaying $\PW^+$ boson,
the pole approximation provides a perfect description of the resonant
regions, except near the Higgs resonance where deviations of
$5$--$8\%$ appear owing to enhanced singly-resonant contributions. As one would expect, larger deviations from the complete
off-shell calculation are observed  where non-resonant configurations become sizeable,
namely away from resonant peaks or in the high-energy tails of some distributions.

With this work we have obtained an independent calculation for semi-leptonic VBS,
that up to now  has received little attention from the theory community, but that might soon become
a relevant piece of the rich search programme at the LHC. We have also provided an entirely new study
of the capability of a LO pole approximation to capture the physics of such a process, whose high-multiplicity and non-trivial
resonance structure render the calculation challenging.
Our results are a crucial step forward in view of applications
of the pole approximation to further LHC processes, and pave the way to extensions of our formalism at NLO,
which is the minimal accuracy requirement of
any theoretical calculation nowadays.

\appendix

\acknowledgments

We acknowledge financial support by the German Federal Ministry for Education and Research (BMBF) under contract no.\ 05H21WWCAA and the German Research Foundation (DFG) under reference number DE 623/8-1.

\bibliography{inspire-bibliography}{}
\bibliographystyle{JHEPmod}

\end{document}